# Exchange Rate Predictability in a Changing World[*]

Joseph P. Byrne[a], Dimitris Korobilis[b], and Pinho J. Ribeiro[c]

February 14, 2014

**Abstract**

An expanding literature articulates the view that Taylor rules are helpful in predicting exchange rates. In a changing world however, Taylor rule parameters may be subject to structural instabilities, for example during the Global Financial Crisis. This paper forecasts exchange rates using such Taylor rules with Time Varying Parameters (TVP) estimated by Bayesian methods. In core out-of-sample results, we improve upon a random walk benchmark for at least half, and for as many as eight out of ten, of the currencies considered. This contrasts with a constant parameter Taylor rule model that yields a more limited improvement upon the benchmark. In further results, Purchasing Power Parity and Uncovered Interest Rate Parity TVP models beat a random walk benchmark, implying our methods have some generality in exchange rate prediction.

[*]Corresponding author at: Department of Economics, University of Glasgow, UK. Email address: dimitris.korobilis@glasgow.ac.uk, Tel: +44 (0)141 330 2950. Fax.: +44 (0)141 330 4940.
[a] Department of Economics, Heriot-Watt University, Edinburgh, UK.
[b] Department of Economics, Adam Smith Business School, University of Glasgow, Glasgow, UK.
[c] Department of Economics, Adam Smith Business School, University of Glasgow, Glasgow, UK.

# 1. Introduction

Academics and market practitioners have long sought to predict exchange rate fluctuations. A long held view, initiated by Meese and Rogoff (1983), proposed that forecasts based upon macroeconomic fundamentals could not improve upon a random walk benchmark, especially at short horizons. Rossi (2013) provides a survey of a subsequent literature that achieved successes in improving upon the benchmark, using theoretical and empirical innovations. Theoretical improvements have included utilising asset pricing models and Taylor rules and, separately, empirical advances have included nonlinear methods.[1] This paper seeks to combine these theoretical and empirical innovations in predicting exchange rates, in a changing world.

Engel and West (2005) and Engel et al. (2008) illustrate that models that can be cast in the standard present-value asset pricing framework imply that exchange rates are approximately random walks. This holds under the assumptions of non-stationary fundamentals and a near unity discount factor. However, Engel and West (2004) present evidence that even when the discount factor is near one, a class of models based on observable fundamentals can still account for a fairly large fraction of the variance in exchange rates. An example in this class includes structural exchange rate models in which monetary policy follows the Taylor (1993) rule. Indeed, Engel et al. (2008), Molodtsova and Papell (2009) and Rossi (2013) find that empirical exchange rate models conditioned on an information set from Taylor rules outperform the random walk benchmark in out-of-sample forecasting, especially at short-horizons.

Despite the optimism instilled by this emerging research, one area remains unresolved. This is the frequent regularity that exchange rate predictability is often sample-dependent, and hence forecasting ability appears in some periods but not in others. Rogoff and Stavrakeva (2008) and Rossi (2013) examine these issues thoroughly. Rogoff and Stavrakeva (2008) show for instance that Molodtsova and Papell's (2009) results may change in a different forecast window. Rossi (2013) also finds that models' predictive power is specific to some currencies in some periods but not others. In fact, she concludes by questioning whether instabilities can be explored to improve exchange rates forecasts.

---

[1] For nonlinear models, see Wolff (1987), Sarno et al. (2004), Rossi (2006), Bacchetta et al. (2010), Balke et al. (2013) and Park and Park (2013). Other empirical approaches have included: long-horizon methods, see Mark (1995); panel models, see for example Papell (1997), Groen (2000), MacDonald and Nagayasu (2000), Mark and Sul (2001) and Engel et al. (2008); and factor exchange rate models, see Engel et al. (2012).

There are several reasons to examine the hypothesis that exchange rate predictability is dependent on instabilities in regression and policy coefficients. Firstly, research shows that macroeconomic conditions and policy actions evolve, often suddenly.[2] Boivin (2006), Kim and Nelson (2006) and Cogley et al. (2010) present evidence that the U.S. Federal Reserve's conduct of monetary policy is better characterized by a changing-coefficients Taylor rule. Trecoci and Vassali (2010) present similar evidence for the U.S., U.K., Germany, France and Italy. Secondly, there is widespread evidence of a time-evolving relationship between exchange rates and fundamentals. Bacchetta and van Wincoop (2004), for example, explain this relationship on the basis of a scapegoat theory. Foreign exchange traders often seek explanations for fluctuations in the exchange rate, such that even when an unobservable is responsible for the actual change, it is common to attribute it to an observable macro variable or the scapegoat. Subsequently, this scapegoat variable influences trading behaviour and the exchange rate. Over time, fluctuations in exchange rates are then explained by time-varying weights attributed to scapegoat variables. In a recent application, Balke et al. (2013) and Park and Park (2013) show that allowing for such type of coefficient adaptivity in an monetary model improves in-sample fit and out-of sample predictive power for exchange rates.

It is timely and topical to exploit non-linear Taylor rules when predicting exchange rates. While an extensive literature focuses on linear and non-linear models with standard fundamentals based models, there is limited research focusing on the predictive ability of non-linear Taylor rules.[3] Non-linear methods are pertinent given the nature of the world economy during the last decade. Taylor (2009) argues that before the Global Financial Crisis the U.S. Fed's conduct of monetary policy was characterized by deviations from a linear Taylor rule. After the Crisis, Central Banks around the world have adopted unconventional monetary policy, also inconsistent with linear Taylor rules. Hence we look afresh at Taylor rules predictive content against a random walk.

---

[2] See for example, Stock and Watson (1996) for evidence on structural instabilities in macroeconomic time series in general.

[3] Rossi (2013) provides an excellent survey of recent work using linear and non-linear Purchasing Power Parity, Monetary Model and Uncovered Interest Rate Parity. Papers that focus on Taylor rule predictive content in a linear modelling framework include, Engel and West (2004, 2005, 2006), Engel et al. (2008), Rogoff and Stavrakeva (2008) Molodtsova et al. (2008) and Molodtsova and Papell (2009, 2013). For non-linear modelling of Taylor rules, Mark (2009) is a notable contribution. He employs a Vector Autoregressive model and least-squares learning techniques to update Taylor rules estimates, inflation and output gap which are then then used to compute the exchange rate value. Using in-sample evidence, he finds that allowing for time-variation in parameters is relevant to account for the volatility of the Deutschemark and the Euro, relative to the U.S dollar. Our approach differs from Mark (2009) in that we focus upon out-of-sample predictability of non-linear Taylor rules.

This paper's main contribution is to predict exchange rates accounting for parameter instabilities in Taylor rules by using Bayesian methods. Previous studies, such as Molodtsova and Papell (2009), Engel et al. (2008), Rossi (2013), among others, assumed constant coefficients in the Taylor rules, along with constant coefficients in the forecasting regression. These restrictions about the degree of parameter adaptivity may rationalize the difficulty in pining down model's forecasting performance over-time. Our hypothesis is that the predictive content might be time-varying because fundamentals themselves and their interaction with exchange rates change over time. In light of this, we estimate time-varying parameter Taylor rules and examine their predictive content in a framework that also allows for the parameters of the forecasting regression to change over time.[4] In a major break with the earlier non-linear exchange rate literature, we estimate time varying parameters using information in the likelihood based upon Bayesian methods. Therefore, we do not rely on calibration (e.g. Wolff, 1987; Bacchetta et al., 2010), which can be subjective and may give less accurate parameter estimates and inferior forecasting performance.[5]

In particular, this paper's dataset consists of quarterly exchange rates from 1973Q1 to 2013Q1, on up to 17 OECD countries relative to the U.S. dollar. We calculate Theil's U statistic from Root Mean Squared Forecast Error (RMSFE) recursively out-of-sample, whilst using three forecasting windows. To preview our results, allowing for time-varying Taylor rules improves upon the driftless random walk at both short and long horizons. In fact, in most forecast windows our approach yields a lower RMSFE than the benchmark for at least half of the currencies in the sample. We improve upon the benchmark for as many as 11 out of 17 currencies in our earlier forecast window, and eight out of 10 in our latest forecast window. Hence, the predictive ability is particularly robust to the recent Financial Crisis.

This paper also contributes to the literature by forecasting using panel methods and Bayesian time-varying parameters regressions conditioned on the standard predictors from Purchasing Power Parity (PPP), a Monetary Model (MM), Uncovered Interest Rate Parity (UIRP) and Engel et al. (2012) factor model. The TVP forecasting regression also performs relatively well for over half of the currencies in most windows, when conditioned on PPP at all horizons and UIRP at long-horizon. The panel model generates lower RMSFE than the benchmark for half or more currencies across windows when based on PPP and factors at all horizons, and Taylor rules and UIRP for long-horizon forecasts. However, results for the panel regression are only robust for PPP at all horizons and factors at longer horizons. The predictive content of the MM is less promising for our quarterly sample period, regardless of the forecasting model.

---

[4] Although in principle forecasting using a rolling regression scheme as in Molodtsova and Papell (2009, 2013) allows for instability to be taken into account, a TVP model allows for instabilities to be updated systematically. We also note that while formal tests of parameter instabilities could be conducted in-sample, our approach relies on verifying the plausibility of time-variation in the relationships by means of out-of-sample forecast evaluation.

[5] Giannone (2010) provides a helpful critique of the results based on Bacchetta's et al.(2010) calibration, and shows how using the full maximum likelihood setup in a Bayesian framework is important in accounting for instabilities. Balke et al. (2013) also use Bayesian methods and focus upon modelling exchange rates in-sample with monetary fundamentals.

The rest of the paper proceeds as follows. The next section sets out the Time-Varying Parameter regression we consider. Section 3 discusses the choice of fundamentals, and Section 4 covers data description and the mechanics of our forecasting exercise. The main empirical results are reported in Section 5. Section 6 deals with robustness checks and the final Section concludes.

## 2. The Time-Varying Parameter Regression

A common practice in forecasting exchange rates is to model the change in the exchange rate as a function of its deviations from its fundamental implied value. As put forward by Mark (1995), this accords with the notion that exchange rates frequently deviate from their fundamental implied value, particularly in the short-run. More precisely, define $s_{t+h} - s_t \equiv \Delta s_{t+h}$ as the h-step-ahead change in the log of exchange rate, and $\Omega_t$ a set of exchange rate fundamentals. Then,

$$\Delta s_{t+h} = \beta_{0t} + \beta_{1t} z_t + \varepsilon_{t+h} \tag{1}$$

where,
$$z_t \equiv \Omega_t - s_t \tag{2}$$

As (1) suggests, $\Omega_t$ signals the exchange rate's fundamental value and hence $z_t$, is the deviation from the fundamental's implied level. When the spot exchange rate is lower than the level implied by the fundamentals, i.e., $s_t < \Omega_t$, then the spot rate is expected to increase.

In equation (2), the time-subscripts $t$ attached to the coefficients $\beta_t = [\beta_{0t}, \beta_{1t}]$, make it evident that the regression allows the coefficients to change over time. The exact coefficient's law of motion is inspired, among others, by Stock and Watson (1996), Rossi (2006), Boivin (2006) and Mumtaz and Sunder-Plassmann (2012). We assume a Random Walk Time-Varying Parameter (RW-TVP). Thus, for $\beta_t = [\beta_{0t}, \beta_{1t}]$, the process is:

$$\beta_t = \beta_{t-1} + \nu_t \qquad (3)$$

where, the error term $\nu_t$ is assumed homoscedastic, uncorrelated with $\varepsilon_{t+h}$ in equation (1) and with a diagonal covariance matrix $Q$. Putting together equations (1) and (3) results in a state-space model, where (1) is the measurement equation and (3) the transition equation.

We use Bayesian methods to estimate the parameters of the state-space model. While the use of the Kalman filter with maximum likelihood is another potential method, the evaluation of a large number of likelihood functions in this case might undermine the estimates (Kim and Nelson, 1999). That is, with the method of maximum likelihood there is potential for accumulation of errors, as estimation of the state variables is conditional on maximum likelihood estimates of the other parameters of the system. In addition, there are is also the issue of identifying objective priors to initialize the Kalman filter. Whilst to address the latter the approach in the literature often involves setting diffuse priors or using a training sample, solving the problem of obtaining efficient parameter estimates is a more cumbersome task. By contrast, Bayesian methods treat all the unknown parameters in the system as jointly distributed random variables, such that the estimate of each of them reflects uncertainty about the others (Kim and Nelson, 1999).

In particular, we rely on the Carter and Kohn (1994) algorithm and the Gibbs sampler to simulate draws from the parameters' posterior distribution. The Gibbs sampler, which falls within the category of Markov Chain Monte Carlo (MCMC) methods, is a numerical method that uses draws from conditional distributions to approximate joint and marginal distributions. More precisely, to fully implement the Bayesian method, we need to (i) elicit priors for the unknown parameters, (ii) specify the form of their posterior conditional distributions, and finally (iii) draw samples from the specified conditional posterior distribution. To parameterize the prior distributions we use pre-sample information. We do so largely because we are comparing the forecasting performance of various models, at a number of forecast windows and horizons. By setting priors based on a training sample we aim at ensuring that all the models are based on the same prior elicitation setting, and hence their performance is not influenced by the model's particular prior parameterization choice. This approach also provides natural shrinkage based on evidence in the likelihood, which in turn ensures that TVP estimates will be more accurate, with smaller variance, resulting in a sharper inference and potentially more precise forecasts.[6] The remainder of the details about priors' elicitation and all the other steps are provided in Appendix B.

## 3. Choice of Fundamentals

Having defined the form and the method to estimate the parameters of our main forecasting regression, an additional modelling issue relates to the exact specification of the fundamental information contained in $\Omega_t$. In this regard, our approach is broadly consistent with models that relate the exchange rate to macroeconomic variables within the asset pricing framework. In this framework, the exchange rate is expressed as the present-value of a linear combination of economic fundamentals and unexpected shocks. Under the assumptions of rational expectations and a random walk time series process for the fundamentals, the framework implies that the spot exchange rate is determined by current observable fundamentals and unobservable noise (Engel and West, 2005). We focus primarily on observable fundamentals derived from the Taylor (1993) rule, MM, PPP, UIRP, and the co-movement among exchange synthetized by factors from exchange rates.

### 3.1.1 Taylor Rule Fundamentals

The Taylor (1993) rule postulates that monetary authorities should set the target for the policy interest rate considering the recent inflation path, inflation deviation from its target, output deviation from its potential level, and the equilibrium real interest rate. Then, it follows that they increase the short-term interest rate when inflation is above the target and/or output is above its potential level. Note that the Taylor principle presupposes an increase in the nominal policy rate more than the rise in inflation rate to stabilize the economy.

An emerging research considers the implications of this policy setting for exchange rates, including Engel and West (2005), Engel et al. (2008), Mark (2009), and Molodtsova and Papell (2009, 2013). The premise is that the home and the foreign central banks conduct monetary policy following the Taylor principle. In line with this principle, the foreign monetary authority, taken as the U.S. in our empirical section, is concerned with inflation and output deviations from their target values. In addition to these targets and consistent with historical evidence, Engel and West (2005) assume that the home country also targets the real exchange rate. It is also common, following Clarida et al. (1998), to consider that central banks adjust the actual interest rate to eliminate a fraction of the gap between the current interest rate target and its recent past level, known as interest rate smoothing. By subtracting the foreign Taylor rule from the home, the following interest rate differential equation is obtained:

$$i_t - i_t^* = \phi_0 + \phi_1 \pi_t - \phi_1^* \pi_t^* + \phi_2 \bar{y}_t - \phi_2^* \bar{y}_t^* + \phi_3 q_t + \phi_4 i_{t-1} - \phi_4^* i_{t-1}^* + \mu_t \tag{4}$$

where $i_t$, is the short-term nominal interest rate set by the central bank, asterisks denote foreign (U.S.) variables; $\pi_t$, is inflation; $\bar{y}_t$, denotes output gap; $q_t$ is the real exchange rate defined as $q_t = s_t + p_t^* - p_t$; $s_t$, is the log nominal exchange rate, defined as the home price of foreign currency; $p_t$, is the log of the price level; $\phi_i$ for $i = 1, ..., 4$, are coefficients, and $\mu_t$ is the unexpected disturbance term, which is assumed to be Gaussian. A full derivation of equation (4) is provided in Appendix A.

---

[6] In our empirical exercise we also experimented using the Kalman filter with Maximum Likelihood (ML). However unlike papers that employ diffuse priors, such as Rossi (2006), we also used data-based priors to initialize the Kalman filter recursions. Our rationale for employing these priors in this case is that while diffuse priors when using ML estimates are objective, they result in larger uncertainty about the TVP estimates, which may lead to loss of forecast power.

The link from monetary policy actions to exchange rates occurs through UIRP and the forward premium puzzle. Molodtsova and Papell (2009) discuss at length such mechanisms. They note for example that under UIRP and rational expectations, any circumstance that causes the home central bank to increase its policy rate relative to the foreign, will lead to an expected depreciation of its currency relative to the foreign country. Such circumstances include for example an increase in inflation above the target, a rise in the output gap or a deviation of the real exchange from the target. Conversely, foreign country's policy actions characterized by an increase in the interest rate will be followed by an expected depreciation of its currency. However, the empirical evidence frequently rejects the UIRP condition and this is known as the forward premium puzzle (Engel, 1996). In fact, the evidence is that high interest rate currencies tend to appreciate, rather than depreciate as UIRP posits. This suggests that while we can substitute out the interest rate differential by the expected change in exchange rate in equation (4) to obtain the exchange rate forecasting regression, we impose no restrictions on the effect of monetary policy on exchanges rates.

Using equation (4) to derive the forecasting regression or estimate Taylor rule fundamentals is valid when parameters are constant over time. In a dynamic world however, Taylor rule parameters may be subject to structural instabilities. Therefore, rather than estimating or assuming Taylor rules fundamentals from models with constant or calibrated parameters, we allow for the possibility of monetary policies that respond to macroeconomic conditions in a time-varying fashion. Hence, we estimate fundamentals from Taylor rules using a TVP regression of the following form:

$$i_t - i_t^* = \phi_{0t} + \phi_{1t}\pi_t - \phi_{1t}^*\pi_t^* + \phi_{2t}\bar{y}_t - \phi_{2t}^*\bar{y}_t^* + \phi_{3t}q_t \\ + \phi_{4t}i_{t-1} - \phi_{4t}^*i_{t-1}^* + \mu_t \tag{5}$$

from which we compute the fundamentals as:[7]

$$\Omega_t \equiv \hat{\phi}_{0t} + \hat{\phi}_{1t}\pi_t - \hat{\phi}_{1t}^*\pi_t^* + \hat{\phi}_{2t}\bar{y}_t - \hat{\phi}_{2t}^*\bar{y}_t^* + \hat{\phi}_{3t}q_t \\ + \hat{\phi}_{4t}i_{t-1} - \hat{\phi}_{4t}^*i_{t-1}^* + s_t \tag{6}$$

where the time-subscript, $t$, attached to the coefficients defines time-varying parameters and the symbol "^" indicates parameter estimates. Note that this is identical to equation (4) except for the time-varying coefficients. This suggests that both, the information set from Taylor rules and the exchange rate forecasts, are generated from TVP regressions.

The exact form of the Taylor rule and hence of equation (5) varies depending on a number of assumptions. We focus on three popular variants.[8] In all variants the equilibrium real interest rate and the inflation target of the home and foreign country are assumed to be identical. Thus, in equation (5) the term $\phi_{0t}$ equals zero.[9] In addition, all specifications are asymmetric; that is, apart from the inflation and output gaps which both countries target, the home country also targets the real exchange rate.

With the above assumptions maintained, the first Taylor rule specification further assumes homogeneous coefficients and no interest rate smoothing – abbreviated as TR$_{on}$. This signifies that it imposes equality in the coefficients on inflation ($\phi_{1t} = \phi_{1t}^*$) and the output gap ($\phi_{2t} = \phi_{2t}^*$) of the home and foreign country Taylor rules. In addition, central banks do not smooth interest rates ($\phi_{4t} = \phi_{4t}^* = 0$). Engel and West (2006) find that it is reasonable to assume parameter homogeneity across countries. The assumption of no interest rate smoothing accords with Engel and West's (2005) formulation. Molodtsova and Papell (2009) use an identical Taylor rule.

---

[7] We can equivalently express the predictor in terms of $z_t$ - see equation (2). In this case we would have: $z_t \equiv \hat{\phi}_{0t} + \hat{\phi}_{1t}\pi_t - \hat{\phi}_{1t}^*\pi_t^* + \hat{\phi}_{2t}\bar{y}_t - \hat{\phi}_{2t}^*\bar{y}_t^* + \hat{\phi}_{3t}q_t + \hat{\phi}_{4t}i_{t-1} - \hat{\phi}_{4t}^*i_{t-1}^*$.
[8] The three variants we consider constitute counterpart to the constant-parameter specifications denoted TR$_{on}$ and TR$_{os}$ and TR$_{en}$ in Appendix A.
[9] This is a typical assumption in this literature including in Engel and West (2005), Engel et.al (2008), Rogoff and Stavrakeva (2008), and Molodtsova and Papell (2013). As Molodtsova and Papell (2013) note, whether to include a constant that capture differences in the equilibrium real interest rate and inflation target is irrelevant, because the forecasting regression includes a constant. We also opted to drop the constant term following our empirical experiment with Taylor featuring this term. In all cases the coefficient was very small and not significantly different from zero.

A second specification is similar to the above, except that it includes lagged interest rates. Therefore, it is an asymmetric rule, with homogeneous coefficients and interest rate smoothing (TR$_{os}$). Since the assumption of coefficients' homogeneity between the home and the foreign country is maintained, then we also have $\phi_{4t} = \phi_{4t}^*$ in equation (5). The inclusion of lagged interest rates implies that central banks limit volatility in the interest rates and is in the spirit of Engel et al. (2008), Mark (2009) and Molodtsova and Papell (2009).

The third variant relaxes the assumption of homogeneous coefficients across countries made above and central banks do not smooth interest rates. Therefore, in equation (5), $\phi_{4t} = \phi_{4t}^* = 0$; and is an asymmetric rule, with heterogeneous coefficients and no interest rate smoothing (TR$_{en}$). Molodtsova and Papell (2009) find that models of this type exhibit a strong forecasting performance.

To estimate each of these variants we set up a state-space representation as in Section 2, but here the measurement equation is defined by (5) and the transition process also follows a random walk. That is, as in equation (3) but with $\beta_t$ replaced by $\phi_t$. The estimation procedure is also based on Bayesian methods and details about priors' elicitation, posterior distributions, and sampling algorithm are provided in Appendix B. However, we note here that like in the forecasting regression, our results rely on data-based information to parameterize priors and the initial conditions.

### 3.1.2 Monetary, PPP, UIRP and Factor Fundamentals

The TVP forecasting regression also uses the content of four alternative sets of information. These are from the monetary model (MM), PPP condition, UIRP hypothesis, and factors from exchange rates. In notation:

$$\text{MM fundamentals: } \Omega_t \equiv (m_t - m_t^*) - (y_t - y_t^*) \tag{7}$$

$$\text{PPP fundamentals: } \Omega_t \equiv p_t - p_t^* \tag{8}$$

$$\text{UIRP fundamentals: } \Omega_t \equiv (i_t - i_t^*) + s_t \tag{9}$$

$$\text{Factor fundamentals: } \Omega_{it} \equiv \sum_{r=1}^{R} \gamma_{r,i} \times f_{r,t} \tag{10}$$

where, in addition to the variables previously defined, $m_t$ is the log of money supply; $y_t$ denotes the log of income; $R$ is the number of factors; $\gamma_{r,i}$ is the loading of factor $r$ in the currency $i$; and $f_{r,t}$ is the estimated $r^{th}$ factor.

While fundamentals given by identities (7), (8), and (9) are standard in the exchange rate literature, those represented by the co-movement among exchange rates as in identity (10) have been recently propounded by Engel et al. (2012). Their basic presumption is that the exchange rate of country $i$ follows the process:

$$s_{it} = \gamma_i f_{it} + v_{it} \quad (11)$$

where, $f_{it}$ is the effect of the factor in country's $i$ exchange rate; $\gamma_i$ is the respective factor loading, and $v_{it}$ is a country specific shock, which is uncorrelated with the factors. Engel et al. (2012) show that under plausible assumptions, for example that the common factor follows a random walk process, the RMSFE of the factor model is lower relative to the RMSFE of the random walk. In our empirical procedure we follow Engel et al. (2012) and allow for one, two or three factors, estimated via principal components.[10]

To obtain initial conditions for the forecasting regression, we simply compute for the initial 20 data-points the series defined by the identity (7) for MM, (8) for PPP, (9) for UIRP, and extract factors to obtain fundamentals given by (10). We then use these fundamentals to estimate a constant-parameter model akin to our forecasting regression, from which we parameterize the priors, initial states, and the covariance matrices of the TVP forecasting regression. The observations used to parameterize priors are discarded, and we use the remaining sample period and the same identities above, (7)-(10), to compute fundamentals that constitute our predictors in the TVP forecasting regression.

Apart from our main forecasting regression which allows the coefficients to vary over time, we also forecast with a second regression which maintains them constant; i.e., $\beta_{it} = \beta_i$, for $i = \{0, 1\}$ in equation (1).[11] Engel et al. (2008) find that panel data methods forecast better than single-equation methods. Accordingly, we also use a Fixed-effect (FE) panel regression as in Engel et al. (2008, 2012). In this case, except for the Taylor rules, the set of information from the MM, PPP, UIRP, and factors is computed exactly as in the TVP forecasting approach, i.e., as in identities, (7)-(10). The information set from Taylor rules specifications is obtained by estimating, via OLS, a single-equation fixed-parameter model similar to equation (4). Table 1 summarises all these aspects.

---

[10] Engel et al. (2012) estimate the factors using maximum likelihood or principal components, and report evidence of fairly comparable results.
[11] As should be clearer in the next Section, allowing for time-variation in the parameters in a recursive forecasting approach implies that there are potentially two sources of variation that will ultimately impact upon the parameters. The first is due to the recursive algorithm when computing the optimal parameter at each time of the in-sample period. The second source arises from extending the sample as observations are added to end of the in-sample period (recursions). Therefore, a TVP model allows for more flexibility and presumably more consistent estimates as the sample is extended. We note as well that the second effect is also prevalent in the constant-parameter forecasting regression.

**Table 1. Empirical Exchange Rate Models and Forecasting Approaches**

| Fundamentals-based Exchange Rate Model | TVP Approach | | Constant Parameter Approach | | Forecast Windows and Number of Currencies Considered (N) |
|---|---|---|---|---|---|
| | Information set (Fundamental) | Forecasting model | Information set (Fundamental) | Forecasting model | |
| Taylor Rule (TR) | Estimated with a random walk Time-Varying Parameter model using Bayesian methods: <br><br> $TR_{on}$: <br> $z_t \equiv i_t - i_t^* = \alpha_{1t}(\pi_t - \pi_t^*) + \alpha_{2t}(\bar{y}_t - \bar{y}_t^*) + \phi_{3t}q_t + \mu_t$ <br><br> $TR_{os}$: <br> $z_t \equiv i_t - i_t^* = \alpha_{1t}(\pi_t - \pi_t^*) + \alpha_{2t}(\bar{y}_t - \bar{y}_t^*) + \alpha_{3t}(i_{t-1} - i_{t-1}^*) + \phi_{3t}q_t + \mu_t$ <br><br> $TR_{en}$: <br> $z_t \equiv i_t - i_t^* = \phi_{1t}\pi_t - \phi_{1t}^*\pi_t^* + \phi_{2t}\bar{y}_t - \phi_{2t}^*\bar{y}_t^* + \phi_{3t}q_t + \mu_t$ | Random walk Time-Varying Parameter (TVP) model, estimated using Bayesian methods: <br><br> $\Delta s_{t+h} = \beta_{0t} + \beta_{1t}z_t + \varepsilon_{t+h}$ | Estimated with a single-equation fixed-parameter model, via Ordinary Least Square estimator: <br><br> $TR_{on}$: <br> $z_{it} = i_{it} - i_t^* = \alpha_1(\pi_{it} - \pi_t^*) + \alpha_2(\bar{y}_{it} - \bar{y}_t^*) + \phi_3 q_{it} + \mu_t$ <br><br> $TR_{os}$: <br> $z_{it} = i_{it} - i_t^* = \alpha_1(\pi_{it} - \pi_t^*) + \alpha_2(\bar{y}_{it} - \bar{y}_t^*) + \alpha_3(i_{it-1} - i_{t-1}^*) + \phi_3 q_{it} + \mu_t$ <br><br> $TR_{en}$: <br> $z_{it} = i_{it} - i_t^* = \phi_1\pi_{it} - \phi_1^*\pi_t^* + \phi_2\bar{y}_{it} - \phi_2^*\bar{y}_t^* + \phi_3 q_{it} + \mu_t$ | Fixed-effect Panel model, estimated via Least Square Dummy Variable estimator: <br><br> $\Delta s_{it+h} = \omega_i + \beta z_{it} + \varepsilon_{it+h}$ | A: 1995Q1-1998Q4; N=17 (all currencies in the sample); <br><br> B: 1999Q1-2013Q1; N=10 (non-Euro area currencies and the Euro); <br><br> C: 2007Q1-2013Q1; N=10 (non-Euro area currencies and the Euro). |
| Monetary Model | Computed as: <br> $z_t \equiv (m_t - m_t^*) - (y_t - y_t^*) - s_t$ | | Computed as: <br> $z_{it} \equiv (m_{it} - m_t^*) - (y_{it} - y_t^*) - s_{it}$ | | |
| PPP | Computed as: <br> $z_t \equiv p_t - p_t^* - s_t$ | | Computed as: <br> $z_{it} \equiv p_{it} - p_t^* - s_{it}$ | | |
| UIRP | Computed as: <br> $z_t \equiv (i_t - i_t^*)$ | | Computed as: <br> $z_{it} \equiv (i_{it} - i_t^*)$ | | |
| Factors | Factors estimated through principal components analysis. | | Factors estimated through principal components analysis. | | |

Notes: This Table summarizes the models considered and the forecasting approaches. The definition of the variables is as follows: $i$ = interest rate; $\pi_t$ = inflation rate; $y_t$ = output; $\bar{y}_t$ = output gap; $q_t$ = real exchange rate; $m_t$ = money; $p_t$ = price level; $s_t$ = nominal exchange rate. The subscripts $i$ and $t$ denote country and time, respectively. Asterisk defines the foreign country. Three variants of Taylor rules (TR) are considered: (i) $TR_{on}$: asymmetric rule with homogeneous coefficients and no interest rate smoothing, (ii) $TR_{os}$: asymmetric rule with homogeneous coefficients and interest rate smoothing) and (iii) $TR_{en}$: asymmetric rule with heterogeneous coefficients and no interest rate smoothing. See Appendix A for derivations. The factor model allows for one (F1), two (F2), or three (F3) factors. The forecasts are computed for one-, four-, eight-, and 12-quarters-ahead forecasts.

## 4. Data and Forecast Mechanics

### 4.1 Data

The data comprises quarterly figures spanning 1973Q1:2013Q1 from 18 OECD countries: United States, United Kingdom, Switzerland, Sweden, Spain, Norway, Netherlands, Korea, Italy, Japan, Germany, France, Finland, Denmark, Canada, Belgium, Austria and Australia. The main source is the IMF's (2012) *International Financial Statistics* (IFS). Some of the countries in our sample period moved from their national currencies to the Euro. To generate the exchange rate series for these countries, the irrevocable conversion factors adopted by each country on the 1$^{st}$ of January 1999 were employed, in the spirit of Engel et al. (2012). The money supply is measured by the aggregate M1 or M2.[15]

To estimate Taylor rules we need the short-run nominal interest rates set by central banks, inflation rates and the output gap or the unemployment gap.[16] We use the central bank's policy rates when available for the entire sample period, or alternatively the discount rate or the money market rate. The proxy for quarterly output is the industrial production in the last month of the quarter. The output gap and unemployment gap are obtained by applying the Hodrick and Prescott (1997) filter recursively, to the output and unemployment series. The price level consists of the consumer price index (CPI) and the inflation rate is defined as the (log) CPI quarterly change. The data on money supply, industrial production, unemployment rate and CPI were seasonally adjusted by taking the mean over four quarters following Engel et al. (2012).

### 4.2 Forecast Implementation and Evaluation

As noted in the previous Sub-section, the sample covers the period 1973Q1 to 2013Q1. We use the sample period from 1974Q1 to 1978Q4 to parameterize the priors and initial conditions for the TVP regressions. The in-sample estimation period begins in 1979Q1 for all models, including the fixed-parameter ones.

---

[15] Exceptions are for Sweden, where M3 is used; Australia -M3; and the UK -M4. See Appendix C for extra details.

[16] In estimating Taylor rules and due to possible endogeneity issues, several authors emphasize the timing of the data employed. The discussion centres on the idea that Taylor rules are forward-looking, and hence ex-post data might reflect policy actions taken in the past. Kim and Nelson (2006) note two approaches that can be employed to account for this. The first comprises using historical real-time forecasts that were available to policy-makers. The second consists in using ex-post data to directly model the policy-makers' expectations. Since historical real-time forecasts are unavailable for our sample of countries, we follow Molodtsova and Papell's (2009) approach, and use data that were observed (as opposed to the real-time forecasts) at time t, while forecasting t + h period.

Rogoff and Stavrakeva (2008) argue that the predictive ability of fundamentals-based exchange rate models is often sample-dependent. To verify models' forecasting performance in alternative forecast windows we consider three sub-samples.[17] The first out-of-sample forecasts are for the period 1995Q1-1998Q4. This corresponds to the pre-Euro period. Forecasts for all the 17 countries' currencies are generated and models' forecast accuracy evaluated. A second forecasting window covers the post-Euro period: 1999Q1-2013Q1. Since we have extended the exchange rates of the Euro-area countries throughout, the forecast of the Euro currency is computed as an average of the forecasts of the Euro-area countries in our sample. The forecast error is constructed as the difference between each of the country's realized value and the computed average. We therefore generate forecasts for the nine non-Euro area countries plus the Euro. These procedures draw from Engel et al. (2012). The last out-of-sample forecast window begins just before the recent financial turmoil and extends to the end of the sample, i.e., 2007Q1-2013Q1. Considering this window is particularly important, given the substantial instabilities that characterized the period, with consequences for the monetary policy reaction functions and the variance of the exchange rate. In this window, we also compute forecasts for 10 currencies, following the procedure just described above.

Our forecasting horizons cover the short and the long horizons. Specifically, we use a direct rather than an iterative method to forecast the h-quarter-ahead change in the exchange rates for h = 1, 4, 8, 12. The benchmark model is the driftless random walk. Since the seminal contribution by Meese and Rogoff (1983) it has been found that it is challenging to improve upon this benchmark.

The forecasting exercise is based on a recursive approach using lagged fundamentals. For concreteness, let $T + h \equiv R + P$ be the sample size comprising a proportion of R observations for in-sample estimation, and P for prediction at h-forecast horizon. Thus, $T + h$ constitutes the total number of observations after discarding data-points used to parameterize priors for the TVP models. We first use R observations to estimate or compute the information set and to generate the parameters of the exchange rate forecasting regression. With these parameters we generate the first h-step-ahead forecast and compute the forecast error. We then add one observation at a time to the end of the in-sample period and repeat the same procedure until all P observations are used.

---

[17] The forecast windows are summarised in the last column of Table 1.

To compare the out-of-sample forecasting performance of our models we employ the sample RMSFE as our metric. We compute the ratio of the RMSFE of the fundamentals-based exchange rate model (FEXM) relative to RMSFE of the driftless random walk, known as the Theil's-U statistic. Hence, models that perform better than the benchmark have a Theil's U less than one. To test the null hypothesis of no difference in the accuracy of the forecasts of FEXM relative to the forecasts of the random walk we compute one-sided Diebold Mariano (1995) (DM) test-statistic.[18] Diebold and Mariano (1995) show that under the null, the test follows a standard normal distribution. We reject the null hypothesis of equal forecast accuracy if the DM statistic is greater than 1.282 at 10% significance level and conclude that the forecast from the FEXM is better than that of the random walk. The appealing feature of the DM test is that we need not make any assumption about the model that generates the forecast. It can be used to evaluate forecasts generated from linear or non-linear models, either nested or non-nested. This contrasts with the typical Clark and West's (2006, 2007) (CW) test-statistic, which is suitable for comparison of (linear) nested models. Additionally, Rogoff and Stavrakeva (2008) make the case for using the DM test, rather than the CW test, arguing that the latter does not always test for minimum mean square forecast error.

## 5. Empirical Results

We summarise the results for each forecast window and horizon by reporting the number of U's less than one (No. of U's < 1), the Median U, and the number of DM test-statistics greater than 1.282 (No. of DM > 1.282). Recall that a value of U less than one suggests that the RMSFE of the fundamentals-based exchange rate model (FEXM) is lower than that of the RW; hence on average, the FEXM forecasts better than the benchmark driftless random walk (RW). Thus, the No. of U's < 1, provides the number of currencies for which the FEXM improves upon the RW. For instance, when the No. of U's < 1 corresponds to half of the currencies in a particular window, then the FEXM improves upon the RW for half of the currencies in that window. The Median U provides the value of the middle U-statistic across the sample of *N* currencies. At this value, the U-statistic of N/2 currencies is less than the *Median U* and the U-statistic of the other N/2 currencies is greater than the same M*edian U*. Therefore, a Median U less than or equal to one along with a No. of U's < 1 for half or more currencies in the window, is also consistent with a better average performance of the FEXM relative to the benchmark. The number of DM statistics greater than 1.282 (No. of DM > 1.282), corresponds to cases in which the null hypothesis under the Diebold and Mariano (1995) test of equal forecast accuracy is rejected, at 10% significance level. The higher the number of rejections across the number of currencies in the window, the better is the average accuracy of the forecasts of the FEXM relative to the forecasts of the benchmark.

---

[18] The Diebold and Mariano (1995) test is computed as: $DM = \frac{\bar{d}}{(\widehat{LRV}_{\bar{d}}/T)^{1/2}}$; where, $\bar{d} = \frac{1}{T_0}\sum d_t$; $\widehat{LRV}_{\bar{d}}$ is the estimated long-run variance of $\sqrt{T}\bar{d}$; and $d_t$ is the difference between the RMSFE of the random walk and the RMSFE of the FEXM.

### 5.1 Taylor Rules Results

Table 2 presents the summary results from the TVP forecasting regression and the Fixed-effect (FE) panel regression, both conditioned on Taylor rules.[19] Focusing first on the TVP regression, the results indicate improvements upon the benchmark for short (h=4) and longer-horizon forecasts (h=8 and h=12) in most forecast windows. For instance, in the first window and at four-quarter-ahead horizon, the TVP regression conditioned on fundamentals from Taylor rules with homogenous coefficients and interest rate smoothing ($TR_{os}$) outperform the RW for 11 out of 17 currencies. As the forecast horizon increases to eight and 12-quarters ahead, it still outperforms the benchmark for nine currencies and 10 currencies, respectively. The regression conditioned on Taylor rules with homogenous coefficients but no interest rate smoothing ($TR_{on}$) shows a similar performance as well. In the last forecast window, the TVP predictive regression improves upon the benchmark for over half of the currencies at four-, eight- or 12-quarter-ahead forecasts. In particular, when conditioned on Taylor rules with heterogeneous coefficients and no interest rate smoothing ($TR_{en}$), it shows the strongest performance. It outperforms the RW for at least half of the currencies at all horizons, reaching as many as seven out of 10 at h=4, and eight out of 10 at h=12. Our regressions performed unsatisfactorily in the forecast window spanning 1999Q1-20013Q1.[20]

Table 2 also illustrates that the statistical significance of the forecasts accuracy of the TVP regression based on Taylor rule's information set stands out for long-horizon forecasts. For example, consider the model conditioned on $TR_{on}$ in the last forecast window. Here we can reject the null of equally forecast accuracy for four out of five currencies that had U less than one at the 12-quarters-ahead forecasts. In contrast, for four-quarter-ahead forecasts the null is rejected only once. A similar observation holds for other Taylor rule specifications and forecast windows. This suggests that although the RMSFE of the Taylor rule- based exchange rate models are smaller than that of the RW for short-run forecasts, they become significantly lower for long-run forecasts. As we noted earlier, this is unsurprising. Since exchange rates frequently deviate from their implied fundamental level in the short-run but return to that level in the long-run, one should expect the forecast accuracy to increase with the forecast horizon.

---
[19]Detailed results by currency are provided in a results appendix. Also, recall that the information set for the TVP Regression is estimated from a TVP Taylor rule, while for the fixed-effect panel regression is estimated from a single-equation constant-parameter model.

[20] In this window, the median U's are all above one, implying that for more than half of the currencies in the window the U's are greater than one.

Shifting the focus to the FE panel regression, results in Table 2 show that in most windows it outperforms the RW benchmark for over half of the currencies, only for long-horizon forecasts, i.e., for h=8. This is the case in the window spanning from 1995Q1 to 1998Q4 where, for instance, the regression with $TR_{os}$ perform well for 13 out of 17 exchange rates. In the last window, it still out-forecasts the RW for most currencies at h=8, regardless of the Taylor rule specification. However, in this particular window most Taylor rule specifications also do well at 12-quarter-ahead forecast. Note as well that like in the TVP regression, the statistical significance of forecasts accuracy of the panel regression stands out as the forecast horizon increases.

The performance of the FE panel regression in our sample is partially similar to the results in Engel et al. (2008).[21] Using a FE panel regression that includes a time effect and a fixed effects, they find that the driftless RW outperforms the Taylor-rule based regression at both, short (h=1quarter) and long (h=16 quarters) forecast horizons. Here, while the findings for the short-run forecasts are similar, for long-run forecasts we find improvement upon the RW benchmark. Of course there are a number of differences between their analysis and ours. Probably the most significant are: (i) the differences in the forecast windows considered and the sample span,[22] and (ii) their use of a Taylor rule specification ($TR_{on}$) with posited coefficients, whereas here we estimate the coefficients.

---

[21] As we noted in the Introduction, there are other papers that condition on Taylor rules to forecast exchange rates in a linear modelling setup. However, we leave the comparison of their results to our Robustness check section, as they employ a single-equation forecasting regression with monthly data, rather than a FE panel regression with quarterly data. Here we compare with Engel et al. (2008) as they also employ a FE panel regression and quarterly data.

[22] Engel's et al. (2008) sample covers the period 1973Q1-2005Q4, while our sample extends for an extra eight years from 2005Q4.

**Table 2. Forecast Evaluation: Taylor Rules**

|  |  | TVP Regression | | | | Fixed-effect Panel Regression | | | |
|---|---|---|---|---|---|---|---|---|---|
| Fundamentals: | | h=1 | h=4 | h=8 | h=12 | h=1 | h=4 | h=8 | h=12 |
| | | Forecast Window: 1995Q1-1998Q4; N=17 | | | | | | | |
| $TR_{on}$ | | | | | | | | | |
| | No. of U's <1 | 5 | **9** | **11** | **10** | 4 | 5 | 4 | 7 |
| | No. of DM >1.282 | 1 | 4 | 9 | 8 | 1 | 1 | 3 | 3 |
| | Median U | 1.023 | 0.989‡ | 0.853‡ | 0.939‡ | 1.012 | 1.044 | 1.076 | 1.048 |
| $TR_{os}$ | | | | | | | | | |
| | No. of U's <1 | 5 | **11** | **9** | **10** | 5 | 7 | **13** | 8 |
| | No. of DM >1.282 | 1 | 4 | 6 | 10 | 0 | 1 | 6 | 7 |
| | Median U | 1.031 | 0.980‡ | 0.999‡ | 0.965‡ | 1.011 | 1.007 | 0.922‡ | 1.099 |
| $TR_{en}$ | | | | | | | | | |
| | No. of U's <1 | 6 | 8 | 8 | 7 | 4 | 5 | **9** | 5 |
| | No. of DM >1.282 | 0 | 3 | 8 | 7 | 0 | 1 | 4 | 2 |
| | Median U | 1.021 | 1.043 | 1.035 | 1.292 | 1.013 | 1.030 | 0.994‡ | 1.142 |
| | | Forecast Window: 1999Q1-2013Q1; N=10 | | | | | | | |
| $TR_{on}$ | | | | | | | | | |
| | No. of U's <1 | 3 | 2 | 3 | 3 | 0 | 1 | 2 | 2 |
| | No. of DM >1.282 | 1 | 1 | 2 | 3 | 0 | 0 | 0 | 1 |
| | Median U | 1.007 | 1.031 | 1.082 | 1.179 | 1.009 | 1.033 | 1.069 | 1.112 |
| $TR_{os}$ | | | | | | | | | |
| | No. of U's <1 | 1 | 2 | 3 | 3 | 2 | 2 | 2 | 2 |
| | No. of DM >1.282 | 0 | 0 | 2 | 2 | 0 | 1 | 1 | 1 |
| | Median U | 1.010 | 1.043 | 1.099 | 1.219 | 1.007 | 1.033 | 1.065 | 1.156 |
| $TR_{en}$ | | | | | | | | | |
| | No. of U's <1 | 1 | 3 | 3 | 1 | 0 | 1 | 2 | 2 |
| | No. of DM >1.282 | 0 | 0 | 1 | 1 | 0 | 0 | 1 | 1 |
| | Median U | 1.007 | 1.036 | 1.083 | 1.226 | 1.010 | 1.040 | 1.088 | 1.159 |
| | | Forecast Window: 2007Q1-2013Q1; N=10 | | | | | | | |
| $TR_{on}$ | | | | | | | | | |
| | No. of U's <1 | 4 | **5** | **7** | **5** | 4 | 4 | **8** | **6** |
| | No. of DM >1.282 | 0 | 1 | 4 | 4 | 0 | 1 | 2 | 3 |
| | Median U | 1.004 | 1.003 | 0.973‡ | 1.007 | 1.003 | 1.003 | 0.984‡ | 0.992‡ |
| $TR_{os}$ | | | | | | | | | |
| | No. of U's <1 | 2 | 4 | **6** | **5** | 3 | 4 | **6** | **5** |
| | No. of DM >1.282 | 0 | 0 | 2 | 5 | 0 | 0 | 5 | 4 |
| | Median U | 1.004 | 1.003 | 0.955‡ | 0.930‡ | 1.010 | 1.001 | 0.945‡ | 0.942‡ |
| $TR_{en}$ | | | | | | | | | |
| | No. of U's <1 | **5** | **7** | **6** | **8** | **5** | 4 | **5** | 4 |
| | No. of DM >1.282 | 1 | 2 | 3 | 4 | 1 | 2 | 3 | 3 |
| | Median U | 0.999‡ | 0.991‡ | 0.912‡ | 0.828‡ | 0.999‡ | 1.004 | 0.972‡ | 1.052 |

Notes: This Table summarises the forecasting performance of the TVP forecasting regression and the Fixed-effect panel regression with Taylor rule fundamentals specified as $TR_{on}$, $TR_{os}$ and $TR_{en}$. See Table 1 for details about the form of the forecasting regressions and how fundamentals are computed or estimated. The benchmark model for both forecasting regressions is the driftless Random Walk (RW). For each regression, set of fundamentals, forecast window and quarterly horizon (h), the "No. of U's < 1" (number of U-statistics less than one), provides the number of currencies for which the model improves upon the RW, since it indicates cases where the RMSFE of the fundamental-based regression is lower than that of the RW. When the U's are less than one for at least half of the currencies in the forecast window, marked in **bold**, then on average, the fundamental-based regression outperforms the benchmark in that window. The "No. of DM > 1.282" (number of DM statistics greater than 1.282) shows cases of rejections of the null hypothesis under the Diebold and Mariano (1995) test of equal forecast accuracy at 10% level of significance. The higher the No. of DM > 1.282, the better the average accuracy of the forecasts of the fundamental-based regression relative to the benchmark is. The "Median U" indicates the middle value of the U-statistic across the sample of $N$ currencies for each forecast window and horizon. When "Median U" is less than or equal to one - marked with the symbol "‡", and U's are less than one for at least half of the currencies in the window, this is also consistent with a better average forecasting performance of the fundamental-based regression relative to the benchmark.

**Figure 1. Recursive U-statistic for the TVP Regression with Taylor Rule Fundamentals**

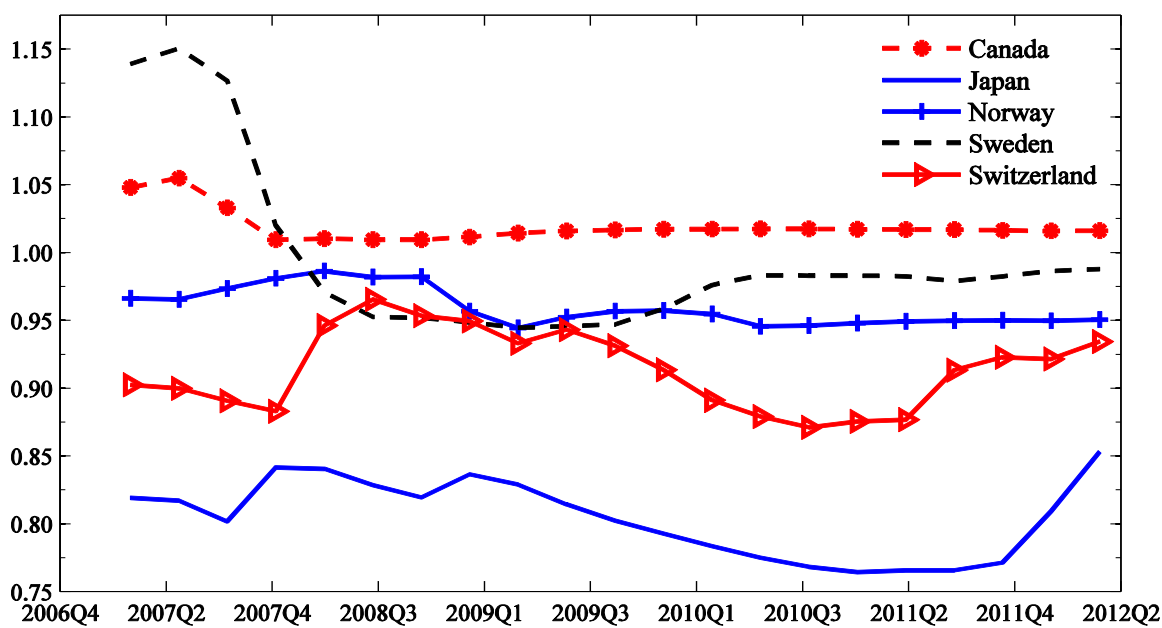

Notes: This Figure illustrates the recursive U-statistics of the TVP regression with Taylor rule-information set. The Taylor rule specification imposes homogenous coefficients between home and foreign country and no smoothing ($TR_{on}$). The forecast window is 2007Q1-2013Q1, at four-quarter-ahead horizon. The recursive U's are computed as successive U's at each point in the forecast window such that the last U includes all forecasts in the window. A recursive U less than one for the most or all part of the forecast window indicates that the corresponding RMSFE of the TVP regression conditioned on $TR_{on}$ is lower than the RMSFE of the RW. Hence, the regression forecasts consistently better than the RW across the window.

On balance and across forecasting regressions, forecasting windows, and horizons, the TVP regression had a better average performance. This is particularly notable at four, eight and 12-quarters-ahead forecast horizons, and in the window that encompasses the recent financial crisis, i.e., 2007Q1-2013Q1. To illustrate the variability of the RMSFE of the TVP regression and check its sensitiveness across the forecast window, Figure 1 plots the recursive U, for h=4, $TR_{en}$ and six exchange rates.[23] The recursive U's are constructed as successive U's at each point in the forecast window, such that the final recursion provides the U for the entire window.

At least two aspects can be observed in Figure 1. First and unsurprisingly, the U's vary considerable at the beginning of the forecast window (up to 2008Q1), since they are computed from few forecasts. Second, as the number of forecasts increases, the pattern of the U-statistics becomes much clear such that there is no obvious sensitivity to the sample:

---

[23] We have selected only this windows and these currencies to make the graph readable, as the pattern of the recursive U's of the other exchange rates in this and other forecast windows and horizons is similar to that displayed in Figure 1. In fact, the same pattern is also observed when we define the GBP as the base currency (not reported), rather than the U.S. dollar in Figure 1.

in those cases where the TVP regression performs unsatisfactorily than the RW it does so for the entire sample; whereas when it does better, it is also consistent. In the Figure, the recursive U's of the TVP regression are slightly above one for Canada for the most part of the window. This indicates that at each additional point forecast, the recursive RMSFE of the TVP regression is greater than the recursive RMSFE of the RW for the Canadian dollar exchange rate, consistent with a poor forecasting performance of the TVP regression. By contrast, the U's are less than one for the UK, Japan, Norway, Sweden and Switzerland. Therefore, the average good performance that we noted for 70% of the currencies in this window and horizon holds across the entire sample period.

To illustrate what determines a U-statistic of certain magnitude for each forecasting regression, Figure 2 depicts the predictive content of the TVP Taylor rule fundamentals, vis-à-vis those resulting from a single-equation constant parameter regression, along with the subsequent h-quarter change in the exchange rate. Recall that the former fundamentals are employed in the TVP forecasting regression, while the latters in the FE panel forecasting regression. The example is based on the UK, for the last forecast window, at h=1 and h=12, and the Taylor rule specification with heterogeneous coefficients and no smoothing ($TR_{en}$). The U-statistics are 1.004 (h=1) and 0.722 (h=12) for the TVP regression, and 0.997 (h=1) and 1.091 (h=12) for the FE panel regression.

The top-left graph shows the case of the TVP Taylor rule fundamentals at one-quarter forecast horizon. As depicted, at this horizon TVP Taylor rule fundamentals fail to predict the direction of the subsequent one-quarter change in the Pound sterling/USD exchange rate for the most part of forecast window, resulting in a U-statistic above one. For instance, while the TVP regression predicts a fall in the Pound sterling from 2007Q2 up to 2008Q4, the data shows an opposite path. In the following periods the regression predicts the correct movements until 2009Q4, failing subsequently until 2010Q3. In the remaining periods it does reasonably well, except between 2011Q1 and 2011Q3. In contrast, Taylor rule fundamentals from the constant parameter regression, depicted in the graph at the bottom left, provide a relative better signal of the subsequent change in the one-quarter Pound sterling exchange rate for the most part of the forecast window. However, since there are also some periods were these fundamentals fail, for example between 2009Q1- 2009Q3 and 2011Q3-2012Q2, the U-static is less one by a narrow margin (U=0.997).

**Figure 2. Predictive content of the TVP Taylor rules vs Constant-Parameter Taylor rules**

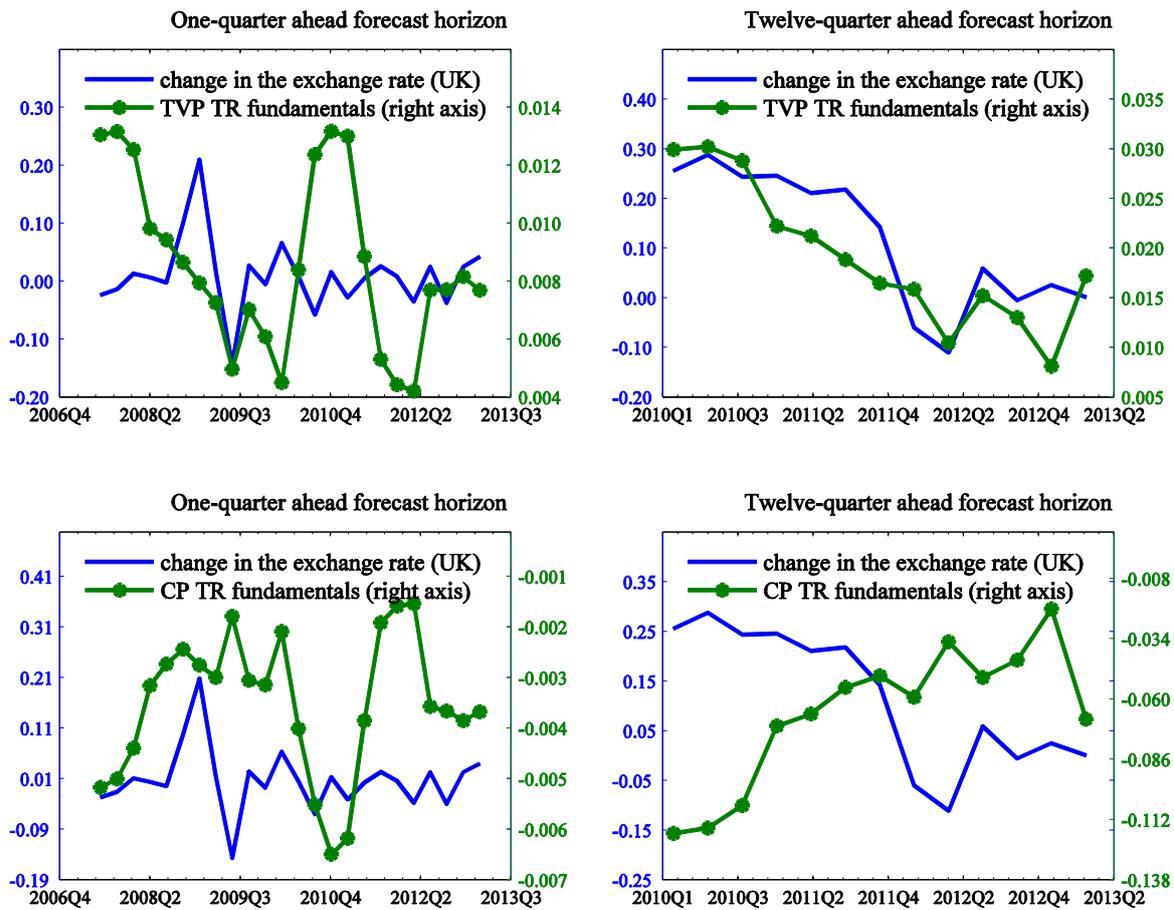

Notes: This Figure depicts the predictive content of the TVP Taylor rule fundamentals, vis-à-vis those resulting from a constant parameter (CP) regression, along with the subsequent h-quarter change in the exchange rate. The Taylor rule specification assumes heterogeneous coefficients and no smoothing (TR$_{en}$). The fundamentals, or more precisely the interest rate differentials, are estimated recursively to nest the forecast method. The in-sample period spans 1979Q1-2006Q4 and the out-of-sample period comprise 2007Q1-2013Q1.

The graphs depicted at the right-hand side of Figure 2 illustrate a similar comparison at the twelve-quarter-ahead forecast horizon. In the top-right graph, the TVP Taylor rule fundamentals predicts almost all the subsequent movements in the exchange rate, yielding a U-statistic significantly less than one (U=0.722). By contrast, fundamentals from the constant-parameter Taylor rules, illustrated in the bottom-left side, are able to correctly signal the changes in the Pound sterling exchange rate in a relatively few cases, resulting in a U-statistic greater than one (1.091). To sum up, we can infer that the relative good performance of the TVP regression across windows and horizons, is associated with a relative better predictive content of the TVP Taylor rule fundamentals, employed in a setting that also allows for the parameters of the forecasting regression to change over time.

## 5.2 Monetary Model, PPP and UIRP Results

Table 3 illustrates the overall performance of the TVP and the FE panel forecasting regressions with classic predictors: the MM, PPP and UIRP.[24] At a glance, regressions based on PPP perform better than the RW benchmark in all forecast windows and horizons. That is, the U's are less than one for over half of the currencies in all forecast windows and nearly all horizons, regardless of the forecasting regression. Note though that the FE panel regression yields an outstanding performance in the first forecast window. In this window, the values of the Median U are substantially below one, and the regression outperforms the RW for a minimum of 14 out of 17, and as many as 15 exchange rates. By contrast and in the same window, the TVP forecasting regression does well for a minimum of nine exchange rates, and a maximum of 11.

Regressions conditioned on fundamentals from the MM fail to improve upon the benchmark for at least half of the currencies in most windows and horizons. For instance, while the FE panel regression outperforms the RW for over 14 out of 17 exchanges rates in the first forecast window and all horizons, in the remaining windows it fails completely. Likewise, the TVP regression performs relatively well for over half of the currencies across forecast horizons mainly in the first window. In the other forecast windows it barely outperforms the RW for at least half of the currencies, except in the last window, at h=8.

Finally, regressions based on UIRP improve upon the RW only for long-horizon forecasts and often with the TVP forecasting approach. In fact, at h=8 and h=12, the TVP regression based on UIRP yields a median U below one in the first and last windows. Consistent with this, the number of U's less than one reach as many as 10 out of 17 in first window and five out of 10 in the last window. By contrast, for the FE panel regression, the Median U is below one in most forecast windows only at h=8. At this forecast horizon, it out-forecasts the RW for 11 out of 17 exchanges in the first window and five out of 10 in the last forecast window. We also note that in cases where the RMSFE of the TVP or FE panel model is lower than that of the RW, the differences in the RMSFE are statistically significant for long-horizon but not for short-horizon forecasts.

Our results are not unusual in the exchange rate literature. For instance, Rossi (2013) also reports a poor performance of FE panel models based on the MM at any horizon. Engel et al. (2008) find improvement over the RW with PPP implied fundamentals at short and most significantly at long-horizon. Cheung et al. (2005) find

---

[24] Detailed results by currency are provided in a results appendix (Appendix E).

**Table 3. Forecast Evaluation: Monetary Model, PPP and UIRP**

| Fundamentals: | TVP Regression | | | | Fixed-effect Panel Regression | | | |
|---|---|---|---|---|---|---|---|---|
| | h=1 | h=4 | h=8 | h=12 | h=1 | h=4 | h=8 | h=12 |
| | Forecast Window: 1995Q1-1998Q4; N=17 | | | | | | | |
| MM | | | | | | | | |
| No. of U's <1 | **11** | **13** | **14** | **13** | **14** | **15** | **15** | **14** |
| No. of DM >1.282 | 5 | 7 | 10 | 11 | 4 | 9 | 12 | 11 |
| Median U | 0.974‡ | 0.923‡ | 0.703‡ | 0.646‡ | 0.956‡ | 0.785‡ | 0.552‡ | 0.623‡ |
| PPP | | | | | | | | |
| No. of U's <1 | **9** | **10** | **11** | **9** | **14** | **15** | **15** | **14** |
| No. of DM >1.282 | 3 | 7 | 7 | 5 | 9 | 12 | 12 | 11 |
| Median U | 0.998‡ | 0.935‡ | 0.978‡ | 0.977‡ | 0.974‡ | 0.866‡ | 0.717‡ | 0.759‡ |
| UIRP | | | | | | | | |
| No. of U's <1 | 5 | **10** | 9 | **10** | **10** | **11** | **11** | **11** |
| No. of DM >1.282 | 1 | 2 | 5 | 8 | 1 | 5 | 10 | 11 |
| Median U | 1.007 | 0.981‡ | 0.985‡ | 0.986‡ | 0.979‡ | 0.969‡ | 0.856‡ | 0.888‡ |
| | Forecast Window: 1999Q1-2013Q1; N=10 | | | | | | | |
| MM | | | | | | | | |
| No. of U's <1 | 1 | 1 | 1 | 1 | 1 | 1 | 1 | 1 |
| No. of DM >1.282 | 0 | 0 | 0 | 0 | 0 | 0 | 1 | 1 |
| Median U | 1.019 | 1.087 | 1.216 | 1.359 | 1.021 | 1.100 | 1.303 | 1.633 |
| PPP | | | | | | | | |
| No. of U's <1 | **8** | **7** | **7** | **7** | **8** | **8** | **7** | 5 |
| No. of DM >1.282 | 1 | 0 | 2 | 4 | 1 | 2 | 3 | 3 |
| Median U | 0.994‡ | 0.971‡ | 0.932‡ | 0.857‡ | 0.994‡ | 0.985‡ | 0.972‡ | 0.955‡ |
| UIRP | | | | | | | | |
| No. of U's <1 | 1 | 1 | 3 | 3 | 2 | 2 | 2 | 2 |
| No. of DM >1.282 | 0 | 0 | 1 | 2 | 0 | 1 | 1 | 1 |
| Median U | 1.008 | 1.030 | 1.096 | 1.204 | 1.009 | 1.037 | 1.070 | 1.137 |
| | Forecast Window: 2007Q1-2013Q1; N=10 | | | | | | | |
| MM | | | | | | | | |
| No. of U's <1 | 2 | 2 | **6** | 4 | 2 | 3 | 4 | 3 |
| No. of DM >1.282 | 0 | 1 | 2 | 1 | 0 | 0 | 0 | 0 |
| Median U | 1.012 | 1.045 | 0.972‡ | 1.021 | 1.007 | 1.015 | 1.057 | 1.301 |
| PPP | | | | | | | | |
| No. of U's <1 | **8** | **8** | **7** | 3 | **7** | **8** | **6** | 4 |
| No. of DM >1.282 | 2 | 4 | 5 | 3 | 3 | 4 | 4 | 4 |
| Median U | 0.989‡ | 0.924‡ | 0.845‡ | 1.029 | 0.991‡ | 0.931‡ | 0.948‡ | 1.260 |
| UIRP | | | | | | | | |
| No. of U's <1 | 2 | 4 | **5** | **5** | 3 | 4 | **5** | 4 |
| No. of DM >1.282 | 0 | 0 | 2 | 4 | 0 | 0 | 3 | 2 |
| Median U | 1.007 | 1.009 | 0.991‡ | 0.944‡ | 1.012 | 1.016 | 0.990‡ | 1.026 |

Notes: This Table summarises the forecasting performance of the TVP forecasting regression and the Fixed-effect panel regression with Monetary (MM), PPP and UIRP fundamentals. See Table 1 for details about the form of the forecasting regressions and how fundamentals are computed or estimated. The benchmark model for both forecasting regressions is the driftless Random Walk (RW). For each regression, set of fundamentals, forecast window and quarterly horizon (h), the "No. of U's < 1" (number of U-statistics less than one), provides the number of currencies for which the model improves upon the RW, since it indicates cases where the RMSFE of the fundamental-based regression is lower than that of the RW. When the U's are less than one for at least half of the currencies in the forecast window, marked in **bold**, then on average, the fundamental-based regression outperforms the benchmark in that window. The "No. of DM > 1.282" (number of DM statistics greater than 1.282) shows cases of rejections of the null hypothesis under the Diebold and Mariano (1995) test of equal forecast accuracy at 10% level of significance. The higher the No. of DM > 1.282, the better the average accuracy of the forecasts of the fundamental-based regression relative to the benchmark is. The "Median U" indicates the middle value of the U-statistic across the sample of N currencies for each forecast window and horizon. When "Median U" is less than or equal to one - marked with the symbol "‡", and U's are less than one for at least half of the currencies in the window, this is also consistent with a better average forecasting performance of the fundamental-based regression relative to the benchmark.

positive results for UIRP at long-horizons. Thus, given the earlier sample period considered in the above studies, our results reinforce the validity of their results for more recent periods.

**5.3 Factor Model Results**

Table 4 reports the performance of our forecasting regressions based on factors from exchange rates.[25] The FE panel regression conditioned on either two or three factors outperforms the RW benchmark in most windows and forecast horizons. For instance, in the first window it performs better for nine out of 17 currencies with F=2 at any h=1, h=4, or h=8. In the last window and with the same number of factors, it improves upon the RW for 50% to 70% of the currencies at h=1, h=4, or h=8. By contrast, the TVP regression with two or three factors improves upon the benchmark for at least half of the currencies across forecast horizons mainly in the window spanning 1995Q1-1998Q4. In the other forecast windows it barely outperforms the RW, except in the last window at h=4.

The Table also shows lack of statistical difference in the accuracy of the forecasts of the models conditioned on factors, relative to the forecasts of the RW. In particular, in all but one forecast window, the null under the DM-test of no difference in the forecast accuracy cannot be rejected for most exchange rate forecasts. The exception occurs in the first forecast window. In this window, the test does reject in favour of the TVP regression for as many as 11 out of 17 currencies, and for the FE panel regression for nine out of 17 currencies. In all other windows and horizons the highest number of rejections never exceeds two.

On balance, the FE panel regression generated lower RMSFE than the RW across all forecast horizons and most forecast windows. In comparison with Engel's et al. (2012) findings, which are based on a FE panel forecasting approach, our results differ from theirs.[26] For all forecast-horizons, Engel's et al. (2012) results with factors estimated via principal components analysis are unsatisfactory for most currencies in all forecast windows they consider. Hence it appears that at least for our last forecast window, the updated sample period is responsible for the satisfactory performance that we find, as Engel et al. (2012) conjectured.

---

[25] Currency by currency results are in a results appendix (Appendix E).
[26] Engel's et al. (2012) results with factors estimated by principal components are presented in their Appendix A., Table A.2B. We note however, that these results cannot be thoroughly compared with ours mainly because the forecast windows and sample period do not completely overlap. Our sample period ends in 2013Q1, while theirs ends in 2007Q4.

**Table 4. Forecast Evaluation: Factor Model**

| | | TVP Regression | | | | Fixed-effect Panel Regression | | | |
|---|---|---|---|---|---|---|---|---|---|
| Fundamentals: | | h=1 | h=4 | h=8 | h=12 | h=1 | h=4 | h=8 | h=12 |
| | | \multicolumn{8}{c}{Forecast Window: 1995Q1-1998Q4} | | | | | | | |
| F1 | | | | | | | | | |
| | No. of U's <1 | 8 | **9** | 8 | 6 | 6 | 7 | 8 | 8 |
| | No. of DM >1.282 | 5 | 6 | 7 | 5 | 2 | 4 | 3 | 4 |
| | Median U | 1.000 | 0.989‡ | 1.035 | 1.126 | 1.035 | 1.061 | 1.032 | 1.091 |
| F2 | | | | | | | | | |
| | No. of U's <1 | **12** | **12** | **13** | **11** | **9** | **9** | **9** | 8 |
| | No. of DM >1.282 | 8 | 9 | 11 | 9 | 7 | 8 | 7 | 6 |
| | Median U | 0.974‡ | 0.883‡ | 0.771‡ | 0.782‡ | 0.985‡ | 0.936‡ | 0.985‡ | 1.073 |
| F3 | | | | | | | | | |
| | No. of U's <1 | **12** | **12** | **12** | **9** | **10** | **11** | **10** | **9** |
| | No. of DM >1.282 | 8 | 9 | 10 | 9 | 8 | 9 | 7 | 7 |
| | Median U | 0.977‡ | 0.904‡ | 0.754‡ | 0.855‡ | 0.980‡ | 0.879‡ | 0.904‡ | 0.951‡ |
| | | \multicolumn{8}{c}{Forecast Window: 1999Q1-2013Q1} | | | | | | | |
| F1 | | | | | | | | | |
| | No. of U's <1 | 1 | 2 | 2 | 1 | 2 | 2 | 2 | 4 |
| | No. of DM >1.282 | 0 | 0 | 0 | 0 | 0 | 0 | 0 | 0 |
| | Median U | 1.009 | 1.032 | 1.121 | 1.340 | 1.010 | 1.048 | 1.153 | 1.209 |
| F2 | | | | | | | | | |
| | No. of U's <1 | 2 | 3 | 1 | 2 | 4 | 4 | 4 | **6** |
| | No. of DM >1.282 | 0 | 0 | 0 | 0 | 0 | 0 | 0 | 0 |
| | Median U | 1.008 | 1.035 | 1.152 | 1.424 | 1.005 | 1.016 | 1.012 | 0.987‡ |
| F3 | | | | | | | | | |
| | No. of U's <1 | 1 | 3 | 2 | 2 | 2 | 3 | 4 | 4 |
| | No. of DM >1.282 | 0 | 0 | 0 | 0 | 0 | 0 | 0 | 1 |
| | Median U | 1.007 | 1.035 | 1.163 | 1.324 | 1.004 | 1.019 | 1.051 | 1.046 |
| | | \multicolumn{8}{c}{Forecast Window: 2007Q1-2013Q1} | | | | | | | |
| F1 | | | | | | | | | |
| | No. of U's <1 | 2 | 3 | 4 | 1 | 4 | **7** | 2 | 2 |
| | No. of DM >1.282 | 0 | 1 | 1 | 1 | 0 | 0 | 1 | 1 |
| | Median U | 1.005 | 1.017 | 1.030 | 1.672 | 1.001 | 0.984‡ | 1.139 | 1.682 |
| F2 | | | | | | | | | |
| | No. of U's <1 | 1 | 4 | 2 | 1 | **5** | **7** | **5** | 3 |
| | No. of DM >1.282 | 0 | 0 | 0 | 0 | 1 | 2 | 1 | 0 |
| | Median U | 1.006 | 1.015 | 1.095 | 1.650 | 0.999‡ | 0.955‡ | 1.011 | 1.416 |
| F3 | | | | | | | | | |
| | No. of U's <1 | 2 | **5** | 2 | 1 | 4 | **7** | 3 | 2 |
| | No. of DM >1.282 | 0 | 1 | 1 | 1 | 0 | 1 | 2 | 1 |
| | Median U | 1.003 | 1.005 | 1.101 | 1.635 | 1.001 | 0.987‡ | 1.064 | 1.498 |

Notes: This Table summarises the forecasting performance of the TVP forecasting regression and the Fixed-effect panel regression with factors (F) extracted from exchange rates. See Table 1 for details about the form of the forecasting regressions. Factors are obtained via principal component analysis. The benchmark model for both forecasting regressions is the driftless Random Walk (RW). For each regression, set of fundamentals, forecast window and quarterly horizon (h), the "No. of U's < 1" (number of U-statistics less than one), provides the number of currencies for which the model improves upon the RW, since it indicates cases where the RMSFE of the fundamental-based regression is lower than that of the RW. When the U's are less than one for at least half of the currencies in the forecast window, marked in **bold**, then on average, the fundamental-based regression outperforms the benchmark in that window. The "No. of DM > 1.282" (number of DM statistics greater than 1.282) shows cases of rejections of the null hypothesis under the Diebold and Mariano (1995) test of equal forecast accuracy at 10% level of significance. The higher the No. of DM > 1.282, the better the average accuracy of the forecasts of the fundamental-based regression relative to the benchmark is. The "Median U" indicates the middle value of the U-statistic across the sample of $N$ currencies for each forecast window and horizon. When "Median U" is less than or equal to one - marked with the symbol "‡", and U's are less than one for at least half of the currencies in the window, this is also consistent with a better average forecasting performance of the fundamental-based regression relative to the benchmark.

**Table 5. Overall Model's Ability to Outperform the Benchmark across Forecast Windows**

| Fundamentals: | TVP Regression | | Fixed-effect Panel Regression | |
|---|---|---|---|---|
| | Short-Run Forecasts | Long-Run Forecasts | Short-Run Forecasts | Long-Run Forecasts |
| TR | Yes | Yes | No | Yes |
| MM | No | Yes | No | No |
| PPP | Yes | Yes | Yes | Yes |
| UIRP | No | Yes | No | Yes |
| Factors | Yes | No | Yes | Yes |

Notes: This Table summarises the overall performance of the TVP regression and the Fixed-effect Panel regression conditioned on TR, MM, PPP, UIRP or factors (F). Refer to Table 1 for details about the form of the forecasting regressions and how fundamentals are computed or estimated. The benchmark model for all regressions is the driftless Random Walk (RW). The Table provides the answer to the question: "Does the regression conditioned on any of the fundamentals outperform the benchmark for at least half of the currencies in most forecast windows, at short or long-horizon forecasts?" The short-horizon comprises h=1 or h=4 quarters, while the long-horizon includes h=8 or h=12 quarters.

We sum up our empirical results in Table 5. There we provide the answer to the question: "Does the regression conditioned on any of the fundamentals, TR, MM, PPP, UIRP, and factors outperform the RW benchmark for at least half of the currencies in most forecast windows, at short or long-horizon forecasts?" Note that the short-horizon is defined as h=1 or h=4 quarters-ahead, and the long-run as h=8 or h=12 quarters-ahead. It turns out that the answer is a "Yes" for the TVP forecasting regression conditioned on TR or PPP at any forecast horizon. The answer is also positive for the same forecasting approach with MM and UIRP and at long-horizon, and factors for short-horizon forecasts. The FE panel forecasting approach yields a positive answer with PPP and factors at any horizon, and TR and UIRP for long-horizon forecasts. Thus, given the relative good performance of the TVP regression with Taylor rule fundamentals we deem nonlinearities to be important.[27] We assess the robustness of these findings below.

## 6. Robustness Checks

We verified the robustness of the empirical findings in the previous Section to various situations. These include: (i) change in base currency from the U.S. dollar to the Pound sterling in all models; (ii) use of unemployment gap rather than output gap in the Taylor rule specifications; (iii) use of monthly data, instead of quarterly data – only for Taylor rule fundamentals (iv) comparison with other forecasting regressions and methods, along with alternative approaches to estimate Taylor rule fundamentals; and (v) estimation of factors by maximum likelihood rather than principal components in the factor model. We describe the main findings in what follows and append all the summary results in Appendix D.[28]

---

[27] However, unlike Molodtsova and Papell (2009) who find more precise forecasts with Taylor rule fundamentals relative to PPP fundamentals, we find the opposite, and we attribute these findings to the quarterly frequency of our data.

### 6.1 Change in Base Currency

Chen et al. (2010) and Engel et al. (2012) stress the importance of verifying the sensitiveness of the model's forecasting performance to a different base numeraire. Accordingly, we replace the U.S. dollar base currency by the Pound sterling (GBP), define all the home country variables relative to the United Kingdom (UK), and repeat the forecasting exercise. Table 6 presents the overall findings. The Table is analogous to Table 5 in the empirical Section, except for the base currency. Here, with the GBP as a base currency it provides the answer to the same question: "Does the regression conditioned on any of the fundamentals TR, MM, PPP, UIRP, or factors outperform the RW benchmark for at least half of the currencies in most forecast windows, at short or long-horizon forecasts?" We highlight the main findings below.

**Table 6. Overall Model's Ability to Outperform the Benchmark across Forecast Windows (GBP base currency)**

| Fundamentals: | TVP Regression | | Fixed-effect Panel Regression | |
|---|---|---|---|---|
| | Short-Run Forecasts | Long-Run Forecasts | Short-Run Forecasts | Long-Run Forecasts |
| TR | Yes | Yes | No | No |
| MM | Yes | No | Yes | Yes |
| PPP | Yes | Yes | Yes | Yes |
| UIRP | No | Yes | No | No |
| Factors | No | Yes | No | Yes |

Notes: This Table summarises the overall performance of the TVP regression and the Fixed-effect Panel regression conditioned on TR, MM, PPP, UIRP or factors (F). Refer to Table 1 for details about the form of the forecasting regressions and how fundamentals are computed or estimated. The benchmark model for all regressions is the driftless Random Walk (RW). Here, the base currency is the Pound Sterling (GBP) rather than the U.S. dollar. The Table provides the answer to the question: "Does the regression conditioned on the fundamental considered outperform the benchmark for at least half of the currencies in most forecast windows, at short or long-horizon forecasts?" The short-horizon comprises h=1 or h=4 quarters, while the long-horizon includes h=8 or h=12 quarters.

The ability of the TVP regression conditioned on Taylor rules to improve on the driftless random walk (RW) remains robust in most forecast windows and for both, short and long-horizon forecasts.[29] In contrast, while with the GBP as a base currency the FE panel also performs well for short and long-horizon forecasts, its relative good performance is confined to one window. Moreover, the number of currencies for which it out-forecasts the RW is less than that of the TVP regression relative to the RW. We therefore infer that allowing for the possibility of a monetary policy that responds to macroeconomic conditions in time-evolving manner provides valuable information for exchange rate forecasts, irrespective of the base currency considered.

---

[28] Results by currency are omitted to save space.

With regards to the sets of information from the MM, PPP, UIRP and factors, the main results from the empirical Section above remain largely intact for PPP and UIRP, but they differ for MM and factors. With PPP, either of the forecasting regressions improves upon the RW for at least half of the currencies in the majority of the forecast windows at all horizons; with an outstanding performance of the FE panel regression. With UIRP, the TVP regression still improves upon the benchmark solely for long-horizon forecasts, while previous results for the FE panel regression no longer hold with the GBP base currency. Results from models conditioned on MM do not hold up to a change in the base currency. In fact, all the findings with the GBP as the base currency are opposite to the findings with the USD base currency.[30] For regressions conditioned on factors, only results for the FE panel regression at long-horizon are robust to change in base currency. For the same regression and short horizons, our findings of relative good performance in the empirical Section no longer hold here. Likewise, previous results for the TVP regression are opposite to those we obtain here with the GBP as a base currency.

On balance and across the forecasting approaches we consider, the improvement over the RW benchmark we report in the empirical Section remains robust to the change in base currency mostly with the TVP regression. To be precise, results from the TVP regression still hold with information sets from the (i) Taylor rules at short (h=4) and long (h=8 and h=12) forecast horizons; (ii) PPP at all horizons; and (iii) UIRP at long-horizon (h=12). Conversely, findings from the FE panel regression remain valid with information sets from the PPP at all horizons and factors at long-horizon (h=12). We also note that as in the results in the empirical Section, in cases where models do better, the differences in the forecast accuracy are significant at long-horizon but not short-horizon forecasts, irrespective of the information set.

---

[29] In particular, results in Appendix D, Tables D.A1 show that in the window spanning the recent financial turmoil and with the Pound sterling as the base currency, the regression with any of the Taylor rule specifications out-forecasts the benchmark in about 50% to 80% of the exchange rates in the sample, at short (h=4) and long (h=8, and h=12) forecast horizons.

[30] For instance, from no predictability with the U.S. base currency, we find that, with the Pound sterling the FE panel model outperform the benchmark for at least half of the currencies in the majority of the forecast windows and horizons.

**6.2 Taylor Rules with Unemployment Gap, instead of Output Gap**

Monetary policy rules can focus on the unemployment gap rather than the output gap. Molodtsova and Papell (2013) find that Taylor rules with the unemployment gap outperform specifications with the output gap. Accordingly, we replace the output gap by the unemployment gap and proceed with the forecasting exercise with either the U.S. dollar base currency or the Pound sterling. We focus on the forecasting window spanning 2007Q1-2013Q1. However, due to unavailability of data on unemployment gap for all the countries in the sample, we forecast nine exchange rates. In general, previous results for this window remain robust for both forecasting regressions. That is, the TVP regression still improves upon the benchmark for over half of the currencies in the window and h=4; h=8 and h=12, regardless of the base currency. Here, the strongest performance occurs with Taylor rules specifications with heterogeneous coefficients and no smoothing ($TR_{en}$), as well as Taylor rules with homogenous coefficients and no smoothing ($TR_{on}$). The FE panel regression also improves upon the RW benchmark for at least half of the currencies in the window at short and long-horizon forecasts, irrespective of the base currency. However, the number of currencies for which it out-forecasts the RW is less than that of the TVP regression relative to the RW for most specifications and particularly at h=4; h=8 and h=12.

**6.3 Forecasting with Monthly Data**

To verify how results would vary to the frequency of data used we experimented with monthly data and regressions conditioned on Taylor rules. We concentrate on the last forecast window and five monthly forecast horizons: h=1; h=3; h=12; h=24 and h=36. Comparable results for the last four horizons at quarterly data frequency are in the last forecast window of Table 2. Overall results from the FE panel regression are qualitatively similar for the two frequencies, especially at longer horizons, i.e., h=24 and h=36. At these horizons, most Taylor rule specifications improve upon the RW for a minimum of 50% of the nine currencies in the window. In contrast, figures from the TVP regression are slightly less aligned with those from quarterly data. That is, with monthly data this regression improves upon the RW for up to 44% of the currencies in the sample for horizons of h=12 and over, reaching as many as 5 out of 9 at h=12 and h=36. Thus, in terms of forecasting regressions with monthly data, the FE panel regression does relatively well.

## 6.4 Changes in Forecasting Regression and Forecasting Method

As noted previously, the regressions used to estimate Taylor rule fundamentals and the forecasting regression allow for time-variation in parameters. A number of papers, including Molodstova and Papel (2009, 2013) and Rossi (2013), estimate Taylor rule fundamentals via a single-equation constant-parameter (SECP) model. These fundamentals are then employed as conditioning information for a SECP forecasting regression. In addition, they adopt a rolling window forecasting approach. Accordingly, we explore their methodology. In particular, we define the rolling windows such that the number of forecasts generated using this method matches with the forecasts in the recursive forecasting method. We focus on all forecast windows and the USD base currency.

Succinctly, the results indicate that with the methodology just described, the regression's overall performance over the RW benchmark is similar to that of the FE panel regression as discussed in the empirical Section. That is, it improves upon the benchmark for at least half of the currencies across forecast windows and horizons only for long-horizon (h=8 and h=12) forecasts. Thus, we still find support for the TVP forecasting regression considered in the empirical Section.[31]

In comparison with other studies that employ a SECP forecasting regression conditioned on Taylor rule fundamentals estimated with a SECP, our results differ from theirs. For example, focusing on monthly data up to June 2006, Molodtsova and Papel (2009) find improvement upon the RW benchmark for as many as 10 out of 12 OECD currencies at one-month-ahead forecast horizon. Rossi (2013) uses monthly data up to 2011. She finds improvement over the RW for seven out of 17 currencies at one-month forecast horizon, but for none of the currencies at long horizons. While there are potentially several reasons why our results differ from those in the above studies, the most obvious aspects are the differences in the data-frequency, sample period and forecast windows. Whilst using monthly data to forecast in our framework is in principle feasible with Bayesian MCMC methods, the computational demands of higher frequency data with such methods are significant.

---

[31] Apart from the forecasting regression and methodology just outlined, we have also experimented with other forecasting regressions along with alternative approaches to compute fundamentals. In particular, using a recursive forecasting method we consider the following combinations: (i) computing Taylor rule fundamentals with a TVP regression, but use a single-equation constant-parameter forecasting regression and (ii) the opposite of combination (i). In general, the findings (not reported) suggest that the results for the TVP forecasting approach discussed in the empirical Section remain robust. The findings also suggest that allowing for TVP when estimating Taylor rule fundamentals seems to be more important than allowing for TVP in the forecasting regression. We infer this from the performance of the models in (i) and (ii). When using a combination defined by (i), the performance of the forecasting regression is very close and occasionally better than that we obtain in the empirical Section with our TVP regression approach. In contrast, when using (ii), while the results are also supportive to the importance of accounting for non-linearity, the overall performance is inferior when compared to the option defined in (i) and to our principal TVP regression.

**6.5 Factors Estimated via Maximum Likelihood**

Our previous results for the factor model were based on factors estimated via principal components. To assess the sensitiveness of models' forecasting performance to the method used to estimate factors, we alternatively estimated them via maximum likelihood. As in the baseline results, the base currency is the U.S. dollar and we focus in the last forecast window. Overall, the relative good performance of the FE panel regression reported in the empirical Section remains robust. This regression still improves upon the RW for at least half of the currencies in the window at h=1, h=4, and h=8. However, with factors estimated by maximum likelihood the TVP regression also outperforms the RW for over half of the currencies in the window, reaching as many as eight out of 10 at either h=4 or h=8. In comparison with Engel's et al. (2012) FE panel regression with factors estimated by maximum likelihood, the results are similar at h=8, but they differ for h=1, h=4 and h=12.[32] That is, at h=8 and their last forecast window (1999Q1-2007Q4) they find improvement over the RW for at least half of the currencies. However, they also find improvement upon the benchmark for h=12, but fail for h=1 and h=4. By contrast, here we find improvement over the benchmark for h=1 and h=4, but fail for h=12.

**7. Conclusions**

An expanding literature articulates the view that Taylor rules are helpful in predicting exchange rates, in the sense that structural exchange rate models that incorporate Taylor rule fundamentals exhibit predictive content for exchange rates. However, a number of studies point out that the predictability appears to turn up in some periods but not others. At the same time, an established literature documents time-evolving macroeconomic conditions and relationships among macroeconomic variables. Taken together, these observations raise the possibility that accounting for nonlinearities may be fundamental in pinning down models forecast ability. To explore this possibility we estimate time-varying Taylor rules and examine their predictive content for exchange rates in a framework that also allows for the parameters of the forecasting regression to change over time. We focus in three alternative forecast windows and four quarterly forecast horizons (h). In most forecast windows and horizons, our approach yields a lower Root Mean Squared Forecast Error (RMSFE) than the driftless random walk (RW) for at least half of the currencies in the sample, reaching as many as 11 out of 17 in one of the windows at h=4 or h=8, and eight out of 10 in another. Results are particularly strong in the window that encompasses the recent financial turmoil (2007Q1-2013Q1), where presumably significant changes in the fundamentals occurred. We proceed and experiment with the usual approach in the literature, whereby constant-parameter models are used to compute fundamentals and forecast, but find a limited performance. Moreover, while the Time-Varying Parameter (TVP) approach is robust to various situations, the constant-parameter approach fails some of our robustness checks. Overall, whilst our findings confirm that Taylor rules are relevant in predicting exchange rates, they also reveal the importance of accounting for nonlinearities, especially in the more recent turbulent times.

---

[32] Engel's et al. (2012) results are presented in Table 4 of their paper.

To assess the performance of other predictors in our updated dataset we also attempt using either a TVP forecasting regression or a constant-parameter forecasting regression conditioned on factors from exchange rates (as in Engel et al., 2012), or on an information set from the Monetary Model (MM), Purchasing Power Parity (PPP), and Uncovered Interest Rate Parity (UIRP). As usual in the literature, we compute the information set with calibrated, rather than estimated coefficients. Our results indicate that the TVP forecasting regression generates lower RMSFE than the benchmark for over half of the currencies in most windows, when conditioned on PPP at all horizons and UIRP at long-horizon. The constant-parameter regression outperforms the RW for half or more currencies across windows when based on PPP and factors at all horizons and UIRP at long-horizon. However, results are only robust for PPP at all horizons and factors at long-horizon (h=12). The predictive content of the MM is barely robust, regardless of the forecasting model. Looking at other papers employing identical predictors these results are not unique, especially within the constant-parameter modelling approach. Some examples include, Rossi (2013) for MM, Engel et al. (2008) for PPP, Cheung et al. (2005) for UIRP, and Engel et al. (2012) for factors. However, authors such as Park and Park (2013) suggest that it is also important to allow for nonlinearities when computing the information set from these predictors. Hence, further work along these lines may also be fruitful.

## Appendix A. Derivation of Interest Rate Differentials Implied by Taylor Rules

In this Appendix interest rate differentials implied by Taylor rules under alternative assumptions are derived. Taylor (1993) suggested the following rule for monetary policy:

$$i_t^T = \pi_t + \tau_1(\pi_t - \pi^T) + \tau_2 \bar{y}_t + r^T \tag{A1}$$

where, $i_t^T$, is the target nominal short-term interest rate set by the central bank;[33] $\pi_t$ is the inflation rate; $\pi^T$, is the target inflation rate; $\bar{y}_t = (y_t - y_t^p)$, is the output gap measured as deviation of actual real GDP ($y_t$) from its potential level ($y_t^p$); and $r^T$, is the equilibrium real interest rate. In equation (A1) the central bank increases the short-term interest rate when inflation is above the target and/or output is above its potential level. In Taylor (1993), $\tau_1 = 0.5$, $\pi^T = 2\%$, $\tau_2 = 0.5$, and $r^T = 2\%$.

We can rearrange equation (A1) by combining the constant parameters, $\pi^T$ and $r^T$, and collecting the inflation rate terms $\pi_t$, to obtain:

$$i_t^T = \delta_0 + \delta_1 \pi_t + \delta_2 \bar{y}_t \tag{A2}$$

where, $\delta_0 = r^T - \tau_1 \pi^T$; $\delta_1 = (1 + \tau_1)$; and $\delta_2 = \tau_2$. According to equation (A2), an increase in inflation for instance by 1%, calls for more than 1% augment in the short-term nominal interest rate by the central bank, since $\delta_1 = (1 + \tau_1)$. Thus, the Taylor principle is maintained.

Following the empirical evidence in Clarida et al. (1998), it is typical to assume that most countries, apart from the U.S, target the real exchange rate ($q_t$). Then, from equation (A2) we obtain:

$$i_t^T = \delta_0 + \delta_1 \pi_t + \delta_2 \bar{y}_t + \delta_3 q_t \tag{A3}$$

where: $$q_t = s_t + p_t^* - p_t \tag{A4}$$

and $s_t$ is the log exchange rate, defined as the home price of foreign currency; $p_t$, is the log of the price level; the asterisk (*) indicates the foreign country. The inclusion of the real exchange rate in equation (A3) accords with the assertion that the central bank targets the level of exchange rate to ensure that PPP holds (Engel and West, 2005). Thus, an increase in $q_t$ is associated with a rise in $i_t^T$.

Equation (A3) can be further extended. If the central bank sets monetary policy at each point in time by adjusting the actual interest rate to eliminate a fraction $(1 - \theta)$ of the gap between the current interest rate target and its recent past level, then it limit volatility in interest rates (Clarida et al., 1998). Algebraically:

$$i_t = (1 - \theta)i_t^T + \theta i_{t-1} + \varepsilon_t \tag{A5}$$

Substituting equation (A3) in (A5) yields:

---

[33] Under the assumption that the target for the nominal interest rate is always attained, there is no difference between the actual and the target interest rate (Molodtsova and Papell, 2009).

$$i_t = (1-\theta)(\delta_0 + \delta_1 \pi_t + \delta_2 \bar{y}_t + \delta_3 q_t) + \theta i_{t-1} + \varepsilon_t \tag{A6}$$

and rearranging equation (A6) we obtain:

$$i_t = \phi_c + \phi_1 \pi_t + \phi_2 \bar{y}_t + \phi_3 q_t + \phi_4 i_{t-1} + \varepsilon_t \tag{A7}$$

where, $\phi_c = (1-\theta)\delta_0$; $\phi_1 = (1-\theta)\delta_1$; $\phi_2 = (1-\theta)\delta_2$; $\phi_3 = (1-\theta)\delta_3$ and $\phi_4 = \theta$.

In contrast with the immediate adjustment in the policy interest rate implied by equations (A2) and (A3), in (A7) the change in the interest rate is gradual. In response to an inflation rate that is above the target, the central bank increases the interest rate by $(1-\theta^p)\delta_1$ at each $p$ period, with $p = \{1 \dots P\}$. As $p$ increases, the maximum change in policy interest rate will converge to $\delta_1$ as in equation (A2), satisfying the Taylor principle (see Molodtsova and Papell, 2009).

Define equation (A7) as the home country's Taylor rule. The foreign country is the U.S., and its Taylor rule is:

$$i_t^* = \phi_c^* + \phi_1^* \pi_t^* + \phi_2^* \bar{y}_t^* + \phi_4^* i_{t-1}^* + \varepsilon_t^* \tag{A8}$$

Note that in equation (A8), it is assumed that the U.S. does not target the real exchange rate, and hence the foreign counterpart of the term $\phi_3 q_t$ in equation (A7) is omitted in equation (A8). This is a standard assumption in the literature and it is in the spirit of Clarida et al. (1998), Engel and West (2005), Mark (2009), among others. Subtracting the foreign country's Taylor rule (Eq. (A8)) from the home (Eq. (A7)), we obtain the following interest rate differential equation:

$$i_t - i_t^* = \phi_0 + \phi_1 \pi_t - \phi_1^* \pi_t^* + \phi_2 \bar{y}_t - \phi_2^* \bar{y}_t^* + \phi_3 q_t + \phi_4 i_{t-1} - \phi_4^* i_{t-1}^* + \mu_t \tag{A9}$$

where, $\phi_0 = \phi_c - \phi_c^*$ and $\mu_t = \varepsilon_t - \varepsilon_t^*$.

In equation (A9), the constant parameter $\phi_0$ allows for the equilibrium real interest rates and inflation targets to differ across home and foreign countries. By contrast, if we assume that the equilibrium real interest rate and the inflation target of the home and foreign country are identical, then the constant is excluded. Also in equation (A9), all coefficients are heterogeneous, only the home central bank targets the real exchange rate and both countries limit volatilities in interest rates. In terms of the parameters of equation (A9) we have: $\phi_0 \neq 0$; $\phi_1 \neq \phi_1^*$; $\phi_2 \neq \phi_2^*$; $\phi_4 \neq \phi_4^*$; $\phi_3 \neq 0$.

We can relax some of the assumptions in equation (A9) to derive alternative specifications. For instance, we can assume that the inflation target and the real equilibrium interest rate of the home and foreign country are similar such that their difference is zero. In addition we can impose that the coefficients on inflation and the output gap are equal between home and foreign country; or consider that central banks do not smooth interest rate. In Table A below we consider some Taylor rule variants under such alternative assumptions. The variants are inspired by Engel and West (2005), Molodtsova and Papell (2009) and Engel et al. (2008). [34]

**Table A. Interest Rate Differentials Implied by Taylor Rules under Alternative Assumptions**

| Assumption: | Coefficients in equation (A9) | Model Variant |
|---|---|---|
| (i) The equilibrium real interest rate and the inflation target of the home and foreign country are identical, hence their difference is zero; (ii) The coefficients on inflation and the output gap are equal between home and foreign country; (iii); Central banks do not smooth interest rate; (v) The home central bank targets the real exchange rate. <br><br> *TR$_{on}$: Homogenous rule, asymmetric and without interest rate smoothing.* | $\phi_0 = 0$; <br> $\alpha_1 = \phi_1 = \phi_1^*$; <br> $\alpha_2 = \phi_2 = \phi_2^*$; <br> $\phi_4 = \phi_4^* = 0$; | $i_t - i_t^* = \alpha_1(\pi_t - \pi_t^*)$ <br> $+ \alpha_2(\bar{y}_t - \bar{y}_t^*)$ <br> $+ \phi_3 q_t + \mu_t$ |
| (i) The equilibrium real interest rate and the inflation target of the home and foreign country are identical, hence their difference is zero; (ii) The coefficients on inflation, the output gap and the interest rate smoothing are equal between home and foreign country; and (iii) The home central bank targets the real exchange rate. <br><br> *TR$_{os}$: Homogenous rule, asymmetric and with interest rate smoothing.* | $\phi_0 = 0$; <br> $\alpha_1 = \phi_1 = \phi_1^*$; <br> $\alpha_2 = \phi_2 = \phi_2^*$; <br> $\alpha_3 = \phi_4 = \phi_4^*$; | $i_t - i_t^* = \alpha_1(\pi_t - \pi_t^*)$ <br> $+ \alpha_2(\bar{y}_t - \bar{y}_t^*)$ <br> $+ \alpha_3(i_{t-1} - i_{t-1}^*)$ <br> $+ \phi_3 q_t + \mu_t$ |
| (i) The equilibrium real interest rate and the inflation target of the home and foreign country are identical, hence their difference is zero; (ii) The coefficients on inflation and the output gap are allowed to differ between home and foreign country; (iii) Central banks do not smooth interest rate; and (iv) The home central bank targets the real exchange rate. <br><br> *TR$_{en}$: Heterogeneous rule, asymmetric and without interest rate smoothing.* | $\phi_0 = 0$; <br> $\phi_4 = \phi_4^* = 0$; | $i_t - i_t^* = \phi_1\pi_t - \phi_1^*\pi_t^*$ <br> $+ \phi_2\bar{y}_t - \phi_2^*\bar{y}_t^*$ <br> $+ \phi_3 q_t + \mu_t$ |

Notes: All the assumptions are relative to equation (A9) in this Appendix. That is: $i_t - i_t^* = \phi_0 + \phi_1\pi_t - \phi_1^*\pi_t^* + \phi_2\bar{y}_t - \phi_2^*\bar{y}_t^* + \phi_3 q_t + \phi_4 i_{t-1} - \phi_4^* i_{t-1}^* + \mu_t$. The alternative specifications are then derived in line with the assumptions in the first column of the Table.

---

[34] Engel and West (2005) derive a Taylor rule specification similar to the one denoted TR$_{os}$ in Table A. Molodtsova and Papell (2009) consider 16 alternative specifications, including the three included in our Table. Engel et al. (2008) consider the specification denoted TR$_{on}$ in the Table, with posited coefficients as follows: $\alpha_1 = 1.5$ $\alpha_2 = 0.1$ and $\phi_3 = 0.1$.

**Appendix B. Details of Bayesian Estimation of the Time-Varying Parameter Models**

This Appendix B describes the estimation of the time-varying parameter models considered in the paper. We use Bayesian methods in the spirit of Kim and Nelson (1999), Koop (2003), and Blake and Mumtaz (2012) to estimate Time-Varying Taylor rules and the Time-Varying Parameters of the forecasting regressions. Here we provide details of the prior Hyperparameters, the conditional posterior distributions and the steps or algorithm used to draw from these conditional distributions.

All the Time-Varying Parameter (TVP) models we consider can be cast in a normal linear state-space model with the following representation:

$$y_t = H_t \beta_t + A z_t + e_t, \tag{B1}$$

$$\beta_t = \mu + F \beta_{t-1} + v_t, \tag{B2}$$

$$e_t \sim i.i.d.(0, R); \tag{B3}$$

$$v_t \sim i.i.d.(0, Q); \tag{B4}$$

$$Cov(e_t, v_t) = 0. \tag{B5}$$

Equation (B1) is the observation or measurement equation and equation (B2) is the state or transition equation. In the observation equation $y_t$ is an $n \times 1$ vector of observations on $n$ variables over time; $\beta_t$ is a $k \times 1$ vector of unobserved state variables (e.g. the time-varying coefficients); $H_t$ is an $n \times k$ matrix with elements that are not fixed or given as data (depending on the model) and links the observed variables in $y_t$ to the unobserved state variable $\beta_t$; $z_t$ is an $r \times 1$ vector of exogenous variables with time-invariant coefficients $A$. The state equation describes the dynamics of the unobserved states.

Our models constitute special cases of the general form of the system given by equations (B1) and (B2). In particular, we do not include additional variables other than those in $H_t$ and hence, $A z_t = 0$. Additionally, the state variable evolves according to a random walk, such that $\mu = 0$ and $F$ is an identity matrix ($I_k$).

To proceed in terms of Bayesian inference, we need to (i) elicit priors for the unknown parameters, (ii) specify the form of their posterior conditional distributions and finally (iii) use an algorithm to draw samples from the specified conditional posterior distribution. What follows outlines these steps.

## I. Priors Hyperparameters and Initial Conditions

The form of our TVP models suggests that we need priors for the variance $R$ of the measurement equation and the variance-covariance matrix $Q$ of the transition equation. In addition, to recover the unobserved state variable $\beta_t$ we also need initial conditions or starting values for the Kalman filter. That is the initial state, $\beta_{0/0}$, and its initial variance, $P_{0/0}$. See Box A for details of the Kalman filter.

To parameterize the prior distributions and initial conditions we use pre-sample information. Specifically, we use a training sample of $T_0 = 20$ observations to estimate via OLS estimator a fixed-coefficient model which is a counterpart to equation (B1) in this Appendix. In particular, the estimated coefficients and the respective covariance matrix are set as initial conditions for the Kalman filter. In notation:

$$\beta_{0/0} \equiv \beta_{OLS} = (H'_{0t}H_{0t})^{-1}(H'_{0t}y_{0t}), \tag{B6}$$

$$P_{0/0} \equiv P_{OLS} = \Sigma_0 \otimes (H'_{0t}H_{0t})^{-1}, \tag{B7}$$

where, $\beta_{OLS}$ and $P_{OLS}$ are, respectively, the coefficients' vector and covariance matrix estimated via OLS and finally,

$$\Sigma_0 \equiv \Sigma_{OLS} = \frac{(y_{0t} - H_{0t}\beta_0)'(y_{0t} - H_{0t}\beta_0)}{T_0 - k}, \tag{B8}$$

where here, $k$ is the number of coefficients estimated.

---

**Box A. The Kalman Filter**

Consider our state-space model given by the system of equations (B1) and (B2). The Kalman filter is a recursive algorithm for computing the optimal estimate of $\beta_t$ given an appropriate information set and knowledge of the other parameters of the state-space. Consider for instance that the parameters $H, A, R, u, F, Q$ are known. The algorithm consists in the steps summarised in figure A.

The first step is to define initial conditions. For a stationary state vector, the unconditional mean and its associated covariance matrix may be employed as initial conditions. For non-stationary processes, unconditional means and covariance matrixes do not exist. In this case the initial condition for the state variable $\beta_{0/0}$ may be defined arbitrarily. However, to indicate a high uncertainty surrounding this arbitrary defined value, we must set the diagonal elements of the covariance matrix $P_{0/0}$ to a very large number. For more details on initial conditions see Kim and Nelson, (1999).

In the second step, i.e., for period $t = 1$ we can now form an optimal prediction of $y_{1/0}$ after computing $\beta_{1/0}$ and its associated covariance matrix, $P_{1/0} = FP_{0/0}F' + Q$. Note that the subscripts make it clear that we are conditioning on the information set at $t = 0$, i.e., contained in our prior initial conditions, $\beta_0$ and $P_0$.

In the third step, we use the observed value of $y_t$ at $t = 1$ to compute the prediction error, $n_{1/0} = y_1 - y_{1/0}$ and its covariance matrix $f_{1/0} = HP_{t\backslash t-1}H' + R$. The information contained in the prediction error can be used to improve the initial inference about $\beta_t$. Thus in the fourth and last step, we can compute $\beta_{1/1} = \beta_{1/0} + K_t n_{1/0}$; where $K_t$ is the Kalman gain, which indicates the weight attributed to new information. It constitutes the ratio of the prediction error variance associated with uncertainty about $\beta_{1/0}$ and the prediction error variance of the error term in equation (B.3). Thus, $K_t = P_{1/0}H'(f_{1/0})^{-1}$. A high uncertainty about $\beta_{1/0}$ implies that more weight is attributed to new information in the prediction error.

The second, third, and fourth steps are then repeated for $t = 2,3,4\ldots,T$. The filter provides an optimal estimate of the state variable at each point in time.

**Figure A. State-Space Model and the Kalman Filter Algorithm**

| | |
|---|---|
| Step 1: Define starting values for the state ($\beta_{t-1/t-1}$) and its covariance matrix ($P_{t-1/t-1}$) at $t = t-1$, i.e. Define initial conditions. | $\beta_{t-1\backslash t-1}$ <br> $P_{t-1\backslash t-1}$ |
| Step 2: At $t = 1$, predict the state vector and its associated covariance matrix. | $\beta_{t\backslash t-1} = \mu + F\beta_{t-1\backslash t-1}$ <br> $P_{t\backslash t-1} = FP_{t-1\backslash t-1}F' + Q$ |
| Step 3: Calculate prediction error ($n_{t/t-1}$) and its covariance matrix ($f_{t/t-1}$). | $y_{t\backslash t-1} = H\beta_{t\backslash t-1} - Az_t$ <br> $n_{t\backslash t-1} = y_t - y_{t\backslash t-1}$ <br> $f_{t\backslash t-1} = HP_{t\backslash t-1}H' + R$ |
| Step 4: Refine inference about ($\beta_{t/t}$) via Kalman gain. | $K_t = P_{t\backslash t-1}H'f_{t\backslash t-1}^{-1}$ <br> $\beta_{t\backslash t} = \beta_{t\backslash t-1} + K_t n_{t\backslash t-1}$ <br> $P_{t\backslash t} = P_{t\backslash t-1} - K_t H P_{t\backslash t-1}$ |
| Step 5: Repeat steps two to four for $t = 2,3,\ldots,T$. | |

Notes: This Figure illustrates the Kalman filtering process when the state vectors are the only unknowns. The first step involves defining the initial conditions for the recursions. In the second step the predicted state and its covariance matrix are computed. In the third step, one calculates the prediction error and the associated covariance matrix. The variances computed in the second and third steps are then used to calculate the Kalman gain, which is then employed to update the state vector. This procedure is repeated for each observation in the sample.

The prior for $Q$ is inverse Wishart, with $T_0$ degrees of freedom and $Q_0$ scale matrix, i.e., $P(Q) \sim IW(Q_0, T_0)$. This prior influences the amount of time-variation allowed for in the coefficients. A large value for the scale matrix $Q_0$ is consistent with more fluctuation in the coefficients. The prior scale matrix is set to $Q_0 = P_{0/0} \times T_0 \times \tau$, where $\tau$ is a scaling factor that reflects the researcher belief in the preciseness of $P_{0/0}$. Since our training sample $T_0$ is small, we consider that the estimate of $P_{0/0}$ is very imprecise and hence set $\tau = 3.510^{-6}$ for all models.[35] This reasoning also accords with Blake and Mumtaz (2012).

---

[35] Note also that the initial training sample size reduces with the forecast horizon. For example, the size of the training sample of the fixed-coefficient model used to parameterize the prior for the forecasting

We set an inverse Gamma prior for the variance of the measurement equation. That is, $P(R) \sim IG(R_0, T_0 - k)$, where $R_0 = \sum_{OLS}$ is the scale parameter and $T_0 - k$ is the prior degree of freedom. To initialize the first step of the Gibbs sampling we need starting values for $R$ and $Q$. We set them to $R_0 = \sum_{OLS}$ and $Q_0 = P_{0/0} \times T_0 \times \tau$.

## II. Conditional Posterior Distribution

Having set the priors and initial conditions the next stage is to set up the Gibbs sampling algorithm to draw from the conditional posterior distribution; hence, we need the form of this distribution. The conditional posterior distribution of the state variable $\tilde{\beta}_T$ given the parameters of the state-space model that define our TVP model is given by:

$$H(\tilde{\beta}_T \backslash \tilde{y}_T, R, Q) = H(\beta_T \backslash \tilde{y}_T) \prod_{t=1}^{T-1} H(\beta_t \backslash \beta_{t+1}, \tilde{y}_t), \tag{B9}$$

where, $\tilde{\beta}_T = [\beta_1, \beta_2 \ldots \beta_T]$ and $\tilde{y}_T = [y_1, y_2 \ldots y_T]$.

The conditional posterior distribution of $R$ given a draw of the state variable $\beta_t$ and the other parameters is given by:

$$H(R \backslash \beta_t, y_t, Q) \sim \Gamma^{-1}\left(\frac{T_0 - k + T}{2}, \frac{\theta_1}{2}\right), \tag{B10}$$

where,

$$\theta_1 = R_0 + (y_t - \beta_t H)'(y_t - \beta_t H). \tag{B11}$$

The conditional posterior distribution of $Q$ given a draw of the state variable $\beta_t$ and the other parameters is:

$$H(Q \backslash \beta_t, y_t, R) \sim IW(\bar{Q}, T + T_0), \tag{B12}$$

where $T$ is the sample size and,

$$\bar{Q} = Q_0 + (\beta_t - \beta_{t-1})'(\beta_t - \beta_{t-1}). \tag{B13}$$

## III. Sampling from the Conditional Posterior Distribution

To draw samples from the conditional posterior distributions we use the Carter and Kohn (1994) algorithm along with the Gibbs sampler. The Carter and Kohn algorithm provide us with the draws of the state variable $\tilde{\beta}_T = [\beta_1, \beta_2 \ldots \beta_T]$ from its conditional posterior distribution. The key updating equations are:

$$\beta_{t \backslash t, \beta_{t+1}} = \beta_{t \backslash t} + K^*(\beta_{t+1} - u + F\beta_{t \backslash t}), \tag{B14}$$

$$P_{t \backslash t, \beta_{t+1}} = P_{t \backslash t} - K^* H^* P_{t \backslash t}, \tag{B15}$$

---

regression at 12-quarters-ahead is $T_0 = 20 - 12 = 8$. With two coefficients ($k = 2$) to be estimated, this results in six degrees of freedom.

where $\beta_{t\backslash t}$ and $P_{t\backslash t}$ are obtained from the Kalman filter and $K^* = P_{t\backslash t}H^{*\prime}f^{*-1}_{t+1\backslash t}$. Equations (B14) and (B15) are substituted backwards from $T-1$, and iterating backwards to period 1. In fact, this algorithm constitutes an integral part of the Gibbs sampling framework, which comprises the following steps:

**Step 1**: Conditional on $R$ and $Q$, draw $\beta_t$ from its conditional posterior distribution given in (B9) using the Kalman filter and the Carter and Kohn algorithm. More in detail:

- 1.1: Run the Kalman filter from $t = 1 \ldots T$ to obtain the mean $\beta_{T\backslash T}$ and the variance $P_{T\backslash T}$ of the distribution $H(\beta_T \backslash \tilde{y}_T)$. Also obtain $\beta_{t\backslash t}$ and $P_{t\backslash t}$ for $t = 1 \ldots T$.
- 1.2: Draw $\beta_T$ from the normal distribution with mean $\beta_{T\backslash T}$ and variance $P_{T\backslash T}$. Denote it $\hat{B}_T$.
- 1.3: At time $t = T-1$, use (B14) to obtain $\beta_{T-1\backslash T-1,\beta_T} = \beta_{T-1\backslash T-1} + K^*(\hat{B}_T - u + F\beta_{T-1\backslash T-1})$. Note that $\beta_{T-1\backslash T-1}$ is the Kalman filter estimate of the state variable at time $T-1$, whereas $\hat{B}_T$ is a draw from $N \sim (\beta_{T\backslash T}, P_{T\backslash T})$ at time $T$ (both from step 1.1). Use also equation (B15) to calculate $P_{T-1\backslash T-1,\beta_T} = P_{T-1\backslash T-1} - K^*H^*P_{T-1\backslash T-1}$. Again, $P_{T-1/T-1}$ is obtained from step 1.1 for $T-1$.
- 1.4: Repeat step 1.3 for $t = T-2, T-3, \ldots 1$, to complete the backward recursions. At the end of sub-step 1.4, a first sample of $\beta_t$ from $t = 1 \ldots T$ is obtained. Denote it $\beta^1_{1T}$.

**Step 2**: Conditional on $\beta_t$ sample $R$ from its conditional posterior distributions given in Expression (B10). To do so, use the draw of $\beta_t$, i.e. $\beta^1_{1T}$, to compute the elements necessary to sample from the inverse Gamma distribution. More precisely, compute the scale matrix given by $scale = (y_t - \beta^1_{1T}H)'(y_t - \beta^1_{1T}H)$ and the posterior degrees of freedom defined as $T_1 = T_0 - k + T$. This provides one draw of $R$ from the inverse Gamma distribution with $\theta_1 = R_0 + scale$ as a scale parameter and $T_1$ degrees of freedom.

**Step 3**: Conditional on $\beta_t$ sample $Q$ from its conditional posterior distribution given by the expression (B12). The draw obtained in step 1, $\beta^1_{1T}$, also allows to sample $Q$. To do so, compute the elements necessary to draw $Q$ from the inverse Wishart distribution. That is, compute the scale matrix $(\beta_t - \beta_{t-1})'(\beta_t - \beta_{t-1})$ and add the prior scale parameter $Q_0$ to obtain the posterior scale matrix $\bar{Q}$ as in equation (B13). Then, use $\bar{Q}$ and $T_1 = T + T_0$ to draw $Q$ from the inverse Wishart distribution.

**Step 4**: Repeat steps 1 to 3 a sufficient number of times until convergence is detected. The methods we used to assess convergence indicate that 1700 draws are sufficient.[36] We then discard the first 300 draws and save the last 1400 draws for inference.

We then use the mean of the marginal posterior distribution of $\beta_t$, as the coefficient's point estimate.

---

[36] We use Geweke's convergence test and the Relative Numerical Efficiency (RNE) measure to assess the convergence of the algorithm.

**Appendix C. Data**

This Appendix describes the data used in the empirical estimation. The sample period is 1973Q1:2013Q1, for 18 OECD countries. The data comprises series of exchange rates, industrial production (IP), money supply, short-term interest rates, consumer price index and unemployment rate. The source of the data is indicated in Table C1 below.

**Table C1. Data Used in the Empirical Estimation**

| Country | Nominal exchange rate | Industrial production index, NSA, 2005=100 | Money supply NSA |
|---|---|---|---|
| Australia | IFS, 193..AE.ZF | IFS, 19366..CZF | M3, OECD, MEI |
| Canada | IFS, 156..AE.ZF | IFS, 15666..CZF | M1, OECD, MEI |
| Denmark | IFS, 128..AE.ZF | IFS, 12866..BZF | M1, OECD, MEI |
| UK | IFS, 112..AE.ZF | IFS, 11266..CZF | M4, Bank of England |
| Japan | IFS, 158..AE.ZF | IFS, 15866..CZF | M1, OECD, MEI |
| Korea | IFS, 542..AE.ZF | IFS, 54266..CZF | M1, OECD, MEI |
| Norway | IFS, 142..AE.ZF | IFS, 14266..CZF | M2, OECD, MEI |
| Sweden | IFS, 144..AE.ZF | OECD MEI | M3, OECD, MEI |
| Switzerland | IFS, 146..AE.ZF | IFS, 14666..BZF | M1, OECD, MEI |
| Austria+ | IFS, 122..AE.ZF | IFS, 12266..BZF | M2=34A.NZF + 34B.NZF+35; IFS |
| Belgium+ | IFS, 124..AE.ZF | IFS, 12466..CZF | M1=34A.NZF + 34B.NZF; IFS |
| France+ | IFS, 132..AE.ZF | 13266..CZF | M1 =34A.NZF + 34B.NZF; IFS |
| Germany+ | IFS, 134..AE.ZF | IFS, 13466..CZF | M1 (34A.NZF + 34B.NZF); IFS |
| Spain+ | IFS, 184..AE.ZF | IFS, 18466..CZF | M2=34A.NZF + 34B.NZF+35;IFS |
| Italy+ | IFS, 136..AE.ZF | IFS, 13666..CZF | M2=34A.NZF + 34B.NZF+35; IFS |
| Finland+ | IFS, 172..AE.ZF | IFS, 17266..CZF | M2=34A.NZF + 34B.NZF+35; IFS |
| Netherlands+ | IFS, 138..AE.ZF | IFS, 13866..CZF | M2=34A.NZF + 34B.NZF+35; IFS |
| United States | | IFS, 11166..CZF | M2, OECD, MEI |

Notes: The "+" symbol indicates Euro Area country. The exchange rate is defined as national currency per U.S. dollar at the end of quarter. To generate the exchange rate series for the eight Euro Area countries from 1999 onwards, the irrevocable conversion factors adopted by each country on the 1st of January 1999 were employed. For example, the Mark/U.S. dollar exchange rate is obtained by multiplying the conversion factor 1.95583/EUR by the EUR/U.S. dollar exchange rate at each post 1998Q4 date. The conversion factors for other countries in our sample are (International Monetary Fund, 2012): Austria: 13.7603, Belgium: 40.3399, Finland: 5.94573, France: 6.55957, Italy: 1936.27, Netherlands: 2.20371, and Spain: 166.386. IFS denotes International Financial Statistics as published by the IMF (2012). OECD, MEI denotes the OECD's (2012) Main Economic Indicators database. NSA stands for non-seasonally adjusted. In the unemployment rate, n.a indicates that the series is not available for the entire sample period.

**Table C1. Data Used in the Empirical Estimation (continued)**

| Country | Short-term nominal interest rate (annual rate) | Consumer price index NSA, 2005=100 | Unemployment rate (last month of quarter), NSA |
| --- | --- | --- | --- |
| Australia | IFS, 19360...ZF | IFS, 64...ZF | OECD, MEI |
| Canada | IFS, 15660B..ZF | IFS, 15664...ZF | OECD, MEI |
| Denmark | IFS, 12860...ZF | IFS, 12864...ZF | OECD, MEI |
| UK | IFS, 11260...ZF | IFS, 11264B..ZF | OECD, MEI |
| Japan | IFS, 15860B..ZF | IFS, 15864...ZF | OECD, MEI |
| Korea | 54260B..ZF | IFS, 54264...ZF | n.a |
| Norway | IFS, 14260...ZF | IFS, 14264...ZF | OECD, MEI |
| Sweden | IFS, 14460B..ZF | IFS, 14464...ZF | OECD, MEI |
| Switzerland | IFS, 14660...ZF | IFS, 14664...ZF | OECD, MEI |
| Austria+ | IFS, 12260B..ZF | IFS, 12264...ZF | OECD, MEI |
| Belgium+ | IFS, 12460B..ZF | IFS, 12464...ZF | OECD, MEI |
| France+ | IFS, 13260B..ZF | IFS, 13264...ZF | OECD, MEI |
| Germany+ | IFS, 13460B..ZF | Bundesbank | OECD, MEI |
| Spain+ | IFS, 18460B..ZF | IFS, 18464...ZF | n.a |
| Italy+ | IFS, 13660B..ZF | IFS, 13664...ZF | n.a |
| Finland+ | Central bank rate. OECD, MEI | IFS, 17263EY.ZF | n.a |
| Netherlands+ | IFS, 13860B..ZF | IFS, 13864...ZF | n.a |
| United States | IFS, 11160B..ZF | IFS, 11164...ZF | OECD, MEI |

# Exchange Rate Predictability in a Changing World: Further Results Appendix[*]

Joseph P. Byrne[1], Dimitris Korobilis[2] and Pinho J. Ribeiro[3]

February 14, 2014

This Results appendix contains:

**Appendix D**: Robustness Checks: Additional Empirical Results

**Appendix E**: Forecast Evaluation: Results by Currency

**Appendix F**: Summary Results from TVP Models Estimated using the Method of Maximum Likelihood

## Appendix D. Robustness Checks: Additional Empirical Results

This Appendix reports further results from the robustness checks. Specifically, we assess the robustness of the results reported in the Empirical Section of the paper to: (i) change in base currency from the U.S. dollar to the Pound sterling in all models; (ii) use of unemployment gap rather than output gap in the Taylor rule specifications; (iii) use of monthly data, instead of quarterly data – only for Taylor rule fundamentals; (iv) comparison with other forecasting regressions and methods along with alternative approaches to estimate Taylor rule fundamentals; and (v) estimation of factors by maximum likelihood rather than principal components in the factor model.[4]

### D.A Change in Base Currency

Tables D.A1, D.A2 and D.A3 report results where the base currency is the Pound sterling; hence the exchange rate is defined as the national currency per Pound sterling. Accordingly, all the variables in the regressions are defined relative to the foreign country,

---

[*] Corresponding Author: Email address: dimitris.korobilis@glasgow.ac.uk, Tel: +44 (0)141 330 2950. Fax.: +44 (0)141 330 4940.

[1] Department of Economics, Heriot-Watt University, Edinburgh, UK.

[2] Department of Economics, Adam Smith Business School, University of Glasgow, Glasgow, UK.

[3] Department of Economics, Adam Smith Business School, University of Glasgow, Glasgow, UK.

[4] To save space for all the robustness checks we only present summary results. Currency by currency results are not included in the paper but are available upon request.

i.e. the United Kingdom. Table D.A1 is for Taylor rules, D.A2 for the monetary model, PPP and UIRP, and D.A3 for the factor model.

**D.B Taylor Rules with Unemployment Gap, instead of Output Gap**

Table D.B1 presents the forecasting performance of models when the information set is defined by Taylor rules where monetary authorities target the unemployment gap, rather than the output gap. The forecasting window is 2007Q1-2013Q1. Results are for the U.S. dollar (USD) base currency and the Pound sterling (GBP).

**D.C Taylor Rules with Monthly Data, instead of Quarterly Data**

Table D.C1 reports the forecasting performance of the regressions with monthly data-frequency. The forecasting window is 2007M1-2013M5, to match the 2007Q1-2013Q1 window in the quarterly data. The forecast horizons in months comprise h=1; h=3, h=12; h=24 and h=36. Results are for the U.S. dollar (USD).

**D.D Changes in Forecasting Regression and Forecasting Method**

Table D.D1 presents the forecasting performance of models when the information set from Taylor rules is estimated with a single-equation constant-parameter (SECP) regression via OLS, and the forecasting regression is also SECP regression. This is the approach in Molodtsova and Papell (2009) and Rossi (2013). To match their approach we also use rolling windows, instead of a recursive forecasting method. Results are for the U.S. dollar (USD) base currency and all the forecast windows considered in the main text.

**D.E Factors Estimated via Maximum Likelihood**

Table D.E1 presents the forecasting performance of models when factors are estimated by maximum likelihood, rather than principal components. The forecasting window is 2007Q1-2013Q1. Results are for the USD base currency.

**Table D.A1 Forecast Evaluation: Taylor Rules (GBP base currency)**

| | | TVP Regression | | | | Fixed-effect Panel Regression | | | |
|---|---|---|---|---|---|---|---|---|---|
| Fundamentals: | | h=1 | h=4 | h=8 | h=12 | h=1 | h=4 | h=8 | h=12 |
| | | Forecast Window: 1995Q1-1998Q4; N=17 | | | | | | | |
| $TR_{on}$ | | | | | | | | | |
| | No. of U's <1 | 4 | 3 | 4 | 3 | 4 | 3 | 3 | 4 |
| | No. of DM >1.282 | 0 | 1 | 3 | 3 | 0 | 1 | 2 | 3 |
| | Median U | 1.016 | 1.078 | 1.191 | 1.310 | 1.022 | 1.045 | 1.116 | 1.290 |
| $TR_{os}$ | | | | | | | | | |
| | No. of U's <1 | 3 | 3 | 4 | 3 | 4 | 3 | 3 | 4 |
| | No. of DM >1.282 | 0 | 1 | 3 | 2 | 0 | 1 | 2 | 3 |
| | Median U | 1.041 | 1.118 | 1.193 | 1.385 | 1.029 | 1.050 | 1.123 | 1.339 |
| $TR_{en}$ | | | | | | | | | |
| | No. of U's <1 | 4 | 3 | 4 | 4 | 3 | 4 | 3 | 4 |
| | No. of DM >1.282 | 1 | 2 | 3 | 3 | 0 | 1 | 2 | 3 |
| | Median U | 1.025 | 1.100 | 1.147 | 1.269 | 1.043 | 1.077 | 1.083 | 1.271 |
| | | Forecast Window: 1999Q1-2013Q1; N=10 | | | | | | | |
| $TR_{on}$ | | | | | | | | | |
| | No. of U's <1 | 1 | **5** | 4 | 4 | 2 | 2 | 3 | 4 |
| | No. of DM >1.282 | 0 | 0 | 1 | 1 | 0 | 1 | 2 | 3 |
| | Median U | 1.005 | 0.999‡ | 1.013 | 1.066 | 1.008 | 1.024 | 1.034 | 1.043 |
| $TR_{os}$ | | | | | | | | | |
| | No. of U's <1 | 1 | 2 | 4 | **5** | 2 | 2 | 3 | 4 |
| | No. of DM >1.282 | 0 | 0 | 1 | 3 | 0 | 1 | 2 | 3 |
| | Median U | 1.009 | 1.031 | 1.027 | 1.010 | 1.004 | 1.016 | 1.022 | 1.045 |
| $TR_{en}$ | | | | | | | | | |
| | No. of U's <1 | 3 | **5** | 4 | 3 | 2 | 2 | 3 | 4 |
| | No. of DM >1.282 | 0 | 1 | 1 | 1 | 0 | 1 | 2 | 3 |
| | Median U | 1.005 | 1.000 | 1.042 | 1.116 | 1.005 | 1.022 | 1.025 | 1.046 |
| | | Forecast Window: 2007Q1-2013Q1; N=10 | | | | | | | |
| $TR_{on}$ | | | | | | | | | |
| | No. of U's <1 | **5** | **6** | **6** | **6** | 4 | 4 | 4 | **5** |
| | No. of DM >1.282 | 1 | 2 | 5 | 4 | 0 | 3 | 4 | 4 |
| | Median U | 0.999‡ | 0.961‡ | 0.958‡ | 0.912‡ | 1.004 | 1.014 | 1.009 | 0.995‡ |
| $TR_{os}$ | | | | | | | | | |
| | No. of U's <1 | 3 | **6** | **6** | **8** | 4 | **6** | **5** | **6** |
| | No. of DM >1.282 | 0 | 2 | 3 | 7 | 0 | 3 | 4 | 4 |
| | Median U | 1.003 | 0.990‡ | 0.973‡ | 0.894‡ | 1.002 | 0.996‡ | 0.995‡ | 0.990‡ |
| $TR_{en}$ | | | | | | | | | |
| | No. of U's <1 | 4 | **7** | **6** | **5** | 4 | **5** | **5** | **5** |
| | No. of DM >1.282 | 0 | 2 | 3 | 5 | 0 | 3 | 3 | 4 |
| | Median U | 1.004 | 0.968‡ | 0.979‡ | 0.975‡ | 1.003 | 1.006 | 1.003 | 0.993‡ |

Notes: This Table summarises the forecasting performance of the TVP forecasting regression and the Fixed-effect panel regression with Taylor rule fundamentals defined as $TR_{on}$, $TR_{os}$ and $TR_{en}$. The only difference with Table 2 in the main text is that here the base currency is the Pound sterling (GBP) rather than the USD. Hence, the interpretation is similar to Table 2 in the paper. That is, For each regression, set of fundamentals, forecast window and quarterly horizon (h), the "No. of U's < 1" (number of U-statistics less than one), provides the number of currencies for which the model improves upon the RW, since it indicates cases where the RMSFE of the fundamental-based regression is lower than that of the RW. When the U's are less than one for at least half of the currencies in the window, marked in **bold**, then on average, the fundamental-based regression outperforms the benchmark in that window. The "No. of DM > 1.282" (number of DM statistics greater than 1.282) shows cases of rejections of the null hypothesis under the Diebold and Mariano (1995) test of equal forecast accuracy at 10% level of significance. The higher the No. of DM > 1.282, the better the average accuracy of the forecasts of the fundamental-based regression relative to the benchmark is. The "Median U" indicates the middle value of the U-statistic across the sample of *N* currencies for each forecast window and horizon. When "Median U" is less than or equal to one - marked with the symbol "‡", and U's are less than one for at least half of the currencies in the window, this is also consistent with a better average forecasting performance of the fundamental-based regression relative to the benchmark.

**Table D.A2 Forecast Evaluation: Monetary Model, PPP and UIRP (GBP base currency)**

| Fundamentals: | | TVP Regression | | | | Fixed-effect Panel Regression | | | |
|---|---|---|---|---|---|---|---|---|---|
| | | h=1 | h=4 | h=8 | h=12 | h=1 | h=4 | h=8 | h=12 |
| | | \multicolumn{8}{c}{Forecast Window: 1995Q1-1998Q4; N=17} | | | | | | | |
| MM | | | | | | | | | |
| | No. of U's <1 | 1 | 0 | 3 | 3 | 3 | 3 | 2 | 2 |
| | No. of DM >1.282 | 0 | 0 | 1 | 2 | 1 | 1 | 1 | 1 |
| | Median U | 1.066 | 1.233 | 1.681 | 1.677 | 1.099 | 1.285 | 1.406 | 1.622 |
| PPP | | | | | | | | | |
| | No. of U's <1 | 3 | 5 | 4 | 4 | **9** | **10** | **9** | 6 |
| | No. of DM >1.282 | 1 | 1 | 2 | 3 | 1 | 4 | 3 | 3 |
| | Median U | 1.039 | 1.200 | 1.315 | 1.509 | 0.994‡ | 0.958‡ | 0.997‡ | 1.110 |
| UIRP | | | | | | | | | |
| | No. of U's <1 | 3 | 4 | 4 | 3 | 5 | 3 | 4 | 4 |
| | No. of DM >1.282 | 0 | 1 | 4 | 2 | 0 | 0 | 3 | 4 |
| | Median U | 1.036 | 1.138 | 1.226 | 1.348 | 1.020 | 1.079 | 1.120 | 1.177 |
| | | \multicolumn{8}{c}{Forecast Window: 1999Q1-2013Q1; N=10} | | | | | | | |
| MM | | | | | | | | | |
| | No. of U's <1 | **6** | 3 | 2 | 4 | **7** | **7** | **7** | **8** |
| | No. of DM >1.282 | 0 | 1 | 1 | 1 | 0 | 2 | 3 | 6 |
| | Median U | 0.999‡ | 1.019 | 1.221 | 1.310 | 0.997‡ | 0.980‡ | 0.939‡ | 0.857‡ |
| PPP | | | | | | | | | |
| | No. of U's <1 | **9** | 5 | 4 | **6** | **9** | **8** | **6** | 5 |
| | No. of DM >1.282 | 1 | 0 | 2 | 1 | 1 | 0 | 0 | 3 |
| | Median U | 0.993‡ | 1.004 | 1.029 | 0.995‡ | 0.986‡ | 0.975‡ | 0.941‡ | 0.950‡ |
| UIRP | | | | | | | | | |
| | No. of U's <1 | 1 | 2 | 4 | **5** | 1 | 2 | 3 | 4 |
| | No. of DM >1.282 | 0 | 0 | 1 | 5 | 0 | 1 | 2 | 3 |
| | Median U | 1.008 | 1.031 | 1.032 | 1.037 | 1.005 | 1.016 | 1.024 | 1.040 |
| | | \multicolumn{8}{c}{Forecast Window: 2007Q1-2013Q1; N=10} | | | | | | | |
| MM | | | | | | | | | |
| | No. of U's <1 | **8** | **8** | **6** | **7** | **8** | **9** | **9** | **9** |
| | No. of DM >1.282 | 1 | 3 | 4 | 4 | 3 | 5 | 5 | 6 |
| | Median U | 0.993‡ | 0.972‡ | 0.974‡ | 0.942‡ | 0.983‡ | 0.912‡ | 0.850‡ | 0.756‡ |
| PPP | | | | | | | | | |
| | No. of U's <1 | **9** | **8** | **8** | **6** | **9** | **9** | **9** | **9** |
| | No. of DM >1.282 | 7 | 7 | 6 | 4 | 5 | 6 | 8 | 8 |
| | Median U | 0.980‡ | 0.918‡ | 0.868‡ | 0.861‡ | 0.971‡ | 0.901‡ | 0.791‡ | 0.657‡ |
| UIRP | | | | | | | | | |
| | No. of U's <1 | 2 | **5** | **5** | **9** | 4 | **5** | **5** | **5** |
| | No. of DM >1.282 | 0 | 2 | 3 | 6 | 0 | 3 | 3 | 4 |
| | Median U | 1.003 | 0.998‡ | 0.995‡ | 0.896‡ | 1.005 | 0.999‡ | 0.998‡ | 0.996‡ |

Notes: This Table summarises the forecasting performance of the TVP forecasting regression and the Fixed-effect panel regression with Monetary (MM), PPP and UIRP fundamentals. The formatting and interpretation is similar to Table 3 in the main text, but here the base currency is the Pound sterling (GBP) rather than the USD. Hence, the interpretation is similar to Table 3 in the paper – also repeated in notes to Table D.A1 in this Appendix for convenience.

**Table D.A3 Forecast Evaluation: Factor Model (GBP base currency)**

| Fundamentals: | TVP Regression | | | | Fixed-effect Panel Regression | | | |
|---|---|---|---|---|---|---|---|---|
| | h=1 | h=4 | h=8 | h=12 | h=1 | h=4 | h=8 | h=12 |
| | Forecast Window: 1995Q1-1998Q4; N=17 | | | | | | | |
| F1 | | | | | | | | |
| No. of U's <1 | 4 | 3 | 2 | 4 | 3 | 3 | 4 | 4 |
| No. of DM >1.282 | 0 | 1 | 1 | 1 | 1 | 1 | 4 | 4 |
| Median U | 1.005 | 1.035 | 1.041 | 1.053 | 1.010 | 1.085 | 1.189 | 1.287 |
| F2 | | | | | | | | |
| No. of U's <1 | 8 | 7 | 6 | 5 | **9** | **9** | **9** | 6 |
| No. of DM >1.282 | 3 | 3 | 2 | 3 | 6 | 4 | 2 | 2 |
| Median U | 1.002 | 1.011 | 1.012 | 1.044 | 0.997‡ | 0.986‡ | 0.989‡ | 1.022 |
| F3 | | | | | | | | |
| No. of U's <1 | 6 | 6 | 4 | 6 | 8 | 8 | **9** | **10** |
| No. of DM >1.282 | 3 | 3 | 0 | 5 | 4 | 7 | 6 | 4 |
| Median U | 1.008 | 1.042 | 1.066 | 1.038 | 1.002 | 1.006 | 0.939‡ | 0.956‡ |
| | Forecast Window: 1999Q1-2013Q1; N=10 | | | | | | | |
| F1 | | | | | | | | |
| No. of U's <1 | 1 | 1 | 2 | 3 | 2 | 3 | 3 | 2 |
| No. of DM >1.282 | 0 | 1 | 1 | 0 | 0 | 2 | 2 | 2 |
| Median U | 1.006 | 1.031 | 1.053 | 1.068 | 1.006 | 1.023 | 1.051 | 1.060 |
| F2 | | | | | | | | |
| No. of U's <1 | 0 | 0 | **5** | 3 | 0 | 1 | 3 | **5** |
| No. of DM >1.282 | 0 | 0 | 1 | 2 | 0 | 0 | 0 | 2 |
| Median U | 1.008 | 1.026 | 1.007 | 1.199 | 1.005 | 1.020 | 1.022 | 1.001 |
| F3 | | | | | | | | |
| No. of U's <1 | 1 | 2 | 3 | 4 | 3 | 2 | 4 | 4 |
| No. of DM >1.282 | 0 | 0 | 0 | 1 | 0 | 0 | 0 | 0 |
| Median U | 1.007 | 1.030 | 1.082 | 1.098 | 1.005 | 1.027 | 1.042 | 1.022 |
| | Forecast Window: 2007Q1-2013Q1; N=10 | | | | | | | |
| F1 | | | | | | | | |
| No. of U's <1 | 1 | 4 | **6** | **6** | 4 | 4 | 4 | **6** |
| No. of DM >1.282 | 0 | 1 | 1 | 4 | 1 | 4 | 4 | 3 |
| Median U | 1.006 | 1.011 | 0.999‡ | 0.984‡ | 1.004 | 1.009 | 1.003 | 0.980‡ |
| F2 | | | | | | | | |
| No. of U's <1 | 0 | 1 | **5** | 3 | 2 | 3 | 4 | 4 |
| No. of DM >1.282 | 0 | 0 | 1 | 2 | 1 | 1 | 2 | 1 |
| Median U | 1.008 | 1.021 | 1.002 | 1.031 | 1.007 | 1.026 | 1.037 | 1.020 |
| F3 | | | | | | | | |
| No. of U's <1 | 1 | 2 | 3 | 3 | 3 | 3 | 4 | 4 |
| No. of DM >1.282 | 0 | 0 | 0 | 2 | 0 | 2 | 3 | 3 |
| Median U | 1.012 | 1.038 | 1.104 | 1.094 | 1.009 | 1.042 | 1.077 | 1.057 |

Notes: This Table summarises the forecasting performance of the TVP forecasting regression and the Fixed-effect panel regression with factors (F) from exchange rates. The formatting and interpretation is similar to Table 4 in the main text, except that here the base currency is the Pound sterling (GBP) rather than the USD. Therefore, the interpretation is similar to Table 4 in the paper – also repeated in notes to Table D.A1 in this Appendix for convenience.

**Table D.B1 Forecast Evaluation: Taylor Rules with Unemployment Gap**

| Fundamentals: | TVP-TVP Regression | | | | Fixed-effect Panel Regression | | | |
|---|---|---|---|---|---|---|---|---|
| | h=1 | h=4 | h=8 | h=12 | h=1 | h=4 | h=8 | h=12 |
| | Forecast Window: 2007Q1-2013Q1; N=9; Base Currency=GBP | | | | | | | |
| $TR_{on}$ | | | | | | | | |
|   No. of U's <1 | 4 | **7** | **8** | 7 | 3 | 4 | **5** | 6 |
|   No. of DM >1.282 | 1 | 2 | 5 | 4 | 0 | 3 | 4 | 4 |
|   Median U | 1.001 | 0.984‡ | 0.940‡ | 0.854‡ | 1.002 | 1.001 | 0.988‡ | 0.988‡ |
| $TR_{os}$ | | | | | | | | |
|   No. of U's <1 | 3 | **6** | 5 | 7 | 4 | **6** | 5 | 6 |
|   No. of DM >1.282 | 0 | 2 | 3 | 5 | 0 | 3 | 4 | 4 |
|   Median U | 1.001 | 0.984‡ | 0.976‡ | 0.912‡ | 1.000 | 0.989‡ | 0.977‡ | 0.977‡ |
| $TR_{en}$ | | | | | | | | |
|   No. of U's <1 | 3 | **6** | 5 | 4 | **5** | 4 | 5 | 4 |
|   No. of DM >1.282 | 0 | 3 | 4 | 3 | 0 | 3 | 4 | 4 |
|   Median U | 1.005 | 0.972‡ | 0.998‡ | 1.010 | 0.999‡ | 1.005 | 0.982‡ | 1.001 |
| | Forecast Window: 2007Q1-2013Q1; N=9; Base Currency=USD | | | | | | | |
| $TR_{on}$ | | | | | | | | |
|   No. of U's <1 | 3 | 3 | 4 | 3 | 4 | **5** | 4 | 4 |
|   No. of DM >1.282 | 0 | 1 | 2 | 2 | 0 | 2 | 4 | 3 |
|   Median U | 1.004 | 1.013 | 1.032 | 1.220 | 1.000 | 0.996‡ | 1.007 | 1.017 |
| $TR_{os}$ | | | | | | | | |
|   No. of U's <1 | 3 | 3 | 4 | 4 | 1 | 5 | **6** | 4 |
|   No. of DM >1.282 | 0 | 0 | 2 | 3 | 0 | 0 | 5 | 4 |
|   Median U | 1.009 | 1.005 | 1.059 | 1.227 | 1.013 | 0.994‡ | 0.946‡ | 1.024 |
| $TR_{en}$ | | | | | | | | |
|   No. of U's <1 | **5** | **6** | **6** | **6** | 5 | 4 | 5 | 3 |
|   No. of DM >1.282 | 1 | 2 | 4 | 4 | 1 | 2 | 4 | 3 |
|   Median U | 0.998‡ | 0.967‡ | 0.923‡ | 0.802‡ | 0.999‡ | 1.004 | 0.989‡ | 1.017 |

Notes: This Table presents the summary results of the forecasting performance of the TVP regression and the Fixed-effect panel regression with Taylor rules information set ($TR_{on}$, $TR_{os}$, $TR_{en}$), based on unemployment gap rather than output gap. The forecast window is 2007Q1-2013Q1 and the number of exchange rates (N) is nine. Thus, the "No. of U's <1", the "No. of DM-stat > 1.282" and the "Median U" is relative to N=9 currencies. Otherwise, they are interpreted in a similar fashion as in notes to Table D.A1 in this Appendix.

**Table D.C1 Forecast Evaluation: Taylor Rules with Monthly Data**

| Fundamentals: | TVP Regression | | | | | Fixed-effect Panel Regression | | | | |
|---|---|---|---|---|---|---|---|---|---|---|
| | h=1 | h=3 | h=12 | h=24 | h=36 | h=1 | h=3 | h=12 | h=24 | h=36 |
| | Forecast Window: 2007M1-2013M5; N=9 | | | | | | | | | |
| $TR_{on}$ | | | | | | | | | | |
|   No. of U's <1 | 1 | 1 | **5** | 4 | 3 | 2 | 2 | **5** | **5** | 4 |
|   No. of DM >1.282 | 0 | 0 | 1 | 2 | 3 | 0 | 0 | 1 | 3 | 3 |
|   Median U | 1.002 | 1.005 | **0.999‡** | 1.004 | 1.018 | 1.002 | 1.006 | **0.998‡** | **0.999‡** | 1.037 |
| $TR_{os}$ | | | | | | | | | | |
|   No. of U's <1 | 3 | 3 | 4 | 4 | **5** | 1 | 1 | 4 | **6** | 5 |
|   No. of DM >1.282 | 0 | 0 | 0 | 3 | 4 | 0 | 0 | 0 | 5 | 3 |
|   Median U | 1.003 | 1.010 | 1.005 | 1.000 | **0.914‡** | 1.002 | 1.010 | 1.012 | **0.973‡** | **0.966‡** |
| $TR_{en}$ | | | | | | | | | | |
|   No. of U's <1 | 1 | 2 | 4 | 4 | 4 | 1 | 1 | 3 | **6** | 7 |
|   No. of DM >1.282 | 0 | 0 | 1 | 1 | 3 | 0 | 0 | 1 | 2 | 4 |
|   Median U | 1.003 | 1.006 | 1.016 | 1.007 | 1.018 | 1.004 | 1.014 | 1.002 | **0.977‡** | **0.961‡** |

Notes: This Table presents the summary results of the forecasting performance of the TVP regression and the Fixed-effect panel regression with Taylor rules information set ($TR_{on}$, $TR_{os}$, $TR_{en}$), using monthly data. Thus, the forecast horizon, h, is monthly. The forecast window is 2007M1-2013M5 and the number of exchange rates (N) is nine. Thus, the "No. of U's <1", the "No. of DM-stat > 1.282" and the "Median U" is relative to N=9 currencies. Otherwise, they are interpreted in a similar fashion as in notes to Table D.A1 in this Appendix.

**Table D.D1 Forecast Evaluation: Taylor Rules (OLS Rolling Windows)**

| Fundamentals: | | TVP Regression | | | | Constant-Parameter Regression (OLS) | | | |
|---|---|---|---|---|---|---|---|---|---|
| | | h=1 | h=4 | h=8 | h=12 | h=1 | h=4 | h=8 | h=12 |
| | | Forecast Window: 1995Q1-1998Q4; N=17 | | | | | | | |
| TR$_{on}$ | | | | | | | | | |
| | No. of U's <1 | 5 | **9** | **11** | **10** | 7 | 8 | 8 | **9** |
| | No. of DM >1.282 | 1 | 4 | 9 | 8 | 1 | 2 | 4 | 5 |
| | Median U | 1.023 | 0.989‡ | 0.853‡ | 0.939‡ | 1.005 | 1.003 | 1.003 | 0.996‡ |
| TR$_{os}$ | | | | | | | | | |
| | No. of U's <1 | 5 | **11** | 9 | **10** | 6 | 8 | **11** | **9** |
| | No. of DM >1.282 | 1 | 4 | 6 | 10 | 1 | 2 | 6 | 7 |
| | Median U | 1.031 | 0.980‡ | 0.999‡ | 0.965‡ | 1.009 | 1.002 | 0.951‡ | 0.907‡ |
| TR$_{en}$ | | | | | | | | | |
| | No. of U's <1 | 6 | 8 | 8 | 7 | 4 | 4 | 7 | 8 |
| | No. of DM >1.282 | 0 | 3 | 8 | 7 | 0 | 0 | 5 | 4 |
| | Median U | 1.021 | 1.043 | 1.035 | 1.292 | 1.050 | 1.151 | 1.154 | 1.240 |
| | | Forecast Window: 1999Q1-2013Q1; N=10 | | | | | | | |
| TR$_{on}$ | | | | | | | | | |
| | No. of U's <1 | 3 | 2 | 3 | 3 | **5** | 2 | 1 | 3 |
| | No. of DM >1.282 | 1 | 1 | 2 | 3 | 0 | 0 | 1 | 1 |
| | Median U | 1.007 | 1.031 | 1.082 | 1.179 | 1.000 | 1.032 | 1.085 | 1.148 |
| TR$_{os}$ | | | | | | | | | |
| | No. of U's <1 | 1 | 2 | 3 | 3 | 1 | 4 | 4 | 1 |
| | No. of DM >1.282 | 0 | 0 | 2 | 2 | 0 | 0 | 1 | 1 |
| | Median U | 1.010 | 1.043 | 1.099 | 1.219 | 1.005 | 1.023 | 1.055 | 1.141 |
| TR$_{en}$ | | | | | | | | | |
| | No. of U's <1 | 1 | 3 | 3 | 1 | 2 | 3 | 4 | 4 |
| | No. of DM >1.282 | 0 | 0 | 1 | 1 | 0 | 0 | 0 | 1 |
| | Median U | 1.007 | 1.036 | 1.083 | 1.226 | 1.011 | 1.020 | 1.031 | 1.050 |
| | | Forecast Window: 2007Q1-2013Q1; N=10 | | | | | | | |
| TR$_{on}$ | | | | | | | | | |
| | No. of U's <1 | 4 | **5** | 7 | 5 | 4 | 3 | **5** | 4 |
| | No. of DM >1.282 | 0 | 1 | 4 | 4 | 0 | 0 | 2 | 3 |
| | Median U | 1.004 | 1.003 | 0.973‡ | 1.007 | 1.006 | 1.008 | 0.961‡ | 1.110 |
| TR$_{os}$ | | | | | | | | | |
| | No. of U's <1 | 2 | 4 | 6 | 5 | 1 | 4 | **8** | 5 |
| | No. of DM >1.282 | 0 | 0 | 2 | 5 | 0 | 1 | 3 | 4 |
| | Median U | 1.004 | 1.003 | 0.955‡ | 0.930‡ | 1.010 | 1.006 | 0.967‡ | 1.016 |
| TR$_{en}$ | | | | | | | | | |
| | No. of U's <1 | **5** | **7** | 6 | **8** | 4 | **5** | 7 | 5 |
| | No. of DM >1.282 | 1 | 2 | 3 | 4 | 0 | 1 | 4 | 4 |
| | Median U | 0.999‡ | 0.991‡ | 0.912‡ | 0.828‡ | 1.001 | 1.015 | 0.933‡ | 1.026 |

Notes: The methodology and results for the TVP regression are exactly as in Table 1 and Table 2 in the text. Results from the constant-parameter regression are obtained by first estimating interest rate differentials (via OLS) with a single-equation constant-parameter model. The estimates are then employed as conditioning information for another single-equation-constant parameter forecasting regression. Unlike in the TVP regression, in the constant-parameter regression the forecasts are generated in rolling windows of 64 quarters for 1995Q1-1998Q4; 80 quarters for 1999Q1-2013Q1, and 112 quarters for 2007Q1-2013Q1. These rolling windows were defined such that the number of forecasts generated with the constant-parameter regression matches the forecasts in the recursive forecasting approach. The interpretation is similar to Table 2 main text.

**Table D.E1 Forecast Evaluation: Factors Estimated by Maximum Likelihood**

| Fundamentals: | TVP Regression | | | | Fixed-effect Panel Regression | | | |
|---|---|---|---|---|---|---|---|---|
| | h=1 | h=4 | h=8 | h=12 | h=1 | h=4 | h=8 | h=12 |
| | Forecast Window: 2007Q1-2013Q1; N=10 | | | | | | | |
| F1 | | | | | | | | |
| No. of U's <1 | 4 | **7** | **7** | 5 | 1 | **6** | 2 | 1 |
| No. of DM >1.282 | 0 | 0 | 3 | 4 | 0 | 0 | 0 | 0 |
| Median U | 1.004 | 0.966‡ | 0.923‡ | 0.954‡ | 1.006 | 0.991 | 1.144 | 1.477 |
| F2 | | | | | | | | |
| No. of U's <1 | **6** | **8** | **8** | 6 | 7 | **8** | 5 | 4 |
| No. of DM >1.282 | 0 | 1 | 2 | 3 | 1 | 2 | 1 | 0 |
| Median U | 0.996‡ | 0.951‡ | 0.829‡ | 0.988‡ | 0.995‡ | 0.958‡ | 0.984‡ | 1.296 |
| F3 | | | | | | | | |
| No. of U's <1 | **6** | **8** | **8** | 6 | 7 | **8** | 6 | 4 |
| No. of DM >1.282 | 0 | 1 | 2 | 4 | 2 | 2 | 1 | 0 |
| Median U | 0.997‡ | 0.950‡ | 0.872‡ | 0.866‡ | 0.995‡ | 0.956‡ | 0.958‡ | 1.238 |

Notes: This Table presents the summary results of the forecasting performance of the TVP forecasting regression and the Fixed-effect panel regression with factors (F) estimated by maximum likelihood rather than principal components method. The U.S. dollar is the base currency. For interpretation of the "No. of U's <1" and "No. of DM-stat > 1.282" and "Median U", see notes to Table D.A1 in this Appendix.

**Appendix E. Forecast Evaluation: Results by Currency (USD base currency)**

This Appendix reports the forecasting performance of the models for each of the currency in the forecasting window, for which the summaries are presented in Tables 2, 3 and 4 in the paper. The U.S. dollar is the base currency. Results are reported by forecast window and the relevant set of fundamentals that enter the model. Accordingly, there are 9 tables:

- Table E.A.1: Window 1995Q1-1998Q4 and Taylor rules ($TR_{on}$, $TR_{os}$ and $TR_{en}$);
- Table E.A.2 Window 1999Q1-2013Q1 and Taylor rules ($TR_{on}$, $TR_{os}$ and $TR_{en}$);
- Table E.A.3 Window 2007Q1-2013Q1 and Taylor rules ($TR_{on}$, $TR_{os}$ and $TR_{en}$);

- Table E.B.1: Window 1995Q1-1998Q4 and MM, PPP and UIRP;
- Table E.B.2 Window 1999Q1-2013Q1 and MM, PPP and UIRP;
- Table E.B.3 Window 2007Q1-2013Q1 and MM, PPP and UIRP;

- Table E.C.1: Window 1995Q1-1998Q4 and factors (F1, F2 and F3);
- Table E.C.2 Window 1999Q1-2013Q1 and factors (F1, F2 and F3);
- Table E.C.3 Window 2007Q1-2013Q1 and factors (F1, F2 and F3).

For each currency, forecast window and horizon the Theil's U-statistic is computed and the model's forecast accuracy relative to the benchmark is assessed using the Diebold and Mariano (1995) (DM) test-statistic. The U-statistic is defined as the ratio of the Root Mean Squared Forecast Error (RMSFE) of the fundamentals-based exchange rate model (FEXM), relative to RMSFE of the driftless random walk (RW). Values less than one-in **bold**, indicate that the RMSFE of the FEXM is lower than that of the RW and hence, the FEXM forecasts better. The null hypothesis under the (DM) test-statistic is that of no difference in the accuracy of forecasts of FEXM relative to the forecasts of the random walk. Asterisks indicate that the null hypothesis of equal forecast accuracy is rejected at 10% (*), 5% (**) and 1% (***), signalling a better average accuracy of the forecasts of the FEXM relative to the benchmark. The last three rows at the bottom of the Tables repeat, for convenience, the summary results reported in the main text, i.e., the "No. of U's<1", the "No. of DM > 1.282" and "Median U".

**Table E.A.1 Forecast Evaluation: Taylor Rules (1995Q1-1998Q4)**

| $TR_{on}$ | TVP Regression | | | | Fixed-effect Panel Regression | | | |
|---|---|---|---|---|---|---|---|---|
| | U(1) | U(4) | U(8) | U(12) | U(1) | U(4) | U(8) | U(12) |
| Australia | **0.964** | **0.868** | **0.634**\*\* | **0.635**\*\*\* | 1.050 | 1.107 | 1.071 | 1.177 |
| Canada | 1.046 | 1.007 | **0.691** | **0.795** | **0.984** | **0.993** | 1.170 | 1.042 |
| Denmark | 1.001 | **0.977**\* | **0.838**\*\*\* | 1.039 | 1.008 | 1.038 | 1.012 | 1.119 |
| UK | **0.981** | 1.014 | 1.194 | 1.901 | 1.057 | 1.408 | 2.158 | 2.463 |
| Japan | 1.065 | 1.164 | 1.188 | 1.893 | 1.024 | 1.144 | 1.363 | 1.698 |
| Korea | 1.000 | **0.989** | **0.983** | **0.966** | **0.934** | **0.946** | **0.899**\*\* | **0.843**\*\*\* |
| Norway | **0.979**\*\* | **0.892**\*\*\* | **0.961**\*\*\* | **0.917**\*\*\* | 1.027 | 1.080 | 1.025 | **0.907** |
| Sweden | **0.997** | **0.859** | **0.735**\*\*\* | **0.678**\*\*\* | 1.068 | 1.199 | 1.076 | **0.747** |
| Switzerland | 1.087 | 1.400 | 1.612 | 1.967 | 1.012 | 1.084 | 1.221 | 1.941 |
| Austria | 1.102 | 1.281 | 1.478 | 2.025 | 1.012 | 1.041 | 1.103 | 1.408 |
| Belgium | 1.049 | 1.075 | **0.853**\*\*\* | **0.939**\* | **0.993**\* | **0.941**\*\*\* | **0.867**\*\*\* | 1.048 |
| France | 1.055 | **0.979** | **0.635**\*\*\* | **0.722**\*\*\* | 1.003 | **0.978** | **0.867**\*\*\* | **0.831**\*\*\* |
| Germany | 1.088 | 1.313 | 1.543 | 1.752 | 1.007 | 1.021 | 1.121 | 1.604 |
| Spain | **0.961** | **0.783**\* | **0.473**\*\*\* | **0.511**\*\*\* | **0.961** | **0.971** | **0.938** | **0.670**\* |
| Italy | 1.023 | **0.972** | **0.618**\*\* | **0.552**\*\* | 1.028 | 1.108 | 1.071 | **0.684** |
| Finland | 1.005 | **0.876**\*\*\* | **0.673**\*\*\* | **0.451**\*\*\* | 1.062 | 1.148 | 1.094 | **0.955** |
| Netherlands | 1.083 | 1.253 | 1.330 | 1.597 | 1.008 | 1.044 | 1.112 | 1.390 |
| No. of U's <1 | 5 | **9** | **11** | **10** | 4 | 5 | 4 | 7 |
| No. of DM >1.282 | 1 | 4 | 9 | 8 | 1 | 1 | 3 | 3 |
| Median U | 1.023 | 0.989 | 0.853 | 0.939 | 1.012 | 1.044 | 1.076 | 1.048 |

Notes: This Table presents the forecasting performance of the TVP forecasting regression and the Fixed-effect panel regression with Taylor rule fundamentals defined as $TR_{on}$, $TR_{os}$ and $TR_{en}$ by country's currency. The benchmark model for both forecasting regressions is the driftless Random Walk (RW). The U (h) is the U-statistic for quarterly forecast horizons, h. For example, U (1) is the U-statistic for one-quarter-ahead forecast. Values less than one (in **bold**), indicate that the fundamentals- based regression generates a lower RMSFE than the RW, and hence forecasts better than the RW. The Diebold and Mariano (1995) test is used to compare the model's forecast accuracy relative to the benchmark at h-quarter- forecast horizon. Asterisks indicate that the null hypothesis of equal forecast accuracy is rejected at 10% (\*), 5% (\*\*) and 1% (\*\*\*), signalling a better average accuracy of the forecasts of the fundamentals-based regression relative to the benchmark. The last three rows at the bottom of the Table repeat, for convenience, the summary results reported in the main text, i.e., the "No. of U's<1", the "No. of DM > 1.282" and "Median U".

**Table E.A.1 (Continued)**

| $TR_{os}$ | TVP Regression | | | | Fixed-effect Panel Regression | | | |
|---|---|---|---|---|---|---|---|---|
| | U(1) | U(4) | U(8) | U(12) | U(1) | U(4) | U(8) | U(12) |
| Australia | **0.974** | **0.996** | **0.729*** | **0.273*** | 1.072 | 1.210 | 1.170 | 1.205 |
| Canada | **0.981** | **0.970** | 1.003 | **0.548** | **0.988** | **0.979** | **0.895** | **0.777*** |
| Denmark | 1.006 | **0.944** | **0.891**** | **0.965*** | 1.003 | **0.977** | **0.776**** | 1.099 |
| UK | 1.042 | 1.356 | 1.908 | 1.457 | 1.067 | 1.418 | 2.162 | 2.476 |
| Japan | 1.053 | 1.136 | 1.182 | 1.281 | 1.011 | 1.102 | 1.291 | 1.671 |
| Korea | 1.001 | **0.980** | **0.953** | **0.897**** | **0.922** | **0.948** | **0.906**** | **0.905**** |
| Norway | **0.977*** | **0.907**** | 1.047 | **0.966*** | 1.031 | 1.042 | **0.767**** | **0.707**** |
| Sweden | 1.031 | **0.909** | **0.735**** | **0.654**** | 1.041 | 1.133 | **0.975** | **0.559**** |
| Switzerland | 1.113 | 1.308 | 1.434 | 1.837 | **0.991** | **0.973** | **0.957** | 1.584 |
| Austria | 1.090 | 1.203 | 1.330 | 1.518 | 1.014 | 1.009 | **0.967** | 1.599 |
| Belgium | 1.008 | **0.968** | **0.999** | 1.142 | **0.990** | **0.933**** | **0.776**** | 1.107 |
| France | 1.033 | **0.883*** | **0.583**** | **0.462**** | 1.003 | **0.959** | **0.735**** | **0.710**** |
| Germany | 1.096 | 1.224 | 1.287 | 1.436 | 1.012 | 1.012 | **0.986** | 1.668 |
| Spain | **0.965** | **0.778*** | **0.487**** | **0.245**** | **0.986** | 1.005 | **0.827** | **0.469**** |
| Italy | 1.037 | **0.983** | **0.689** | **0.632*** | 1.015 | 1.084 | 1.231 | **0.695** |
| Finland | **0.987** | **0.819**** | **0.685**** | **0.518**** | 1.014 | 1.007 | **0.817**** | **0.674**** |
| Netherlands | 1.076 | 1.137 | 1.206 | 1.372 | 1.004 | **0.987** | **0.922** | 1.470 |
| | | | | | | | | |
| No. of U's <1 | 5 | **11** | 9 | **10** | 5 | 7 | **13** | 8 |
| No. of DM >1.282 | 1 | 4 | 6 | 10 | 0 | 1 | 6 | 7 |
| Median U | 1.031 | 0.980 | 0.999 | 0.965 | 1.011 | 1.007 | 0.922 | 1.099 |

**Table E.A.1 (Continued)**

| TR$_{en}$ | TVP Regression | | | | Fixed-effect Panel Regression | | | |
|---|---|---|---|---|---|---|---|---|
| | U(1) | U(4) | U(8) | U(12) | U(1) | U(4) | U(8) | U(12) |
| Australia | **0.972** | **0.888** | **0.68\*\*** | **0.696\*\*\*** | 1.042 | 1.065 | **0.994** | 1.012 |
| Canada | **0.995** | **0.945** | **0.629\*** | **0.558\*\*** | **0.981** | **0.967** | 1.039 | **0.932** |
| Denmark | 1.002 | 1.043 | 1.035 | 1.292 | 1.014 | 1.029 | **0.952** | 1.288 |
| UK | 1.026 | 1.238 | 1.298 | 1.440 | 1.052 | 1.291 | 1.699 | 1.600 |
| Japan | 1.066 | 1.210 | 1.185 | 1.333 | 1.013 | 1.091 | 1.292 | 1.664 |
| Korea | **0.994** | **0.962** | **0.938\*** | **0.900\*\*** | **0.895** | **0.948** | **0.942\*\*** | **0.977** |
| Norway | 1.025 | 1.225 | 1.307 | 1.373 | **0.997** | **0.996** | **0.969** | 1.405 |
| Sweden | **0.989** | **0.972** | **0.759\*\*\*** | 1.040 | 1.050 | 1.132 | **0.940** | **0.668\*\*\*** |
| Switzerland | 1.100 | 1.416 | 1.718 | 2.489 | 1.017 | 1.077 | 1.154 | 1.906 |
| Austria | 1.124 | 1.412 | 1.647 | 1.943 | 1.010 | 1.030 | 1.061 | 1.369 |
| Belgium | 1.136 | 1.456 | 1.687 | 2.201 | **0.988** | **0.916\*\*\*** | **0.787\*\*\*** | 1.142 |
| France | 1.009 | **0.881\*\*\*** | **0.635\*\*\*** | **0.706\*\*\*** | 1.018 | **0.939** | **0.638\*\*\*** | 1.026 |
| Germany | 1.081 | 1.229 | 1.322 | 1.478 | 1.004 | 1.011 | 1.085 | 1.697 |
| Spain | **0.967** | **0.805\*** | **0.464\*\*\*** | **0.481\*\*\*** | 1.013 | 1.030 | **0.746\*** | **0.782** |
| Italy | 1.021 | **0.949** | **0.650\*** | **0.444\*\*\*** | 1.024 | 1.089 | 1.044 | **0.621\*** |
| Finland | **0.985** | **0.817\*\*\*** | **0.662\*\*\*** | **0.596\*\*\*** | 1.019 | 1.022 | **0.890** | 1.076 |
| Netherlands | 1.113 | 1.275 | 1.350 | 1.413 | 1.007 | 1.043 | 1.095 | 1.355 |
| No. of U's <1 | 6 | 8 | 8 | 7 | 4 | 5 | **9** | 5 |
| No. of DM >1.282 | 0 | 3 | 8 | 7 | 0 | 1 | 4 | 2 |
| Median U | 1.021 | 1.043 | 1.035 | 1.292 | 1.013 | 1.030 | 0.994 | 1.142 |

**Table E.A.2 Forecast Evaluation: Taylor Rules (1999Q1-2013Q1)**

| $TR_{on}$ | TVP Regression | | | | Fixed-effect Panel Regression | | | |
|---|---|---|---|---|---|---|---|---|
| | U(1) | U(4) | U(8) | U(12) | U(1) | U(4) | U(8) | U(12) |
| Australia | 1.018 | 1.104 | 1.244 | 1.443 | 1.018 | 1.054 | 1.115 | 1.256 |
| Canada | 1.014 | 1.049 | 1.127 | 1.444 | 1.007 | 1.024 | 1.054 | 1.107 |
| Denmark | 1.006 | 1.046 | 1.086 | 1.093 | 1.007 | 1.038 | 1.081 | 1.102 |
| UK | 1.013 | 1.015 | 1.073 | 1.420 | 1.007 | 1.023 | 1.034 | 1.089 |
| Japan | 1.005 | 1.002 | **0.920** | **0.879*** | 1.003 | 1.009 | **0.972** | **0.879** |
| Korea | 1.061 | 1.175 | 1.229 | 1.252 | 1.019 | 1.074 | 1.122 | 1.230 |
| Norway | **0.981**** | **0.964**** | **0.953*** | **0.936**** | 1.010 | 1.035 | 1.100 | 1.223 |
| Sweden | **0.998** | 1.015 | 1.078 | 1.106 | 1.017 | 1.046 | 1.072 | 1.252 |
| Switzerland | **0.997** | **0.951** | **0.864**** | **0.730**** | 1.001 | **0.995** | **0.971** | **0.886**** |
| Euro | 1.009 | 1.062 | 1.184 | 1.381 | 1.011 | 1.032 | 1.066 | 1.117 |
| No. of U's <1 | 3 | 2 | 3 | 3 | 0 | 1 | 2 | 2 |
| No. of DM >1.282 | 1 | 1 | 2 | 3 | 0 | 0 | 0 | 1 |
| Median U | 1.007 | 1.031 | 1.082 | 1.179 | 1.009 | 1.033 | 1.069 | 1.112 |

Notes: See notes to Table E.A.1

**Table E.A.2 (Continued)**

| $TR_{os}$ | TVP Regression | | | | Fixed-effect Panel Regression | | | |
|---|---|---|---|---|---|---|---|---|
| | U(1) | U(4) | U(8) | U(12) | U(1) | U(4) | U(8) | U(12) |
| Australia | 1.018 | 1.052 | 1.176 | 1.407 | 1.014 | 1.038 | 1.117 | 1.309 |
| Canada | 1.017 | 1.052 | 1.158 | 1.495 | 1.013 | 1.043 | 1.118 | 1.221 |
| Denmark | 1.002 | 1.015 | 1.043 | 1.051 | 1.004 | 1.025 | 1.068 | 1.107 |
| UK | 1.010 | 1.035 | 1.121 | 1.232 | 1.007 | 1.027 | 1.058 | 1.137 |
| Japan | 1.002 | **0.996** | **0.974** | **0.944** | **0.999** | **0.987** | **0.959** | **0.906** |
| Korea | 1.032 | 1.090 | 1.109 | 1.206 | 1.017 | 1.046 | 1.063 | 1.174 |
| Norway | 1.005 | 1.003 | **0.958*** | **0.951*** | 1.005 | 1.016 | 1.052 | 1.090 |
| Sweden | 1.012 | 1.052 | 1.089 | 1.305 | 1.023 | 1.103 | 1.264 | 1.635 |
| Switzerland | **0.995** | **0.937** | **0.787**** | **0.592***** | **0.990** | **0.938*** | **0.829****** | **0.708****** |
| Euro | 1.010 | 1.064 | 1.178 | 1.344 | 1.008 | 1.046 | 1.126 | 1.216 |
| No. of U's <1 | 1 | 2 | 3 | 3 | 2 | 2 | 2 | 2 |
| No. of DM >1.282 | 0 | 0 | 2 | 2 | 0 | 1 | 1 | 1 |
| Median U | 1.010 | 1.043 | 1.099 | 1.219 | 1.007 | 1.033 | 1.065 | 1.156 |

| $TR_{en}$ | TVP Regression | | | | Fixed-effect Panel Regression | | | |
|---|---|---|---|---|---|---|---|---|
| | U(1) | U(4) | U(8) | U(12) | U(1) | U(4) | U(8) | U(12) |
| Australia | 1.022 | 1.112 | 1.265 | 1.407 | 1.015 | 1.051 | 1.127 | 1.285 |
| Canada | 1.007 | 1.044 | 1.151 | 1.519 | 1.008 | 1.029 | 1.074 | 1.150 |
| Denmark | 1.004 | 1.024 | 1.080 | 1.068 | 1.010 | 1.037 | 1.079 | 1.081 |
| UK | 1.006 | **0.995** | **0.975** | 1.547 | 1.006 | 1.025 | 1.054 | 1.125 |
| Japan | 1.004 | **0.982** | **0.943** | 1.123 | 1.005 | 1.003 | **0.972** | **0.875** |
| Korea | 1.032 | 1.082 | 1.120 | 1.184 | 1.022 | 1.082 | 1.132 | 1.269 |
| Norway | 1.009 | 1.029 | 1.032 | 1.132 | 1.015 | 1.044 | 1.150 | 1.231 |
| Sweden | 1.012 | 1.051 | 1.086 | 1.267 | 1.018 | 1.066 | 1.141 | 1.415 |
| Switzerland | **0.996** | **0.950** | **0.872**** | **0.793***** | 1.000 | **0.988** | **0.972**** | **0.871****** |
| Euro | 1.005 | 1.050 | 1.177 | 1.459 | 1.011 | 1.044 | 1.098 | 1.167 |
| No. of U's <1 | 1 | 3 | 3 | 1 | 0 | 1 | 2 | 2 |
| No. of DM >1.282 | 0 | 0 | 1 | 1 | 0 | 0 | 1 | 1 |
| Median U | 1.007 | 1.036 | 1.083 | 1.226 | 1.010 | 1.040 | 1.088 | 1.159 |

**Table E.A.3 Forecast Evaluation: Taylor Rules (2007Q1-2013Q1)**

| $TR_{on}$ | TVP Regression | | | | Fixed-effect Panel Regression | | | |
|---|---|---|---|---|---|---|---|---|
| | U(1) | U(4) | U(8) | U(12) | U(1) | U(4) | U(8) | U(12) |
| Australia | 1.023 | 1.145 | 1.477 | 1.789 | 1.008 | 1.025 | 1.096 | 1.295 |
| Canada | 1.004 | 1.014 | 1.022 | 1.723 | 1.003 | 1.005 | **0.990** | **0.985** |
| Denmark | 1.004 | 1.009 | **0.970** | 1.131 | 1.004 | 1.003 | 1.022 | 1.102 |
| UK | **0.999** | **0.962** | **0.863*** | **0.757** | 1.000 | **0.996** | **0.987** | **0.999** |
| Japan | **0.996** | **0.897**** | **0.789**** | **0.523**** | **0.996** | **0.890**** | **0.780**** | **0.715**** |
| Korea | **0.998** | **0.998** | 1.000 | 1.062 | **0.997** | **0.979** | **0.953** | **0.944** |
| Norway | 1.010 | 1.007 | **0.977**** | **0.951**** | 1.004 | 1.003 | **0.957** | **0.903*** |
| Sweden | 1.006 | 1.017 | 1.036 | 1.182 | 1.007 | 1.013 | **0.985** | 1.150 |
| Switzerland | **0.990** | **0.930** | **0.748**** | **0.521**** | **0.992** | **0.961** | **0.790**** | **0.655**** |
| Euro | 1.005 | **0.992** | **0.899** | **0.714**** | 1.003 | 1.007 | **0.983** | 1.024 |
| | | | | | | | | |
| No. of U's <1 | 4 | **5** | **7** | **5** | 4 | 4 | **8** | **6** |
| No. of DM >1.282 | 0 | 1 | 4 | 4 | 0 | 1 | 2 | 3 |
| Median U | 1.004 | 1.003 | 0.973 | 1.007 | 1.003 | 1.003 | 0.984 | 0.992 |

Notes: See notes to Table E.A.1

**Table E.A.3 (Continued)**

| $TR_{os}$ | TVP Regression | | | | Fixed-effect Panel Regression | | | |
|---|---|---|---|---|---|---|---|---|
| | U(1) | U(4) | U(8) | U(12) | U(1) | U(4) | U(8) | U(12) |
| Australia | 1.015 | 1.005 | 1.092 | 1.305 | 1.011 | 1.000 | 1.013 | 1.136 |
| Canada | 1.009 | 1.020 | 1.114 | 1.756 | 1.008 | 1.002 | **0.962**** | 1.015 |
| Denmark | 1.014 | 1.056 | 1.084 | 1.318 | 1.011 | 1.037 | 1.094 | 1.276 |
| UK | 1.001 | **0.978** | **0.885** | **0.832*** | 1.003 | **0.979** | **0.916*** | **0.868**** |
| Japan | 1.004 | **0.950** | **0.821*** | **0.773**** | 0.999 | 0.913 | **0.804**** | **0.748**** |
| Korea | **0.993** | **0.971** | **0.956** | 1.028 | **0.995** | **0.950** | **0.895** | **0.845** |
| Norway | 1.014 | 1.003 | **0.954** | **0.783*** | 1.013 | 1.006 | **0.927**** | **0.841**** |
| Sweden | 1.000 | 1.017 | 1.054 | 1.250 | 1.015 | 1.027 | 1.002 | 1.311 |
| Switzerland | **0.993** | **0.949** | **0.680*** | **0.382**** | **0.992** | **0.939** | **0.706**** | **0.463**** |
| Euro | 1.004 | 1.003 | **0.899** | **0.756**** | 1.011 | 1.026 | 1.030 | 1.043 |
| | | | | | | | | |
| No. of U's <1 | 2 | 4 | **6** | 5 | 3 | 4 | **6** | 5 |
| No. of DM >1.282 | 0 | 0 | 2 | 5 | 0 | 0 | 5 | 4 |
| Median U | 1.004 | 1.003 | 0.955 | 0.930 | 1.010 | 1.001 | 0.945 | 0.942 |

| $TR_{en}$ | TVP Regression | | | | Fixed-effect Panel Regression | | | |
|---|---|---|---|---|---|---|---|---|
| | U(1) | U(4) | U(8) | U(12) | U(1) | U(4) | U(8) | U(12) |
| Australia | 1.021 | 1.144 | 1.456 | 1.882 | 1.010 | 1.048 | 1.159 | 1.414 |
| Canada | 1.008 | 1.016 | 1.015 | 1.666 | 1.005 | 1.012 | 1.010 | 1.029 |
| Denmark | **0.996** | **0.997** | **0.906** | **0.672**** | 1.007 | 1.024 | 1.102 | 1.597 |
| UK | 1.004 | **0.963** | **0.889*** | **0.722*** | **0.997** | 1.005 | 1.032 | 1.091 |
| Japan | **0.994** | **0.853**** | **0.767**** | **0.766**** | **0.996** | **0.876**** | **0.749**** | **0.739**** |
| Korea | 1.004 | 1.006 | 1.007 | **0.933** | **0.990** | **0.963** | **0.928** | **0.925** |
| Norway | **0.972**** | **0.951**** | **0.919** | **0.803** | **0.985**** | **0.954**** | **0.772**** | **0.630*** |
| Sweden | **0.993** | **0.988** | 1.012 | **0.867** | 1.001 | 1.003 | **0.935** | 1.075 |
| Switzerland | **0.991** | **0.936** | **0.661**** | **0.493**** | **0.991** | **0.935** | **0.666**** | **0.713**** |
| Euro | 1.008 | **0.994** | **0.900** | **0.853** | 1.002 | 1.027 | 1.019 | 1.384 |
| | | | | | | | | |
| No. of U's <1 | **5** | **7** | **6** | **8** | 5 | 4 | **5** | 4 |
| No. of DM >1.282 | 1 | 2 | 3 | 4 | 1 | 2 | 3 | 3 |
| Median U | 0.999 | 0.991 | 0.912 | 0.828 | 0.999 | 1.004 | 0.972 | 1.052 |

**Table E.B.1 Forecast Evaluation: Monetary Model (MM), PPP and UIRP (1995Q1-1998Q4)**

| MM | TVP Regression | | | | Fixed-effect Panel Regression | | | |
|---|---|---|---|---|---|---|---|---|
| | U(1) | U(4) | U(8) | U(12) | U(1) | U(4) | U(8) | U(12) |
| Australia | 1.004 | 1.089 | **0.773** | **0.691**** | **0.965** | **0.994** | 0.772 | **0.576**** |
| Canada | **0.973** | **0.938** | **0.786** | **0.934** | **0.938** | **0.942** | 1.096 | 1.138 |
| Denmark | 1.006 | **0.930** | **0.703**** | **0.468**** | **0.937** | **0.715*** | **0.443**** | **0.720**** |
| UK | 1.043 | 1.352 | 9.243 | 1.700 | 1.150 | 2.092 | 3.281 | 3.839 |
| Japan | 1.025 | 1.070 | 1.111 | 1.250 | 1.001 | **0.997** | **0.941** | 1.004 |
| Korea | 1.005 | **0.965** | **0.808** | **0.794** | **0.981** | **0.903** | **0.752*** | **0.485**** |
| Norway | **0.919**** | **0.671**** | **0.850**** | **0.835**** | **0.933**** | **0.688**** | **0.552**** | **0.486**** |
| Sweden | **0.978** | 1.049 | 1.254 | 1.272 | **0.976** | **0.872** | **0.648**** | **0.742**** |
| Switzerland | **0.992** | **0.923** | **0.542**** | **0.196**** | **0.961** | **0.797**** | **0.397**** | **0.302**** |
| Austria | **0.928*** | **0.756**** | **0.469**** | **0.409**** | **0.951*** | **0.772**** | **0.465**** | **0.437**** |
| Belgium | **0.974**** | **0.837**** | **0.594**** | **0.636**** | **0.931*** | **0.738**** | **0.419**** | **0.347**** |
| France | **0.924*** | **0.670*** | **0.675*** | **0.502**** | **0.944** | **0.747**** | **0.455**** | **0.371**** |
| Germany | **0.937** | **0.747*** | **0.374**** | **0.646**** | **0.923** | **0.698*** | **0.474**** | **0.848** |
| Spain | **0.972** | **0.868** | **0.934** | 1.495 | **0.961** | **0.765** | **0.551**** | **0.804** |
| Italy | 1.024 | **0.988** | **0.598**** | **0.564**** | 1.043 | 1.056 | **0.803** | **0.838** |
| Finland | **0.958*** | **0.724**** | **0.474**** | **0.397**** | **0.956**** | **0.783**** | **0.622**** | **0.623**** |
| Netherlands | **0.921** | **0.694**** | **0.318**** | **0.318**** | **0.951** | **0.785**** | **0.448**** | **0.315**** |
| | | | | | | | | |
| No. of U's <1 | **11** | **13** | **14** | **13** | **14** | **15** | **15** | **14** |
| No. of DM >1.282 | 5 | 7 | 10 | 11 | 4 | 9 | 12 | 11 |
| Median U | 0.974 | 0.923 | 0.703 | 0.646 | 0.956 | 0.785 | 0.552 | 0.623 |

Notes: This Table presents the forecasting performance of the TVP and the Fixed-effect panel regressions with information set from the MM, PPP and UIRP by country's currency. The benchmark model for both forecasting regressions is the driftless Random Walk (RW). The U (h) is the U-statistic for quarterly forecast horizons, h. For example, U (1) is the U-statistic for one-quarter-ahead forecast. Values less than one (in **bold**), indicate that the fundamentals- based regression generates a lower RMSFE than the RW, and hence forecasts better than the RW. The Diebold and Mariano (1995) test is used to compare the model's forecast accuracy relative to the benchmark at h-quarter- forecast horizon. Asterisks indicate that the null hypothesis of equal forecast accuracy is rejected at 10% (*), 5% (**) and 1% (***), signalling a better average accuracy of the forecasts of the fundamentals-based regression relative to the benchmark. The last three rows at the bottom of the Table repeat, for convenience, the summary results reported in the main text, i.e., the "No. of U's<1", the "No. of DM > 1.282" and "Median U".

**Table E.B.1 (Continued)**

| PPP | TVP Regression | | | | Fixed-effect Panel Regression | | | |
|---|---|---|---|---|---|---|---|---|
| | U(1) | U(4) | U(8) | U(12) | U(1) | U(4) | U(8) | U(12) |
| Australia | **0.998** | **0.920** | **0.864** | **0.977** | **0.988** | **0.910** | **0.821** | **0.914** |
| Canada | 1.023 | 1.127 | 1.568 | 1.872 | 1.091 | 1.265 | 1.849 | 2.235 |
| Denmark | **0.970**\*\*\* | **0.838**\*\*\* | **0.777**\*\* | **0.811**\*\* | **0.934**\*\* | **0.716**\*\*\* | **0.575**\*\*\* | **0.660**\*\*\* |
| UK | 1.033 | 1.417 | 1.876 | 1.828 | 1.004 | 1.195 | 1.317 | 1.093 |
| Japan | 1.066 | 1.168 | 1.000 | 1.224 | **0.982** | **0.866**\*\* | **0.646**\*\*\* | **0.420**\* |
| Korea | **0.973**\*\* | **0.846**\*\*\* | **0.689**\*\*\* | **0.682**\* | **0.978**\* | **0.883**\*\*\* | **0.747**\*\*\* | **0.666**\*\* |
| Norway | **0.981** | **0.940** | 1.037 | 1.088 | **0.953**\* | **0.788**\*\* | **0.76**\*\* | **0.765**\*\* |
| Sweden | **0.998** | 1.069 | 1.103 | 1.181 | **0.999** | **0.899** | **0.717**\*\* | **0.838**\* |
| Switzerland | **0.973** | **0.853**\*\* | **0.684**\* | **0.806** | **0.957**\*\*\* | **0.787**\*\* | **0.607**\*\* | **0.670**\*\* |
| Austria | 1.003 | **0.935**\*\* | **0.822**\* | **0.930** | **0.953**\*\*\* | **0.789**\*\*\* | **0.677**\*\* | **0.757**\*\* |
| Belgium | 1.026 | 1.025 | **0.983** | 1.110 | **0.937**\*\*\* | **0.765**\*\*\* | **0.67**\*\* | **0.759**\*\* |
| France | **0.981**\*\* | **0.848**\*\* | **0.742**\*\* | **0.859**\* | **0.940**\*\*\* | **0.745**\*\*\* | **0.633**\*\*\* | **0.718**\*\*\* |
| Germany | 1.031 | 1.050 | 1.118 | 1.382 | **0.969**\*\* | **0.851**\*\*\* | **0.796**\* | **0.901** |
| Spain | **0.956** | **0.701**\* | **0.402**\*\*\* | **0.435**\*\*\* | **0.945** | **0.685**\*\* | **0.445**\*\*\* | **0.493**\*\*\* |
| Italy | 1.015 | **0.931** | **0.978** | **0.635**\* | 1.001 | **0.888** | **0.598**\*\*\* | **0.582**\*\*\* |
| Finland | **0.989** | **0.882**\*\* | **0.785**\*\* | **0.875** | **0.987** | **0.930**\* | **0.976** | 1.073 |
| Netherlands | 1.043 | 1.039 | 1.009 | 1.178 | **0.974**\*\* | **0.883**\*\*\* | **0.850** | **0.967** |
| No. of U's <1 | **9** | **10** | **11** | **9** | **14** | **15** | **15** | **14** |
| No. of DM >1.282 | 3 | 7 | 7 | 5 | 9 | 12 | 12 | 11 |
| Median U | 0.998 | 0.935 | 0.978 | 0.977 | 0.974 | 0.866 | 0.717 | 0.759 |

**Table E.B.1 (Continued)**

| UIRP | TVP Regression | | | | Fixed-effect Panel Regression | | | |
|---|---|---|---|---|---|---|---|---|
| | U(1) | U(4) | U(8) | U(12) | U(1) | U(4) | U(8) | U(12) |
| Australia | **0.978** | 1.019 | **0.753** | **0.292**\*\*\* | **0.970** | **0.987** | **0.746**\* | **0.37**\*\*\* |
| Canada | **0.991** | **0.983** | **0.985** | **0.555**\*\* | **0.937** | **0.834** | **0.605**\* | **0.505**\*\*\* |
| Denmark | 1.007 | **0.945** | **0.929**\*\*\* | **0.986** | **0.975** | **0.912**\*\* | **0.856**\*\*\* | **0.894**\*\*\* |
| UK | 1.049 | 1.385 | 2.034 | 1.357 | 1.023 | 1.223 | 1.394 | 1.248 |
| Japan | 1.036 | 1.121 | 1.184 | 1.249 | 1.041 | 1.138 | 1.183 | 1.287 |
| Korea | 1.000 | **0.978** | **0.952** | **0.897**\*\* | **0.999** | **0.981** | **0.959** | **0.90**\*\* |
| Norway | **0.978**\* | **0.926**\*\*\* | 1.062 | **0.990** | **0.956**\* | **0.792**\*\*\* | **0.757**\*\*\* | **0.723**\*\*\* |
| Sweden | 1.011 | **0.902** | **0.716**\*\*\* | **0.671**\*\*\* | **0.965** | **0.867** | **0.447**\*\*\* | **0.400**\*\*\* |
| Switzerland | 1.109 | 1.300 | 1.441 | 1.916 | 1.044 | 1.126 | 1.190 | 1.373 |
| Austria | 1.087 | 1.203 | 1.330 | 1.525 | 1.037 | 1.074 | 1.092 | 1.203 |
| Belgium | 1.003 | **0.977** | 1.062 | 1.245 | **0.979** | **0.921**\*\* | **0.858**\*\*\* | **0.888**\*\*\* |
| France | 1.006 | **0.848** | **0.599**\*\*\* | **0.532**\*\*\* | **0.964** | **0.838**\* | **0.686**\*\*\* | **0.622**\*\*\* |
| Germany | 1.096 | 1.221 | 1.300 | 1.446 | 1.033 | 1.074 | 1.103 | 1.228 |
| Spain | **0.971** | 0.786 | **0.489**\*\*\* | **0.220**\*\* | **0.966** | **0.788** | **0.468**\*\*\* | **0.210**\*\*\* |
| Italy | 1.032 | **0.981** | 0.671 | **0.607**\*\* | 1.026 | **0.969** | **0.489**\*\* | **0.286**\*\*\* |
| Finland | **0.988** | **0.816**\*\*\* | **0.686**\*\*\* | **0.571**\*\*\* | **0.978** | **0.889**\*\* | **0.784**\*\*\* | **0.763**\*\*\* |
| Netherlands | 1.095 | 1.142 | 1.213 | 1.476 | 1.018 | 1.034 | 1.038 | 1.135 |
| | | | | | | | | |
| No. of U's <1 | 5 | **10** | 9 | 10 | **10** | **11** | **11** | **11** |
| No. of DM >1.282 | 1 | 2 | 5 | 8 | 1 | 5 | 10 | 11 |
| Median U | 1.007 | 0.981 | 0.985 | 0.986 | 0.979 | 0.969 | 0.856 | 0.888 |

**Table E.B.2 Forecast Evaluation: Monetary Model (MM), PPP and UIRP (1999Q1-2013Q1)**

| MM | TVP Regression | | | | Fixed-effect Panel Regression | | | |
|---|---|---|---|---|---|---|---|---|
| | U(1) | U(4) | U(8) | U(12) | U(1) | U(4) | U(8) | U(12) |
| Australia | 1.033 | 1.093 | 1.157 | 1.507 | 1.039 | 1.135 | 1.360 | 1.796 |
| Canada | 1.032 | 1.103 | 1.170 | 1.498 | 1.037 | 1.145 | 1.394 | 1.643 |
| Denmark | 1.011 | 1.055 | 1.149 | 1.320 | 1.022 | 1.112 | 1.316 | 1.624 |
| UK | 1.015 | 1.077 | 1.298 | 1.337 | 1.043 | 1.202 | 1.532 | 2.108 |
| Japan | 1.024 | 1.171 | 1.243 | 1.138 | 1.015 | 1.088 | 1.290 | 1.705 |
| Korea | 1.064 | 1.420 | 1.601 | 2.304 | 1.051 | 1.176 | 1.376 | 1.717 |
| Norway | 1.047 | 1.190 | 1.418 | 1.951 | 1.019 | 1.086 | 1.281 | 1.566 |
| Sweden | **0.983** | **0.945** | **0.886** | **0.731** | **0.998** | **0.984** | **0.939*** | **0.864*** |
| Switzerland | 1.006 | 1.045 | 1.188 | 1.380 | 1.006 | 1.052 | 1.188 | 1.399 |
| Euro | 1.009 | 1.081 | 1.283 | 1.319 | 1.016 | 1.085 | 1.238 | 1.460 |
| No. of U's <1 | 1 | 1 | 1 | 1 | 1 | 1 | 1 | 1 |
| No. of DM >1.282 | 0 | 0 | 0 | 0 | 0 | 0 | 1 | 1 |
| Median U | 1.019 | 1.087 | 1.216 | 1.359 | 1.021 | 1.100 | 1.303 | 1.633 |

Notes: See notes to Table E.B.1

**Table E.B.2 (Continued)**

| PPP | TVP Regression | | | | Fixed-effect Panel Regression | | | |
|---|---|---|---|---|---|---|---|---|
| | U(1) | U(4) | U(8) | U(12) | U(1) | U(4) | U(8) | U(12) |
| Australia | 1.027 | 1.087 | 1.145 | 1.153 | 1.026 | 1.089 | 1.161 | 1.229 |
| Canada | 1.006 | 1.015 | 1.044 | 1.029 | **0.997** | **0.999** | 1.059 | 1.005 |
| Denmark | **0.996** | **0.976** | **0.922** | **0.886** | 1.002 | 1.017 | 1.026 | 1.023 |
| UK | **0.999** | 1.008 | 1.108 | 1.103 | **0.991** | **0.965** | **0.967** | 1.078 |
| Japan | **0.992** | **0.945** | **0.787**** | **0.688***** | **0.982** | **0.899*** | **0.730**** | **0.580***** |
| Korea | **0.986** | **0.913** | **0.852** | **0.765**** | **0.992** | **0.942** | **0.832*** | **0.702***** |
| Norway | **0.993** | **0.971** | **0.888*** | **0.718**** | **0.997** | **0.985** | **0.996** | 1.030 |
| Sweden | **0.991*** | **0.962** | **0.896** | **0.655*** | **0.985*** | **0.927*** | **0.792**** | **0.614**** |
| Switzerland | **0.987** | **0.964** | **0.942** | **0.836** | **0.992** | **0.985** | **0.977** | **0.904** |
| Euro | **0.995** | **0.972** | **0.955** | **0.878** | **0.995** | **0.985** | **0.941** | **0.869** |
| No. of U's <1 | **8** | **7** | **7** | **7** | **8** | **8** | **7** | **5** |
| No. of DM >1.282 | 1 | 0 | 2 | 4 | 1 | 2 | 3 | 3 |
| Median U | 0.994 | 0.971 | 0.932 | 0.857 | 0.994 | 0.985 | 0.972 | 0.955 |

| UIRP | TVP Regression | | | | Fixed-effect Panel Regression | | | |
|---|---|---|---|---|---|---|---|---|
| | U(1) | U(4) | U(8) | U(12) | U(1) | U(4) | U(8) | U(12) |
| Australia | 1.021 | 1.058 | 1.178 | 1.429 | 1.015 | 1.044 | 1.122 | 1.318 |
| Canada | 1.016 | 1.052 | 1.225 | 1.475 | 1.014 | 1.041 | 1.094 | 1.182 |
| Denmark | 1.002 | 1.020 | 1.040 | 1.045 | 1.005 | 1.027 | 1.072 | 1.106 |
| UK | 1.010 | 1.035 | 1.133 | 1.198 | 1.009 | 1.034 | 1.059 | 1.119 |
| Japan | 1.001 | 1.005 | **0.978** | **0.955** | **0.999** | **0.981** | **0.948** | **0.896** |
| Korea | 1.031 | 1.086 | 1.103 | 1.211 | 1.022 | 1.052 | 1.061 | 1.156 |
| Norway | 1.006 | 1.005 | **0.968** | **0.959*** | 1.009 | 1.024 | 1.067 | 1.100 |
| Sweden | 1.011 | 1.050 | 1.089 | 1.298 | 1.021 | 1.082 | 1.201 | 1.517 |
| Switzerland | **0.995** | **0.944** | **0.793**** | **0.605***** | **0.989** | **0.940*** | **0.833***** | **0.693***** |
| Euro | 1.004 | 1.024 | 1.134 | 1.258 | 1.008 | 1.045 | 1.123 | 1.209 |
| No. of U's <1 | 1 | 1 | 3 | 3 | 2 | 2 | 2 | 2 |
| No. of DM >1.282 | 0 | 0 | 1 | 2 | 0 | 1 | 1 | 1 |
| Median U | 1.008 | 1.030 | 1.096 | 1.204 | 1.009 | 1.037 | 1.070 | 1.137 |

**Table E.B.3 Forecast Evaluation: Monetary Model (MM), PPP and UIRP (2007Q1-2013Q1)**

| MM | TVP Regression | | | | Fixed-effect Panel Regression | | | |
|---|---|---|---|---|---|---|---|---|
| | U(1) | U(4) | U(8) | U(12) | U(1) | U(4) | U(8) | U(12) |
| Australia | 1.026 | 1.063 | **0.947*** | 1.003 | 1.032 | 1.105 | 1.437 | 1.999 |
| Canada | 1.031 | 1.140 | **0.998** | 1.213 | 1.021 | 1.042 | 1.292 | 2.041 |
| Denmark | 1.012 | 1.012 | **0.933** | **0.641*** | 1.014 | 1.026 | 1.056 | 1.340 |
| UK | 1.013 | 1.080 | 1.046 | 1.448 | **0.995** | **0.915** | **0.724** | **0.757** |
| Japan | 1.013 | 1.122 | 1.021 | 1.040 | 1.002 | 1.041 | 1.187 | 1.338 |
| Korea | **0.990** | **0.932** | **0.868** | 1.258 | **0.990** | **0.939** | **0.811** | **0.674** |
| Norway | 1.038 | 1.094 | 1.633 | 4.479 | 1.012 | 1.014 | 1.134 | 2.181 |
| Sweden | **0.982** | **0.880*** | **0.827*** | **0.933** | 1.005 | **0.999** | **0.987** | 1.264 |
| Switzerland | 1.007 | 1.027 | 1.102 | **0.937** | 1.002 | 1.016 | 1.057 | 1.140 |
| Euro Area | 1.007 | 1.001 | **0.944** | **0.867** | 1.009 | 1.002 | **0.943** | **0.913** |
| | | | | | | | | |
| No. of U's <1 | 2 | 2 | **6** | 4 | 2 | 3 | 4 | 3 |
| No. of DM >1.282 | 0 | 1 | 2 | 1 | 0 | 0 | 0 | 0 |
| Median U | 1.012 | 1.045 | 0.972 | 1.021 | 1.007 | 1.015 | 1.057 | 1.301 |

Notes: See notes to Table E.B.1

**Table E.B.3 (Continued)**

| PPP | TVP Regression | | | | Fixed-effect Panel Regression | | | |
|---|---|---|---|---|---|---|---|---|
| | U(1) | U(4) | U(8) | U(12) | U(1) | U(4) | U(8) | U(12) |
| Australia | 1.035 | 1.117 | 1.405 | 1.642 | 1.038 | 1.143 | 1.509 | 1.850 |
| Canada | 1.004 | **0.975** | 1.027 | 1.347 | 1.004 | **0.964** | 1.186 | 1.682 |
| Denmark | **0.990** | **0.930** | **0.847** | 1.006 | **0.994** | **0.963** | 1.064 | 1.404 |
| UK | **0.971** | **0.825*** | **0.615**** | 1.375 | **0.967** | **0.794*** | **0.421**** | **0.186**** |
| Japan | **0.977** | **0.778**** | **0.650**** | **0.702*** | **0.966*** | **0.754**** | **0.576**** | **0.430**** |
| Korea | **0.959*** | **0.777**** | **0.552**** | **0.286*** | **0.972*** | **0.855**** | **0.650**** | **0.335**** |
| Norway | **0.989** | **0.928** | **0.816*** | 1.029 | **0.992** | **0.935** | **0.999** | 1.791 |
| Sweden | **0.984**** | **0.912**** | **0.844*** | **0.602*** | **0.981**** | **0.887**** | **0.679**** | **0.631*** |
| Switzerland | **0.999** | 1.037 | 1.334 | 1.402 | 1.002 | 1.050 | 1.337 | 1.474 |
| Euro Area | **0.989** | **0.919** | **0.925** | 1.030 | **0.989** | **0.927** | **0.897** | 1.116 |
| No. of U's <1 | **8** | **8** | **7** | 3 | **7** | **8** | **6** | 4 |
| No. of DM >1.282 | 2 | 4 | 5 | 3 | 3 | 4 | 4 | 4 |
| Median U | 0.989 | 0.924 | 0.845 | 1.029 | 0.991 | 0.931 | 0.948 | 1.260 |

| UIRP | TVP Regression | | | | Fixed-effect Panel Regression | | | |
|---|---|---|---|---|---|---|---|---|
| | U(1) | U(4) | U(8) | U(12) | U(1) | U(4) | U(8) | U(12) |
| Australia | 1.018 | 1.014 | 1.089 | 1.312 | 1.014 | 1.020 | 1.033 | 1.136 |
| Canada | 1.006 | 1.020 | 1.187 | 1.731 | 1.011 | 1.012 | **0.97*** | 1.038 |
| Denmark | 1.011 | 1.058 | 1.115 | 1.368 | 1.013 | 1.040 | 1.135 | 1.326 |
| UK | 1.002 | **0.975** | **0.870** | **0.854**** | 1.007 | **0.996** | **0.946** | **0.905** |
| Japan | 1.008 | **0.976** | **0.821*** | **0.788**** | **0.998** | **0.905** | **0.786**** | **0.731**** |
| Korea | **0.994** | **0.968** | **0.952** | 1.033 | **0.999** | **0.966** | **0.909** | **0.855** |
| Norway | 1.012 | 1.004 | **0.974** | **0.842** | 1.016 | 1.025 | 1.010 | 1.014 |
| Sweden | 1.001 | 1.015 | 1.054 | 1.226 | 1.020 | 1.044 | 1.042 | 1.374 |
| Switzerland | **0.999** | **0.964** | **0.697*** | **0.379**** | **0.994** | **0.951** | **0.740**** | **0.463**** |
| Euro Area | 1.010 | 1.041 | 1.008 | **0.835**** | 1.013 | 1.032 | 1.065 | 1.083 |
| No. of U's <1 | 2 | 4 | **5** | 5 | 3 | 4 | **5** | 4 |
| No. of DM >1.282 | 0 | 0 | 2 | 4 | 0 | 0 | 3 | 2 |
| Median U | 1.007 | 1.009 | 0.991 | 0.944 | 1.012 | 1.016 | 0.990 | 1.026 |

**Table E.C.1 Forecast Evaluation: Factors from Exchange Rates (1995Q1-1998Q4)**

| F1 | TVP Regression | | | | Fixed-effect Panel Regression | | | |
|---|---|---|---|---|---|---|---|---|
| | U(1) | U(4) | U(8) | U(12) | U(1) | U(4) | U(8) | U(12) |
| Australia | **0.972** | **0.932**** | **0.916**** | 2.018 | 1.019 | 1.061 | **0.801** | **0.547**** |
| Canada | **0.947** | **0.912** | **0.755*** | **0.456**** | 1.097 | 1.199 | 1.032 | **0.875** |
| Denmark | **0.953**** | **0.838**** | **0.740**** | 1.153 | **0.978** | **0.944** | 1.072 | 1.237 |
| UK | 1.023 | 1.331 | 1.803 | 2.039 | 1.043 | 1.601 | 2.242 | 2.257 |
| Japan | 1.048 | 1.151 | 1.228 | 2.095 | 1.085 | 1.280 | 1.427 | 1.809 |
| Korea | 1.000 | 1.476 | 1.485 | 1.444 | 1.004 | **0.981** | **0.905** | **0.754*** |
| Norway | **0.967**** | **0.750**** | **0.618**** | **0.622**** | **0.988** | **0.849*** | **0.821** | **0.859** |
| Sweden | 1.003 | **0.937** | **0.464**** | **0.508**** | 1.040 | 1.041 | **0.711*** | **0.813*** |
| Switzerland | 1.131 | 1.389 | 1.579 | 1.818 | 1.055 | 1.177 | 1.350 | 1.597 |
| Austria | 1.017 | 1.015 | 1.035 | 1.099 | 1.052 | 1.140 | 1.286 | 1.472 |
| Belgium | **0.962**** | **0.879**** | 1.446 | 1.126 | **0.982** | **0.956** | 1.057 | 1.208 |
| France | **0.954**** | **0.827**** | **0.708**** | **0.934** | **0.968*** | **0.881*** | **0.959** | 1.091 |
| Germany | 1.058 | 1.163 | 1.338 | 1.520 | 1.044 | 1.124 | 1.272 | 1.458 |
| Spain | **0.972** | 1.313 | 1.213 | **0.676**** | **0.957** | **0.714*** | **0.535**** | **0.568**** |
| Italy | 1.024 | **0.989** | **0.904** | 1.073 | 1.071 | 1.146 | **0.969** | **0.931** |
| Finland | **0.953*** | **0.761**** | **0.554**** | **0.471**** | **0.964*** | **0.813**** | **0.808*** | **0.864** |
| Netherlands | 1.034 | 1.089 | 1.226 | 1.258 | 1.035 | 1.106 | 1.245 | 1.428 |
| | | | | | | | | |
| No. of U's <1 | 8 | 9 | 8 | 6 | 6 | 7 | 8 | 8 |
| No. of DM >1.282 | 5 | 6 | 7 | 5 | 2 | 4 | 3 | 4 |
| Median U | 1.000 | 0.989 | 1.035 | 1.126 | 1.035 | 1.061 | 1.032 | 1.091 |

Notes: This Table presents the forecasting performance of the TVP and the Fixed-effect panel regressions with one (F1), two (F2) or three (F3) factors from exchange rates by country's currency. The benchmark model for both forecasting regressions is the driftless Random Walk (RW). The U (h) is the U-statistic for quarterly forecast horizons, h. For example, U (1) is the U-statistic for one-quarter-ahead forecast. Values less than one (in **bold**), indicate that the fundamentals-based regression generates a lower RMSFE than the RW, and hence forecasts better than the RW. The Diebold and Mariano (1995) test is used to compare the model's forecast accuracy relative to the benchmark at h-quarter-forecast horizon. Asterisks indicate that the null hypothesis of equal forecast accuracy is rejected at 10% (*), 5% (**) and 1% (***), signalling a better average accuracy of the forecasts of the fundamentals-based regression relative to the benchmark. The last three rows at the bottom of the Table repeat, for convenience, the summary results reported in the main text, i.e., the "No. of U's<1", the "No. of DM > 1.282" and "Median U".

**Table E.C.1 (Continued)**

| F2 | TVP Regression | | | | Fixed-effect Panel Regression | | | |
|---|---|---|---|---|---|---|---|---|
| | U(1) | U(4) | U(8) | U(12) | U(1) | U(4) | U(8) | U(12) |
| Australia | **0.991** | **0.999** | **0.856**\*\*\* | **0.268**\*\*\* | 1.035 | 1.125 | 1.565 | 1.838 |
| Canada | **0.974** | **0.953** | **0.786**\*\* | **0.454**\*\*\* | 1.169 | 1.535 | 2.282 | 2.830 |
| Denmark | **0.934**\*\* | **0.759**\*\*\* | **0.606**\*\*\* | 1.134 | **0.938**\*\* | **0.744**\*\*\* | **0.636**\*\*\* | **0.714**\*\*\* |
| UK | 1.025 | 1.341 | 1.929 | 2.158 | 1.028 | 1.384 | 1.938 | 2.154 |
| Japan | 1.051 | 1.164 | 1.708 | 2.122 | 1.012 | 1.021 | **0.985** | **0.851** |
| Korea | 1.001 | 1.158 | 1.456 | 1.426 | 1.006 | 1.010 | 1.046 | 1.152 |
| Norway | **0.983**\* | **0.836**\*\*\* | **0.721**\*\*\* | **0.782**\*\*\* | 1.024 | 1.093 | 1.204 | 1.292 |
| Sweden | **0.998** | **0.931** | **0.495**\*\*\* | **0.687**\*\*\* | 1.020 | 1.090 | 1.220 | 1.314 |
| Switzerland | 1.013 | 1.040 | **0.930** | **0.979** | **0.952**\*\* | **0.752**\*\*\* | **0.422**\*\*\* | **0.296**\*\*\* |
| Austria | **0.970**\*\* | **0.883**\*\*\* | **0.771**\*\*\* | **0.969** | **0.936**\*\*\* | **0.725**\*\*\* | **0.452**\*\*\* | **0.481**\*\*\* |
| Belgium | **0.945**\*\* | **0.824**\*\*\* | **0.725**\*\*\* | 1.112 | **0.931**\*\*\* | **0.752**\*\*\* | **0.632**\*\*\* | **0.683**\*\*\* |
| France | **0.942**\*\* | **0.764**\*\*\* | **0.606**\*\*\* | **0.542**\*\*\* | **0.954**\*\*\* | **0.810**\*\*\* | **0.786**\*\* | **0.876** |
| Germany | **0.939**\*\*\* | **0.775**\*\*\* | **0.591**\*\*\* | **0.646**\*\*\* | **0.932**\*\*\* | **0.717**\*\*\* | **0.458**\*\*\* | **0.494**\*\*\* |
| Spain | **0.962** | **0.793**\* | 1.213 | **0.676**\*\* | **0.967** | **0.868**\* | **0.966** | 1.081 |
| Italy | 1.025 | 1.556 | **0.901** | 1.070 | 1.026 | 1.063 | 1.210 | 1.290 |
| Finland | **0.959**\*\* | **0.757**\*\*\* | **0.555**\*\*\* | **0.501**\*\*\* | **0.985** | **0.936** | 1.001 | 1.073 |
| Netherlands | **0.930**\*\*\* | **0.766**\*\*\* | **0.548**\*\*\* | **0.596**\*\*\* | **0.929**\*\*\* | **0.720**\*\*\* | **0.471**\*\*\* | **0.505**\*\*\* |
| | | | | | | | | |
| No. of U's <1 | **12** | **12** | **13** | **11** | 9 | 9 | 9 | 8 |
| No. of DM >1.282 | 8 | 9 | 11 | 9 | 7 | 8 | 7 | 6 |
| Median U | 0.974 | 0.883 | 0.771 | 0.782 | 0.985 | 0.936 | 0.985 | 1.073 |

**Table E.C.1 (Continued)**

| F3 | TVP Regression | | | | Fixed-effect Panel Regression | | | |
|---|---|---|---|---|---|---|---|---|
| | U(1) | U(4) | U(8) | U(12) | U(1) | U(4) | U(8) | U(12) |
| Australia | **0.989** | **0.978** | **0.754**** | **0.293**** | 1.003 | 1.029 | 1.299 | 1.507 |
| Canada | **0.977** | 1.024 | 1.078 | **0.476**** | 1.545 | 2.651 | 4.735 | 5.776 |
| Denmark | **0.929*** | **0.730**** | **0.540**** | 1.072 | **0.932**** | **0.705**** | **0.519**** | **0.580**** |
| UK | 1.001 | 1.130 | 1.368 | 1.484 | **0.999** | 1.136 | 1.383 | 1.530 |
| Japan | 1.050 | 1.176 | 1.670 | 2.060 | 1.008 | **0.987** | **0.904** | **0.743** |
| Korea | 1.000 | **0.904** | 1.386 | 1.321 | 1.000 | **0.987** | 1.032 | 1.230 |
| Norway | **0.960**** | **0.746**** | **0.613**** | **0.614**** | **0.980*** | **0.879*** | 0.893 | 0.951 |
| Sweden | **0.996** | **0.924** | **0.465**** | **0.606**** | 1.011 | 1.044 | 1.089 | 1.191 |
| Switzerland | 1.016 | 1.059 | **0.997** | 1.142 | **0.955**** | **0.771**** | **0.482**** | **0.369**** |
| Austria | **0.977**** | **0.907**** | **0.820**** | 1.086 | **0.943**** | **0.761**** | **0.556**** | **0.595**** |
| Belgium | **0.940**** | **0.802**** | **0.673**** | 1.083 | **0.920**** | **0.704**** | **0.504**** | **0.528**** |
| France | **0.929*** | **0.695**** | **0.450**** | **0.855*** | **0.937**** | **0.710**** | **0.506**** | **0.547**** |
| Germany | **0.952**** | **0.820**** | **0.703**** | **0.781**** | **0.939**** | **0.750**** | **0.554**** | **0.599**** |
| Spain | **0.962** | **0.797*** | 1.217 | **0.683**** | **0.966** | **0.850**** | 0.926 | 1.038 |
| Italy | 1.026 | 1.548 | **0.904** | 1.067 | 1.029 | 1.068 | 1.222 | 1.317 |
| Finland | **0.972**** | **0.772**** | **0.578**** | **0.542**** | 1.009 | 1.067 | 1.235 | 1.351 |
| Netherlands | **0.940**** | **0.791**** | **0.611**** | **0.687**** | **0.933**** | **0.744**** | **0.541**** | **0.582**** |
| | | | | | | | | |
| No. of U's <1 | **12** | **12** | **12** | **9** | **10** | **11** | **10** | **9** |
| No. of DM >1.282 | 8 | 9 | 10 | 9 | 8 | 9 | 7 | 7 |
| Median U | 0.977 | 0.904 | 0.754 | 0.855 | 0.980 | 0.879 | 0.904 | 0.951 |

**Table E.C.2 Forecast Evaluation: Factors from Exchange Rates (1999Q1-2013Q1)**

| F1 | TVP Regression | | | | Fixed-effect Panel Regression | | | |
|---|---|---|---|---|---|---|---|---|
| | U(1) | U(4) | U(8) | U(12) | U(1) | U(4) | U(8) | U(12) |
| Australia | 1.023 | 1.067 | 1.150 | 1.178 | 1.021 | 1.072 | 1.183 | 1.390 |
| Canada | 1.019 | 1.070 | 1.117 | 1.396 | 1.016 | 1.067 | 1.176 | 1.235 |
| Denmark | 1.004 | 1.015 | 1.081 | 1.560 | 1.004 | 1.022 | 1.005 | **0.929** |
| UK | **0.996** | **0.985** | 1.116 | 1.150 | **0.994** | **0.977** | **0.975** | **0.925** |
| Japan | 1.004 | 1.011 | **0.992** | 1.284 | 1.013 | 1.099 | 1.189 | 1.183 |
| Korea | 1.026 | 1.087 | 1.728 | 1.796 | 1.045 | 1.145 | 1.294 | 1.431 |
| Norway | 1.008 | 1.031 | 1.125 | 1.564 | 1.007 | 1.029 | 1.130 | 1.294 |
| Sweden | 1.017 | 1.068 | 1.185 | 1.935 | 1.021 | 1.088 | 1.198 | 1.442 |
| Switzerland | 1.009 | 1.033 | **0.990** | **0.928** | **0.998** | **0.976** | **0.909** | **0.847** |
| Euro | 1.002 | **0.982** | 1.184 | 1.227 | 1.004 | 1.023 | 1.017 | **0.967** |
| No. of U's <1 | 1 | 2 | 2 | 1 | 2 | 2 | 2 | 4 |
| No. of DM >1.282 | 0 | 0 | 0 | 0 | 0 | 0 | 0 | 0 |
| Median U | 1.009 | 1.032 | 1.121 | 1.340 | 1.010 | 1.048 | 1.153 | 1.209 |

Notes: See notes to Table E.C.1

**Table E.C.2 (Continued)**

| F2 | TVP Regression | | | | Fixed-effect Panel Regression | | | |
|---|---|---|---|---|---|---|---|---|
| | U(1) | U(4) | U(8) | U(12) | U(1) | U(4) | U(8) | U(12) |
| Australia | 1.020 | 1.075 | 1.189 | 1.665 | **0.996** | **0.990** | **0.966** | **0.895** |
| Canada | 1.021 | 1.065 | 1.168 | 1.579 | **0.993** | **0.975** | **0.981** | **0.936** |
| Denmark | 1.010 | 1.050 | 1.137 | 1.110 | 1.010 | 1.054 | 1.119 | 1.163 |
| UK | **0.991** | **0.974** | **0.953** | **0.919** | 0.990 | 0.960 | 0.926 | 0.836 |
| Japan | 1.006 | 1.021 | 1.644 | 1.269 | 1.009 | 1.073 | 1.151 | 1.220 |
| Korea | 1.026 | 1.097 | 1.690 | 1.797 | 1.010 | 1.010 | 1.007 | **0.972** |
| Norway | 1.002 | 1.015 | 1.087 | 1.878 | **0.999** | **0.987** | **0.970** | **0.976** |
| Sweden | 1.011 | 1.050 | 1.166 | 2.644 | 1.005 | 1.022 | 1.018 | **0.998** |
| Switzerland | **0.998** | **0.984** | 1.008 | **0.945** | 1.009 | 1.090 | 1.285 | 1.444 |
| Euro | 1.007 | **0.989** | 1.118 | 1.205 | 1.005 | 1.033 | 1.062 | 1.070 |
| | | | | | | | | |
| No. of U's <1 | 2 | 3 | 1 | 2 | 4 | 4 | 4 | **6** |
| No. of DM >1.282 | 0 | 0 | 0 | 0 | 0 | 0 | 0 | 0 |
| Median U | 1.008 | 1.035 | 1.152 | 1.424 | 1.005 | 1.016 | 1.012 | 0.987 |

| F3 | TVP Regression | | | | Fixed-effect Panel Regression | | | |
|---|---|---|---|---|---|---|---|---|
| | U(1) | U(4) | U(8) | U(12) | U(1) | U(4) | U(8) | U(12) |
| Australia | 1.022 | 1.080 | 1.223 | 1.672 | 1.003 | 1.007 | **0.983** | **0.906** |
| Canada | 1.028 | 1.083 | 1.599 | 1.836 | 1.026 | 1.160 | 1.437 | 1.404 |
| Denmark | 1.011 | 1.058 | 1.160 | 1.103 | 1.009 | 1.051 | 1.113 | 1.174 |
| UK | **0.989** | **0.946** | **0.866** | **0.830** | 0.989 | 0.943 | 0.863 | **0.794*** |
| Japan | 1.005 | 1.030 | 1.630 | 1.281 | 1.008 | 1.080 | 1.187 | 1.264 |
| Korea | 1.027 | 1.096 | 1.681 | 1.826 | 1.000 | **0.986** | 1.046 | 1.029 |
| Norway | 1.003 | 1.015 | 1.092 | 1.739 | **0.998** | **0.983** | **0.948** | **0.956** |
| Sweden | 1.008 | 1.041 | 1.167 | 1.367 | 1.002 | 1.006 | **0.984** | **0.954** |
| Switzerland | 1.001 | **0.988** | **0.988** | **0.978** | 1.013 | 1.106 | 1.305 | 1.441 |
| Euro | 1.007 | **0.990** | 1.109 | 1.207 | 1.005 | 1.031 | 1.057 | 1.064 |
| | | | | | | | | |
| No. of U's <1 | 1 | 3 | 2 | 2 | 2 | 3 | 4 | 4 |
| No. of DM >1.282 | 0 | 0 | 0 | 0 | 0 | 0 | 0 | 1 |
| Median U | 1.007 | 1.035 | 1.163 | 1.324 | 1.004 | 1.019 | 1.051 | 1.046 |

**Table E.C.3 Forecast Evaluation: Factors from Exchange Rates (2007Q1-2013Q1)**

| F1 | TVP Regression | | | | Fixed-effect Panel Regression | | | |
|---|---|---|---|---|---|---|---|---|
| | U(1) | U(4) | U(8) | U(12) | U(1) | U(4) | U(8) | U(12) |
| Australia | 1.015 | 1.060 | **0.993** | **0.701*** | 1.013 | 1.048 | 1.326 | 1.745 |
| Canada | 1.012 | 1.063 | 1.014 | 1.524 | 1.009 | **0.990** | 1.383 | 2.145 |
| Denmark | 1.003 | **0.969** | 0.870 | 2.417 | **0.999** | 0.974 | 1.104 | 1.597 |
| UK | **0.983** | 0.912 | 0.840 | 1.132 | **0.975** | 0.858 | 0.820 | **0.954** |
| Japan | **0.997** | **0.902*** | **0.810\*\*\*** | 1.169 | 1.000 | **0.899** | **0.778\*\*\*** | **0.724\*\*\*** |
| Korea | 1.001 | 1.008 | 1.885 | 2.389 | **0.996** | **0.967** | 1.141 | 1.742 |
| Norway | 1.009 | 1.004 | 1.046 | 2.309 | 1.007 | **0.999** | 1.449 | 3.092 |
| Sweden | 1.013 | 1.031 | 1.094 | 1.829 | 1.019 | 1.059 | 1.460 | 3.164 |
| Switzerland | 1.006 | 1.026 | 1.185 | 1.410 | 1.001 | 1.009 | 1.136 | 1.412 |
| Euro | 1.001 | 1.056 | 1.439 | 1.820 | 1.000 | **0.978** | 1.117 | 1.623 |
| | | | | | | | | |
| No. of U's <1 | 2 | 3 | 4 | 1 | 4 | **7** | 2 | 2 |
| No. of DM >1.282 | 0 | 1 | 1 | 1 | 0 | 0 | 1 | 1 |
| Median U | 1.005 | 1.017 | 1.030 | 1.672 | 1.001 | 0.984 | 1.139 | 1.682 |

Notes: See notes to Table E.C.1

**Table E.C.3 (Continued)**

| F2 | TVP Regression | | | | Fixed-effect Panel Regression | | | |
|---|---|---|---|---|---|---|---|---|
| | U(1) | U(4) | U(8) | U(12) | U(1) | U(4) | U(8) | U(12) |
| Australia | 1.010 | 1.028 | 1.149 | 1.691 | **0.989*** | **0.938**** | **0.884**** | **0.989** |
| Canada | 1.012 | 1.026 | 1.016 | 1.915 | **0.994** | **0.923*** | **0.958** | 1.264 |
| Denmark | 1.007 | **0.989** | **0.916** | 2.038 | 1.006 | 1.017 | 1.308 | 1.910 |
| UK | **0.974** | **0.880** | **0.742** | **0.747** | 0.972 | 0.847 | 0.702 | 0.665 |
| Japan | 1.002 | **0.952** | 1.067 | 1.124 | 1.008 | 1.076 | 1.239 | 1.315 |
| Korea | 1.005 | 1.042 | 1.845 | 2.379 | **0.978** | **0.901** | **0.845** | **0.823** |
| Norway | 1.003 | **0.984** | 1.039 | 1.457 | **0.998** | **0.951** | 1.063 | 2.044 |
| Sweden | 1.006 | 1.003 | 1.122 | 1.733 | 1.000 | **0.960** | **0.936** | 1.741 |
| Switzerland | 1.011 | 1.033 | 1.235 | 1.124 | 1.023 | 1.161 | 1.685 | 2.130 |
| Euro | 1.003 | 1.057 | 1.197 | 1.610 | 1.001 | **0.984** | 1.098 | 1.517 |
| | | | | | | | | |
| No. of U's <1 | 1 | 4 | 2 | 1 | **5** | **7** | **5** | 3 |
| No. of DM >1.282 | 0 | 0 | 0 | 0 | 1 | 2 | 1 | 0 |
| Median U | 1.006 | 1.015 | 1.095 | 1.650 | 0.999 | 0.955 | 1.011 | 1.416 |

| F3 | TVP Regression | | | | Fixed-effect Panel Regression | | | |
|---|---|---|---|---|---|---|---|---|
| | U(1) | U(4) | U(8) | U(12) | U(1) | U(4) | U(8) | U(12) |
| Australia | 1.014 | 1.047 | 1.228 | 1.653 | 1.002 | **0.993** | 1.013 | 1.085 |
| Canada | 1.020 | 1.052 | 1.207 | 3.014 | 1.047 | 1.195 | 1.912 | 2.377 |
| Denmark | 1.006 | **0.986** | **0.915** | 2.066 | 1.004 | **0.998** | 1.201 | 1.767 |
| UK | **0.974** | **0.846*** | **0.630**** | **0.522**** | 0.975 | 0.844* | 0.627** | 0.518** |
| Japan | 1.001 | **0.969** | 1.071 | 1.150 | 1.009 | 1.105 | 1.301 | 1.387 |
| Korea | 1.003 | 1.023 | 1.845 | 2.431 | **0.968** | **0.866** | **0.825** | **0.932** |
| Norway | 1.002 | **0.986** | 1.036 | 1.927 | **0.998** | **0.955** | 1.053 | 1.942 |
| Sweden | 1.000 | **0.977** | 1.096 | 1.286 | **0.994** | **0.928** | **0.814*** | 1.491 |
| Switzerland | 1.015 | 1.041 | 1.235 | 1.421 | 1.031 | 1.202 | 1.787 | 2.218 |
| Euro | 1.002 | 1.055 | 1.106 | 1.617 | 1.001 | **0.980** | 1.076 | 1.506 |
| | | | | | | | | |
| No. of U's <1 | 2 | **5** | 2 | 1 | 4 | **7** | 3 | 2 |
| No. of DM >1.282 | 0 | 1 | 1 | 1 | 0 | 1 | 2 | 1 |
| Median U | 1.003 | 1.005 | 1.101 | 1.635 | 1.001 | 0.987 | 1.064 | 1.498 |

**Appendix F. Summary Results from TVP Models Estimated via Maximum Likelihood**

The results for the TVP regressions reported in the Paper and in the previous appendices are based in Bayesian methods. This Appendix presents summary results comparable to those in the Paper (see Tables 2-6), but using the Kalman Filter and Maximum Likelihood. Thus, they are interpreted in a similar fashion as in the Paper. In addition it also reports robustness of these results to the change in base currency from the U.S. dollar to the Pound sterling in all models and to the use of other forecasting methods with alternative approaches to estimate Taylor rule fundamentals. For convenience the results from the Fixed-effect panel regression are repeated. To be precise, the tables in this Appendix are as follows:[5]

Tables with summary results by model comparable to tables 2-4 in the main text:
- Table ML1. Forecast Evaluation: Taylor Rules;
- Table ML2. Forecast Evaluation: Monetary Model, PPP and UIRP;
- Table ML3. Forecast Evaluation: Factor Model.

Tables summarising the overall performance of the models across forecast windows and horizons, comparable to tables 5 and 6 in the main text:
- Table ML4. Overall Model's Ability to Outperform the Benchmark across Forecast Windows ;
- Table ML5. Overall Model's Ability to Outperform the Benchmark across Forecast Windows (GBP base currency).

Tables summarising the results from robustness to change in base currency, comparable to tables D.A1, D.A2 and D.A3 in Appendix D:
- Table ML6. Forecast Evaluation: Taylor Rules (GBP base currency);
- Table ML7. Forecast Evaluation: Monetary Model, PPP and UIRP (GBP base currency);
- Table ML8. Forecast Evaluation: Factor Model (GBP base currency).

Table summarising the results from robustness to the use of different forecasting method and regression:
- Table ML9. Forecast Evaluation: Taylor Rules (OLS Rolling Windows).

---

[5] All currency by currency results are excluded to save space.

**Table ML1. Forecast Evaluation: Taylor Rules (USD base currency)**

|  |  | TVP Regression | | | | Fixed-Effect Panel Regression | | | |
|---|---|---|---|---|---|---|---|---|---|
| Fundamentals: |  | h=1 | h=4 | h=8 | h=12 | h=1 | h=4 | h=8 | h=12 |
|  |  | Forecast Window: 1995Q1-1998Q4; N=17 | | | | | | | |
| TRon |  |  |  |  |  |  |  |  |  |
|  | No. of U's <1 | 4 | **9** | **9** | **9** | 4 | 5 | 4 | 7 |
|  | No. of DM >1.282 | 1 | 4 | 5 | 7 | 1 | 1 | 3 | 3 |
|  | Median U | 1.017 | 0.995‡ | 1.000 | 0.989‡ | 1.012 | 1.044 | 1.076 | 1.048 |
| TRos |  |  |  |  |  |  |  |  |  |
|  | No. of U's <1 | 6 | 8 | **10** | 8 | 5 | 7 | **13** | 8 |
|  | No. of DM >1.282 | 1 | 4 | 7 | 5 | 0 | 1 | 6 | 7 |
|  | Median U | 1.014 | 1.018 | 0.938‡ | 1.192 | 1.011 | 1.007 | 0.922‡ | 1.099 |
| TRen |  |  |  |  |  |  |  |  |  |
|  | No. of U's <1 | 7 | 6 | 6 | 5 | 4 | 5 | **9** | 5 |
|  | No. of DM >1.282 | 1 | 2 | 4 | 5 | 0 | 1 | 4 | 2 |
|  | Median U | 1.006 | 1.077 | 1.102 | 1.074 | 1.013 | 1.030 | 0.994‡ | 1.142 |
|  |  | Forecast Window: 1999Q1-2013Q1; N=10 | | | | | | | |
| TRon |  |  |  |  |  |  |  |  |  |
|  | No. of U's <1 | 1 | 0 | 0 | 2 | 0 | 1 | 2 | 2 |
|  | No. of DM >1.282 | 0 | 0 | 0 | 1 | 0 | 0 | 0 | 1 |
|  | Median U | 1.007 | 1.044 | 1.117 | 1.129 | 1.009 | 1.033 | 1.069 | 1.112 |
| TRos |  |  |  |  |  |  |  |  |  |
|  | No. of U's <1 | 1 | 1 | 1 | 2 | 2 | 2 | 2 | 2 |
|  | No. of DM >1.282 | 0 | 0 | 1 | 2 | 0 | 1 | 1 | 1 |
|  | Median U | 1.010 | 1.046 | 1.110 | 1.172 | 1.007 | 1.033 | 1.065 | 1.156 |
| TRen |  |  |  |  |  |  |  |  |  |
|  | No. of U's <1 | 4 | 1 | 0 | 3 | 0 | 1 | 2 | 2 |
|  | No. of DM >1.282 | 0 | 0 | 0 | 2 | 0 | 0 | 1 | 1 |
|  | Median U | 1.004 | 1.048 | 1.120 | 1.161 | 1.010 | 1.040 | 1.088 | 1.159 |
|  |  | Forecast Window: 2007Q1-2013Q1; N=10 | | | | | | | |
| TRon |  |  |  |  |  |  |  |  |  |
|  | No. of U's <1 | 2 | **5** | **5** | **5** | 4 | 4 | **8** | **6** |
|  | No. of DM >1.282 | 0 | 1 | 2 | 3 | 0 | 1 | 2 | 3 |
|  | Median U | 1.008 | 0.998‡ | 0.993‡ | 0.994‡ | 1.003 | 1.003 | 0.984‡ | 0.992‡ |
| TRos |  |  |  |  |  |  |  |  |  |
|  | No. of U's <1 | 2 | 4 | **6** | **6** | 3 | 4 | **6** | **5** |
|  | No. of DM >1.282 | 0 | 0 | 3 | 5 | 0 | 0 | 5 | 4 |
|  | Median U | 1.007 | 1.008 | 0.981‡ | 0.965‡ | 1.010 | 1.001 | 0.945‡ | 0.942‡ |
| TRen |  |  |  |  |  |  |  |  |  |
|  | No. of U's <1 | **5** | 4 | **5** | **5** | **5** | 4 | **5** | 4 |
|  | No. of DM >1.282 | 1 | 1 | 2 | 2 | 1 | 2 | 3 | 3 |
|  | Median U | 1.002 | 1.008 | 1.000 | 1.030 | 0.999‡ | 1.004 | 0.972‡ | 1.052‡ |

Notes: This Table summarises the forecasting performance of the TVP forecasting regression and the Fixed-Effect Panel regression with Taylor rule fundamentals defined as TRon, TRos and TRen. See Table 1 in the main text for details. The benchmark model for both forecasting regressions is the driftless Random Walk (RW). For each regression, fundamental and quarterly horizon ($h$), the "No. of U's < 1" (number of U-statistics less than one), provides the number of currencies for which the model improves upon the RW, since it indicates cases where the RMSFE of the fundamental-based regression is lower than that of the RW. When the U's are less than one for at least half of the currencies in the forecast window, marked in **bold**, then on average, the fundamental-based regression outperforms the benchmark in that window. The "No. of DM > 1.282" (number of DM statistics greater than 1.282) shows cases of rejections of the null hypothesis under the Diebold and Mariano (1995) test of equal forecast accuracy at 10% level of significance. The higher the No. of DM > 1.282, the better the average accuracy of the forecasts of the fundamental-based regression relative to the benchmark. The "Median U" indicates the middle value of the U-statistic across the sample of $N$ currencies for each forecast window and horizon. When "Median U" is less than one, then the fundamental-based regression outperforms the RW benchmark, for more than half of the currencies in the sample - this is marked with the symbol "‡" in the Table.

**Table ML2. Forecast Evaluation: Monetary Model, PPP and UIRP (USD base currency)**

|  |  | TVP Regression | | | | Fixed-Effect Panel Regression | | | |
| --- | --- | --- | --- | --- | --- | --- | --- | --- | --- |
| Fundamentals: |  | h=1 | h=4 | h=8 | h=12 | h=1 | h=4 | h=8 | h=12 |
|  |  | Forecast Window: 1995Q1-1998Q4; N=17 | | | | | | | |
| MM |  |  |  |  |  |  |  |  |  |
|  | No. of U's <1 | **12** | **13** | **15** | **14** | **14** | **15** | **15** | **14** |
|  | No. of DM >1.282 | 4 | 7 | 12 | 11 | 4 | 9 | 12 | 11 |
|  | Median U | 0.983‡ | 0.901‡ | 0.667‡ | 0.561‡ | 0.956‡ | 0.785‡ | 0.552‡ | 0.623‡ |
| PPP |  |  |  |  |  |  |  |  |  |
|  | No. of U's <1 | **15** | **15** | **14** | **14** | **14** | **15** | **15** | **14** |
|  | No. of DM >1.282 | 8 | 9 | 13 | 12 | 9 | 12 | 12 | 11 |
|  | Median U | 0.969‡ | 0.863‡ | 0.761‡ | 0.694‡ | 0.974‡ | 0.866‡ | 0.717‡ | 0.759‡ |
| UIRP |  |  |  |  |  |  |  |  |  |
|  | No. of U's <1 | **10** | **11** | **11** | **10** | **10** | **11** | **11** | **11** |
|  | No. of DM >1.282 | 1 | 7 | 10 | 9 | 1 | 5 | 10 | 11 |
|  | Median U | 0.995‡ | 0.970‡ | 0.927‡ | 0.876‡ | 0.979‡ | 0.969‡ | 0.856‡ | 0.888‡ |
|  |  | Forecast Window: 1999Q1-2013Q1; N=10 | | | | | | | |
| MM |  |  |  |  |  |  |  |  |  |
|  | No. of U's <1 | 1 | 1 | **0** | **0** | 1 | 1 | 1 | 1 |
|  | No. of DM >1.282 | **0** | **0** | **0** | **0** | 0 | 0 | 1 | 1 |
|  | Median U | 1.022 | 1.073 | 1.214 | 1.470 | 1.021 | 1.100 | 1.303 | 1.633 |
| PPP |  |  |  |  |  |  |  |  |  |
|  | No. of U's <1 | **5** | **6** | 1 | 4 | **8** | **8** | **7** | **5** |
|  | No. of DM >1.282 | **0** | **0** | **0** | 2 | 1 | 2 | 3 | 3 |
|  | Median U | 0.999‡ | 0.986‡ | 1.020 | 1.027 | 0.994‡ | 0.985‡ | 0.972‡ | 0.955‡ |
| UIRP |  |  |  |  |  |  |  |  |  |
|  | No. of U's <1 | 1 | 1 | 2 | 2 | 2 | 2 | 2 | 2 |
|  | No. of DM >1.282 | **0** | **0** | 1 | 1 | 0 | 1 | 1 | 1 |
|  | Median U | 1.009 | 1.041 | 1.086 | 1.127 | 1.009 | 1.037 | 1.070 | 1.137 |
|  |  | Forecast Window: 2007Q1-2013Q1; N=10 | | | | | | | |
| MM |  |  |  |  |  |  |  |  |  |
|  | No. of U's <1 | 2 | 1 | 4 | 3 | 2 | 3 | 4 | 3 |
|  | No. of DM >1.282 | **0** | **0** | **0** | 3 | 0 | 0 | 0 | 0 |
|  | Median U | 1.013 | 1.024 | 1.036 | 1.185 | 1.007 | 1.015 | 1.057 | 1.301 |
| PPP |  |  |  |  |  |  |  |  |  |
|  | No. of U's <1 | **8** | **5** | 2 | **5** | **7** | **8** | **6** | 4 |
|  | No. of DM >1.282 | 1 | 1 | 1 | 3 | 3 | 4 | 4 | 4 |
|  | Median U | 0.993‡ | 1.002 | 1.084 | 0.931‡ | 0.991‡ | 0.931‡ | 0.948‡ | 1.260 |
| UIRP |  |  |  |  |  |  |  |  |  |
|  | No. of U's <1 | 3 | 4 | **7** | **5** | 3 | 4 | **5** | 4 |
|  | No. of DM >1.282 | **0** | 1 | 4 | 4 | 0 | 0 | 3 | 2 |
|  | Median U | 1.004 | 1.002 | 0.983‡ | 0.995‡ | 1.012 | 1.016 | 0.990‡ | 1.026 |

Notes: This Table summarises the forecasting performance of the TVP forecasting regression and the Fixed-effect panel regression with Monetary (MM), PPP and UIRP fundamentals. See Table 1 in the main text for details about the form of the forecasting regressions and how fundamentals are computed or estimated. The benchmark model for both forecasting regressions is the driftless Random Walk (RW). For each regression, set of fundamentals, forecast window and quarterly horizon (h), the "No. of U's < 1" (number of U-statistics less than one), provides the number of currencies for which the model improves upon the RW, since it indicates cases where the RMSFE of the fundamental-based regression is lower than that of the RW. When the U's are less than one for at least half of the currencies in the forecast window, marked in **bold**, then on average, the fundamental-based regression outperforms the benchmark. The "No. of DM > 1.282" (number of DM statistics greater than 1.282) shows cases of rejections of the null hypothesis under the Diebold and Mariano (1995) test of equal forecast accuracy at 10% level of significance. The higher the No. of DM > 1.282, the better the average accuracy of the forecasts of the fundamental-based regression relative to the benchmark is. The "Median U" indicates the middle value of the U-statistic across the sample of *N* currencies for each forecast window and horizon. When "Median U" is less than or equal to one - marked with the symbol "‡", and U's are less than one for at least half of the currencies in the window, this is also consistent with a better average forecasting performance of the fundamental-based regression relative to the benchmark.

**Table ML3. Forecast Evaluation: Factor Model (USD base currency)**

|  | TVP Regression | | | | Fixed-Effect Panel Regression | | | |
|---|---|---|---|---|---|---|---|---|
| Fundamentals: | h=1 | h=4 | h=8 | h=12 | h=1 | h=4 | h=8 | h=12 |
| | | | | Forecast Window: 1995Q1-1998Q4 | | | | |
| **F1** | | | | | | | | |
| No. of U's <1 | **9** | **10** | **11** | **11** | 6 | 7 | 8 | 8 |
| No. of DM >1.282 | 5 | 2 | 6 | 5 | 2 | 4 | 3 | 4 |
| Median U | 0.995‡ | 0.976‡ | 0.876‡ | 0.762‡ | 1.035 | 1.061 | 1.032 | 1.091 |
| **F2** | | | | | | | | |
| No. of U's <1 | **10** | **13** | **14** | **12** | **9** | **9** | **9** | 8 |
| No. of DM >1.282 | 7 | 5 | 10 | 7 | 7 | 8 | 7 | 6 |
| Median U | 0.967‡ | 0.913‡ | 0.824‡ | 0.849‡ | 0.985‡ | 0.936‡ | 0.985‡ | 1.073 |
| **F3** | | | | | | | | |
| No. of U's <1 | **11** | **13** | **14** | **12** | **10** | **11** | **10** | **9** |
| No. of DM >1.282 | 6 | 5 | 11 | 7 | 8 | 9 | 7 | 7 |
| Median U | 0.975‡ | 0.918‡ | 0.860‡ | 0.827‡ | 0.980‡ | 0.879‡ | 0.904‡ | 0.951‡ |
| | | | | Forecast Window: 1999Q1-2013Q1 | | | | |
| **F1** | | | | | | | | |
| No. of U's <1 | 3 | **5** | 2 | **5** | 2 | 2 | 2 | 4 |
| No. of DM >1.282 | **0** | **0** | **0** | 4 | 0 | 0 | 0 | 0 |
| Median U | 1.003 | 1.001 | 1.083 | 0.975‡ | 1.010 | 1.048 | 1.153 | 1.209 |
| **F2** | | | | | | | | |
| No. of U's <1 | 4 | 4 | 3 | **6** | 4 | 4 | 4 | **6** |
| No. of DM >1.282 | **0** | **0** | **0** | 4 | 0 | 0 | 0 | 0 |
| Median U | 1.003 | 1.017 | 1.072 | 0.957‡ | 1.005 | 1.016 | 1.012 | 0.987‡ |
| **F3** | | | | | | | | |
| No. of U's <1 | 4 | 4 | 2 | **6** | 2 | 3 | 4 | 4 |
| No. of DM >1.282 | **0** | **0** | **0** | 4 | 0 | 0 | 0 | 1 |
| Median U | 1.002 | 1.012 | 1.065 | 0.940‡ | 1.004 | 1.019 | 1.051 | 1.046 |
| | | | | Forecast Window: 2007Q1-2013Q1 | | | | |
| **F1** | | | | | | | | |
| No. of U's <1 | **5** | 2 | 1 | 4 | 4 | **7** | 2 | 2 |
| No. of DM >1.282 | **0** | 1 | 1 | 3 | 0 | 0 | 1 | 1 |
| Median U | 1.001 | 1.034 | 1.280 | 1.146 | 1.001 | 0.984‡ | 1.139 | 1.682 |
| **F2** | | | | | | | | |
| No. of U's <1 | **5** | 3 | 1 | 4 | **5** | **7** | **5** | 3 |
| No. of DM >1.282 | **0** | 1 | 1 | 1 | 1 | 2 | 1 | 0 |
| Median U | 0.997‡ | 1.035 | 1.241 | 1.049 | 0.999‡ | 0.955‡ | 1.011 | 1.416 |
| **F3** | | | | | | | | |
| No. of U's <1 | **6** | 3 | 0 | 3 | 4 | **7** | 3 | 2 |
| No. of DM >1.282 | **0** | **0** | **0** | 2 | 0 | 1 | 2 | 1 |
| Median U | 0.997‡ | 1.037 | 1.249 | 1.100 | 1.001 | 0.987‡ | 1.064 | 1.498 |

Notes: This Table summarises the forecasting performance of the TVP forecasting regression and the Fixed-effect panel regression with factors (F) extracted from exchange rates. See Table 1 in the main text for details about the form of the forecasting regressions. Factors are obtained via principal component analysis. The benchmark model for both forecasting regressions is the driftless Random Walk (RW). For each regression, set of fundamentals, forecast window and quarterly horizon (h), the "No. of U's < 1" (number of U-statistics less than one), provides the number of currencies for which the model improves upon the RW, since it indicates cases where the RMSFE of the fundamental-based regression is lower than that of the RW. When the U's are less than one for at least half of the currencies in the forecast window, marked in **bold**, then on average, the fundamental-based regression outperforms the benchmark. The "No. of DM > 1.282" (number of DM statistics greater than 1.282) shows cases of rejections of the null hypothesis under the Diebold and Mariano (1995) test of equal forecast accuracy at 10% level of significance. The higher the No. of DM > 1.282, the better the average accuracy of the forecasts of the fundamental-based regression relative to the benchmark is. The "Median U" indicates the middle value of the U-statistic across the sample of $N$ currencies for each forecast window and horizon. When "Median U" is less than or equal to one - marked with the symbol "‡", and U's are less than one for at least half of the currencies in the window, this is also consistent with a better average forecasting performance of the fundamental-based regression relative to the benchmark.

**Table ML4. Overall Model's Ability to Outperform the Benchmark across Forecast Windows (USD base currency)**

|  | TVP Regression | | Fixed-Effect Panel Regression | |
| --- | --- | --- | --- | --- |
|  | Short-Run Forecasts | Long-Run Forecasts | Short-Run Forecasts | Long-Run Forecasts |
| TR | Yes | Yes | No | Yes |
| MM | No | No | No | No |
| PPP | Yes | Yes | Yes | Yes |
| UIRP | No | Yes | No | Yes |
| Factors | Yes | Yes | Yes | Yes |

Notes: This Table summarises the overall performance of the TVP regression and the Fixed-effect Panel regression conditioned on TR, MM, PPP, UIRP or factors (F). Refer to Table 1 in the main text for details about the form of the forecasting regressions and how fundamentals are computed or estimated. The benchmark model for all regressions is the driftless Random Walk (RW). The TVP regression is estimated using the method of Maximum Likelihood. The Table provides the answer to the question: "Does the regression conditioned on any of the fundamentals outperform the benchmark for at least half of the currencies in most forecast windows, at short or long-horizon forecasts?" The short-horizon comprises h=1 or h=4 quarters, while the long-horizon is h=8 or h=12 quarters.

**Table ML5. Overall Model's Ability to Outperform the Benchmark across Forecast Windows (GBP base currency)**

|  | TVP Regression | | Fixed-Effect Panel Regression | |
| --- | --- | --- | --- | --- |
|  | Short-Run Forecasts | Long-Run Forecasts | Short-Run Forecasts | Long-Run Forecasts |
| TR | No | No | No | No |
| MM | Yes | Yes | Yes | Yes |
| PPP | Yes | Yes | Yes | Yes |
| UIRP | No | No | No | No |
| Factors | Yes | Yes | No | Yes |

Notes: This Table summarises the overall performance of the TVP regression and the Fixed-effect Panel regression conditioned on TR, MM, PPP, UIRP or factors (F). Refer to Table 1 in the main text for details about the form of the forecasting regressions and how fundamentals are computed or estimated. The benchmark model for all regressions is the driftless Random Walk (RW). Here, the base currency is the Pound Sterling (GBP) rather than the U.S. dollar and the TVP regression is estimated using the method of Maximum Likelihood. The Table provides the answer to the question: "Does the regression conditioned on the fundamental considered outperform the benchmark for at least half of the currencies in most forecast windows, at short or long-horizon forecasts?" The short-horizon comprises h=1 or h=4 quarters, while the long-horizon is h=8 or h=12 quarters.

**Table ML6. Forecast Evaluation: Taylor Rules (GBP base currency)**

| Fundamentals: | TVP Regression | | | | Fixed-Effect Panel Regression | | | |
|---|---|---|---|---|---|---|---|---|
| | h=1 | h=4 | h=8 | h=12 | h=1 | h=4 | h=8 | h=12 |
| | Forecast Window: 1995Q1-1998Q4; N=17 | | | | | | | |
| TRon | | | | | | | | |
| No. of U's <1 | 3 | 2 | 5 | 4 | 4 | 3 | 3 | 4 |
| No. of DM >1.282 | 1 | 1 | 4 | 4 | 0 | 1 | 2 | 3 |
| Median U | 1.017 | 1.087 | 1.118 | 1.261 | 1.022 | 1.045 | 1.116 | 1.290 |
| TRos | | | | | | | | |
| No. of U's <1 | 5 | 4 | 4 | 5 | 4 | 3 | 3 | 4 |
| No. of DM >1.282 | 1 | 2 | 3 | 3 | 0 | 1 | 2 | 3 |
| Median U | 1.021 | 1.089 | 1.324 | 1.339 | 1.029 | 1.050 | 1.123 | 1.339 |
| TRen | | | | | | | | |
| No. of U's <1 | 4 | 2 | 5 | 3 | 3 | 4 | 3 | 4 |
| No. of DM >1.282 | 1 | 1 | 2 | 3 | 0 | 1 | 2 | 3 |
| Median U | 1.024 | 1.087 | 1.189 | 1.207 | 1.043 | 1.077 | 1.083 | 1.271 |
| | Forecast Window: 1999Q1-2013Q1; N=10 | | | | | | | |
| TRon | | | | | | | | |
| No. of U's <1 | 3 | 3 | 3 | 4 | 2 | 2 | 3 | 4 |
| No. of DM >1.282 | 0 | 1 | 2 | 2 | 0 | 1 | 2 | 3 |
| Median U | 1.001 | 1.022 | 1.058 | 1.057 | 1.008 | 1.024 | 1.034 | 1.043 |
| TRos | | | | | | | | |
| No. of U's <1 | 2 | 2 | 3 | 3 | 2 | 2 | 3 | 4 |
| No. of DM >1.282 | 0 | 0 | 2 | 3 | 0 | 1 | 2 | 3 |
| Median U | 1.006 | 1.018 | 1.058 | 1.056 | 1.004 | 1.016 | 1.022 | 1.045 |
| TRen | | | | | | | | |
| No. of U's <1 | 1 | 3 | 3 | 4 | 2 | 2 | 3 | 4 |
| No. of DM >1.282 | 1 | 0 | 1 | 3 | 0 | 1 | 2 | 3 |
| Median U | 1.007 | 1.022 | 1.043 | 1.063 | 1.005 | 1.022 | 1.025 | 1.046 |
| | Forecast Window: 2007Q1-2013Q1; N=10 | | | | | | | |
| TRon | | | | | | | | |
| No. of U's <1 | **6** | **6** | **5** | **5** | 4 | 4 | 4 | **5** |
| No. of DM >1.282 | 1 | 3 | 4 | 4 | 0 | 3 | 4 | 4 |
| Median U | 1.000 | 0.989‡ | 0.990‡ | 1.001 | 1.004 | 1.014 | 1.009 | 0.995‡ |
| TRos | | | | | | | | |
| No. of U's <1 | 4 | 4 | 4 | **5** | 4 | **6** | **5** | **6** |
| No. of DM >1.282 | 0 | 2 | 4 | 4 | 0 | 3 | 4 | 4 |
| Median U | 1.001 | 1.009 | 1.015 | 1.001 | 1.002 | 0.996‡ | 0.995‡ | 0.990‡ |
| TRen | | | | | | | | |
| No. of U's <1 | 3 | **6** | **5** | **5** | 4 | **5** | **5** | **5** |
| No. of DM >1.282 | 0 | 1 | 4 | 2 | 0 | 3 | 3 | 4 |
| Median U | 1.005 | 0.993‡ | 0.998‡ | 1.016 | 1.003 | 1.006 | 1.003 | 0.993‡ |

Notes: This Table summarises the forecasting performance of the TVP forecasting regression and the Fixed-Effect Panel regression with Taylor rule fundamentals defined as TRon, TRos and TRen. The only difference with Table 2 in the main text is that here the base currency is the Pound sterling (GBP) rather than the USD, and the TVP regression is estimated using the method of Maximum Likelihood. Hence, the interpretation is similar to Table 2 in the paper. For each regression, fundamental and quarterly horizon (*h*), the "No. of U's < 1" (number of U-statistics less than one), provides the number of currencies for which the model improves upon the RW, since it indicates cases where the RMSFE of the fundamental-based regression is lower than that of the RW. When the U's are less than one for at least half of the currencies in the forecast window, marked in **bold**, then on average, the fundamental-based regression outperforms the benchmark. The "No. of DM > 1.282" (number of DM statistics greater than 1.282) shows cases of rejections of the null hypothesis under the Diebold and Mariano (1995) test of equal forecast accuracy at 10% level of significance. The higher the No. of DM > 1.282, the better the average accuracy of the forecasts of the fundamental-based regression relative to the benchmark. The "Median U" indicates the middle value of the U-statistic across the sample of *N* currencies for each forecast window and horizon. When "Median U" is less than one, then the fundamental-based regression outperforms the RW benchmark, for more than half of the currencies in the sample - this is marked with the symbol "‡" in the Table.

**Table ML7. Forecast Evaluation: Monetary Model, PPP and UIRP (GBP base currency)**

| Fundamentals: | TVP Regression | | | | Fixed-Effect Panel Regression | | | |
|---|---|---|---|---|---|---|---|---|
| | h=1 | h=4 | h=8 | h=12 | h=1 | h=4 | h=8 | h=12 |
| | Forecast Window: 1995Q1-1998Q4; N=17 | | | | | | | |
| MM | | | | | | | | |
| No. of U's <1 | 2 | 5 | 4 | 2 | 3 | 3 | 2 | 2 |
| No. of DM >1.282 | 0 | 2 | 2 | 2 | 1 | 1 | 1 | 1 |
| Median U | 1.062 | 1.200 | 1.273 | 1.379 | 1.099 | 1.285 | 1.406 | 1.622 |
| PPP | | | | | | | | |
| No. of U's <1 | **10** | **12** | **9** | 4 | **9** | **10** | **9** | 6 |
| No. of DM >1.282 | 2 | 5 | 6 | 3 | 1 | 4 | 3 | 3 |
| Median U | 0.993‡ | 0.902‡ | 0.997‡ | 1.187 | 0.994‡ | 0.958‡ | 0.997‡ | 1.110 |
| UIRP | | | | | | | | |
| No. of U's <1 | 6 | 4 | 4 | 5 | 5 | 3 | 4 | 4 |
| No. of DM >1.282 | 0 | 2 | 4 | 4 | 0 | 0 | 3 | 4 |
| Median U | 1.013 | 1.063 | 1.136 | 1.169 | 1.020 | 1.079 | 1.120 | 1.177 |
| | Forecast Window: 1999Q1-2013Q1; N=10 | | | | | | | |
| MM | | | | | | | | |
| No. of U's <1 | **7** | 4 | **5** | **5** | **7** | **7** | **7** | **8** |
| No. of DM >1.282 | 0 | 1 | 2 | 4 | 0 | 2 | 3 | 6 |
| Median U | 0.998‡ | 1.007 | 0.984‡ | 1.027 | 0.997‡ | 0.980‡ | 0.939‡ | 0.857‡ |
| PPP | | | | | | | | |
| No. of U's <1 | **6** | **7** | **5** | **8** | **9** | **8** | **6** | **5** |
| No. of DM >1.282 | 0 | 0 | 1 | 3 | 1 | 0 | 0 | 3 |
| Median U | 0.997‡ | 0.974‡ | 1.024 | 0.906‡ | 0.986‡ | 0.975‡ | 0.941‡ | 0.950‡ |
| UIRP | | | | | | | | |
| No. of U's <1 | 2 | 2 | 3 | 4 | 1 | 2 | 3 | 4 |
| No. of DM >1.282 | 0 | 1 | 2 | 3 | 0 | 1 | 2 | 3 |
| Median U | 1.004 | 1.020 | 1.025 | 1.035 | 1.005 | 1.016 | 1.024 | 1.040 |
| | Forecast Window: 2007Q1-2013Q1; N=10 | | | | | | | |
| MM | | | | | | | | |
| No. of U's <1 | **7** | **8** | **7** | **5** | **8** | **9** | **9** | **9** |
| No. of DM >1.282 | 4 | 6 | 6 | 3 | 3 | 5 | 5 | 6 |
| Median U | 0.985‡ | 0.919‡ | 0.908‡ | 1.005 | 0.983‡ | 0.912‡ | 0.850‡ | 0.756‡ |
| PPP | | | | | | | | |
| No. of U's <1 | **8** | **9** | **8** | **8** | **9** | **9** | **9** | **9** |
| No. of DM >1.282 | 3 | 6 | 6 | 6 | 5 | 6 | 8 | 8 |
| Median U | 0.982‡ | 0.877‡ | 0.824‡ | 0.735‡ | 0.971‡ | 0.901‡ | 0.791‡ | 0.657‡ |
| UIRP | | | | | | | | |
| No. of U's <1 | 3 | 4 | 4 | **5** | 4 | **5** | **5** | **5** |
| No. of DM >1.282 | 2 | 3 | 4 | 4 | 0 | 3 | 3 | 4 |
| Median U | 1.002 | 1.007 | 1.009 | 0.999‡ | 1.005 | 0.999‡ | 0.998‡ | 0.996‡ |

Notes: This Table summarises the forecasting performance of the TVP forecasting regression and the Fixed-effect panel regression with Taylor rule fundamentals defined as $TR_{on}$, $TR_{os}$ and $TR_{en}$. The only difference with Table 2 in the main text is that here the base currency is the Pound sterling (GBP) rather than the USD, and the TVP regression is estimated using the method of Maximum Likelihood. Hence, the interpretation is similar to Table 3 in the paper. That is, For each regression, set of fundamentals, forecast window and quarterly horizon (h), the "No. of U's < 1" (number of U-statistics less than one), provides the number of currencies for which the model improves upon the RW, since it indicates cases where the RMSFE of the fundamental-based regression is lower than that of the RW. When the U's are less than one for at least half of the currencies in the forecast window, marked in **bold**, then on average, the fundamental-based regression outperforms the benchmark. The "No. of DM > 1.282" (number of DM statistics greater than 1.282) shows cases of rejections of the null hypothesis under the Diebold and Mariano (1995) test of equal forecast accuracy at 10% level of significance. The higher the No. of DM > 1.282, the better the average accuracy of the forecasts of the fundamental-based regression relative to the benchmark is. The "Median U" indicates the middle value of the U-statistic across the sample of *N* currencies for each forecast window and horizon. When "Median U" is less than or equal to one - marked with the symbol "‡", and U's are less than one for at least half of the currencies in the window, this is also consistent with a better average forecasting performance of the fundamental-based regression relative to the benchmark.

**Table ML8. Forecast Evaluation: Factor Model (GBP base currency)**

| Fundamentals: | TVP Regression | | | | Fixed-Effect Panel Regression | | | |
|---|---|---|---|---|---|---|---|---|
| | h=1 | h=4 | h=8 | h=12 | h=1 | h=4 | h=8 | h=12 |
| | Forecast Window: 1995Q1-1998Q4; N=17 | | | | | | | |
| F1 | | | | | | | | |
| No. of U's <1 | 6 | 6 | 4 | 4 | 3 | 3 | 4 | 4 |
| No. of DM >1.282 | 1 | 3 | 3 | 3 | 1 | 1 | 4 | 4 |
| Median U | 1.012 | 1.059 | 1.088 | 1.248 | 1.010 | 1.085 | 1.189 | 1.287 |
| F2 | | | | | | | | |
| No. of U's <1 | **10** | 7 | 5 | 3 | **9** | **9** | **9** | 6 |
| No. of DM >1.282 | 3 | 1 | 3 | 2 | 6 | 4 | 2 | 2 |
| Median U | 0.997‡ | 1.019 | 1.030 | 1.217 | 0.997‡ | 0.986‡ | 0.989‡ | 1.022 |
| F3 | | | | | | | | |
| No. of U's <1 | 5 | 7 | 5 | 3 | 8 | 8 | **9** | **10** |
| No. of DM >1.282 | 3 | 4 | 4 | 1 | 4 | 7 | 6 | 4 |
| Median U | 1.026 | 1.019 | 1.142 | 1.214 | 1.002 | 1.006 | 0.939‡ | 0.956‡ |
| | Forecast Window: 1999Q1-2013Q1; N=10 | | | | | | | |
| F1 | | | | | | | | |
| No. of U's <1 | 3 | 4 | 2 | 2 | 2 | 3 | 3 | 2 |
| No. of DM >1.282 | 0 | 0 | 1 | 2 | 0 | 2 | 2 | 2 |
| Median U | 1.004 | 1.010 | 1.043 | 1.119 | 1.006 | 1.023 | 1.051 | 1.060 |
| F2 | | | | | | | | |
| No. of U's <1 | 2 | 4 | **6** | **6** | 0 | 1 | 3 | **5** |
| No. of DM >1.282 | 0 | 0 | 2 | 4 | 0 | 0 | 0 | 2 |
| Median U | 1.006 | 1.012 | 0.988‡ | 0.969‡ | 1.005 | 1.020 | 1.022 | 1.001 |
| F3 | | | | | | | | |
| No. of U's <1 | 2 | 4 | 4 | **8** | 3 | 2 | 4 | 4 |
| No. of DM >1.282 | 0 | 1 | 2 | 4 | 0 | 0 | 0 | 0 |
| Median U | 1.006 | 1.012 | 1.017 | 0.955‡ | 1.005 | 1.027 | 1.042 | 1.022 |
| | Forecast Window: 2007Q1-2013Q1; N=10 | | | | | | | |
| F1 | | | | | | | | |
| No. of U's <1 | 3 | **6** | 4 | 2 | 4 | 4 | 4 | **6** |
| No. of DM >1.282 | 1 | 2 | 3 | 2 | 1 | 4 | 4 | 3 |
| Median U | 1.003 | 0.994‡ | 1.015 | 1.077 | 1.004 | 1.009 | 1.003 | 0.980‡ |
| F2 | | | | | | | | |
| No. of U's <1 | 2 | **6** | **8** | 6 | 2 | 3 | 4 | 4 |
| No. of DM >1.282 | 1 | 4 | 4 | 4 | 1 | 1 | 2 | 1 |
| Median U | 1.005 | 0.978‡ | 0.955‡ | 0.991‡ | 1.007 | 1.026 | 1.037 | 1.020 |
| F3 | | | | | | | | |
| No. of U's <1 | 3 | **7** | **7** | **7** | 3 | 3 | 4 | 4 |
| No. of DM >1.282 | 0 | 4 | 5 | 6 | 0 | 2 | 3 | 3 |
| Median U | 1.006 | 0.980‡ | 0.925‡ | 0.925‡ | 1.009 | 1.042 | 1.077 | 1.057 |

Notes: This Table summarises the forecasting performance of the TVP forecasting regression and the Fixed-effect panel regression with Monetary (MM), PPP and UIRP fundamentals. The formatting and interpretation is similar to Table 3 in the main text, but here the base currency is the Pound sterling (GBP) rather than the USD, and the TVP regression is estimated using the method of Maximum Likelihood. Hence, the interpretation is similar to Table 3 in the paper – also repeated in notes to Table D.A1 in this Appendix for convenience.

**Table ML9. Forecast Evaluation: Taylor Rules (OLS Rolling Windows)**

| | TVP Regression | | | | Constant-Parameter Regression (OLS) | | | |
|---|---|---|---|---|---|---|---|---|
| Fundamentals: | h=1 | h=4 | h=8 | h=12 | h=1 | h=4 | h=8 | h=12 |
| | \multicolumn{8}{c}{Forecast Window: 1995Q1-1998Q4; N=17} | | | | | | | |
| **TRon** | | | | | | | | |
| No. of U's <1 | 4 | **9** | **9** | 9 | 7 | 8 | 8 | **9** |
| No. of DM >1.282 | 1 | 4 | 5 | 7 | 1 | 2 | 4 | 5 |
| Median U | 1.017 | 0.995‡ | 1.000 | 0.989‡ | 1.005 | 1.003 | 1.003 | 0.996‡ |
| **TRos** | | | | | | | | |
| No. of U's <1 | 6 | 8 | **10** | 8 | 6 | 8 | **11** | **9** |
| No. of DM >1.282 | 1 | 4 | 7 | 5 | 1 | 2 | 6 | 7 |
| Median U | 1.014 | 1.018 | 0.938‡ | 1.192 | 1.009 | 1.002 | 0.951‡ | 0.907‡ |
| **TRen** | | | | | | | | |
| No. of U's <1 | 7 | 6 | 6 | 5 | 4 | 4 | 7 | 8 |
| No. of DM >1.282 | 1 | 2 | 4 | 5 | 0 | 0 | 5 | 4 |
| Median U | 1.006 | 1.077 | 1.102 | 1.074 | 1.050 | 1.151 | 1.154 | 1.240 |
| | \multicolumn{8}{c}{Forecast Window: 1999Q1-2013Q1; N=10} | | | | | | | |
| **TRon** | | | | | | | | |
| No. of U's <1 | 1 | 0 | 0 | 2 | **5** | 2 | 1 | 3 |
| No. of DM >1.282 | 0 | 0 | 0 | 1 | 0 | 0 | 1 | 1 |
| Median U | 1.007 | 1.044 | 1.117 | 1.129 | 1.000 | 1.032 | 1.085 | 1.148 |
| **TRos** | | | | | | | | |
| No. of U's <1 | 1 | 1 | 1 | 2 | 1 | 4 | 4 | 1 |
| No. of DM >1.282 | 0 | 0 | 1 | 2 | 0 | 0 | 1 | 1 |
| Median U | 1.010 | 1.046 | 1.110 | 1.172 | 1.005 | 1.023 | 1.055 | 1.141 |
| **TRen** | | | | | | | | |
| No. of U's <1 | 4 | 1 | 0 | 3 | 2 | 3 | 4 | 4 |
| No. of DM >1.282 | 0 | 0 | 0 | 2 | 0 | 0 | 0 | 1 |
| Median U | 1.004 | 1.048 | 1.120 | 1.161 | 1.011 | 1.020 | 1.031 | 1.050 |
| | \multicolumn{8}{c}{Forecast Window: 2007Q1-2013Q1; N=10} | | | | | | | |
| **TRon** | | | | | | | | |
| No. of U's <1 | 2 | **5** | **5** | 5 | 4 | 3 | **5** | 4 |
| No. of DM >1.282 | 0 | 1 | 2 | 3 | 0 | 0 | 2 | 3 |
| Median U | 1.008 | 0.998‡ | 0.993‡ | 0.994‡ | 1.006 | 1.008 | 0.961‡ | 1.110 |
| **TRos** | | | | | | | | |
| No. of U's <1 | 2 | 4 | **6** | **6** | 1 | 4 | **8** | 5 |
| No. of DM >1.282 | 0 | 0 | 3 | 5 | 0 | 1 | 3 | 4 |
| Median U | 1.007 | 1.008 | 0.981‡ | 0.965‡ | 1.010 | 1.006 | 0.967‡ | 1.016 |
| **TRen** | | | | | | | | |
| No. of U's <1 | **5** | 4 | 5 | 5 | 4 | **5** | 7 | 5 |
| No. of DM >1.282 | 1 | 1 | 2 | 2 | 0 | 1 | 4 | 4 |
| Median U | 1.002 | 1.008 | 1.000 | 1.030 | 1.001 | 1.015 | 0.933‡ | 1.026 |

Notes: This Table summarises the forecasting performance of the TVP forecasting regression and the Fixed-effect panel regression with factors (F) from exchange rates. The formatting and interpretation is similar to Table 4 in the main text, except that here the base currency is the Pound sterling (GBP) rather than the USD, and the TVP regression is estimated using the method of Maximum Likelihood. Therefore, the interpretation is similar to Table 4 in the paper – also repeated in notes to Table ML1 in this Appendix for convenience.